\documentclass[aps,pra,reprint,superscriptaddress]{revtex4-1}
\usepackage{amsmath}
\usepackage{amssymb}
\usepackage{mathrsfs}
\usepackage{graphicx}
\usepackage[caption=false]{subfig}
\usepackage{enumitem}
\usepackage{color}
\usepackage[percent]{overpic}
\usepackage{bm}
\usepackage{rotating}
\usepackage{hyperref}
\usepackage{cancel}
\usepackage{verbatim}
\usepackage{mathtools}
\usepackage{graphicx}
\usepackage{tensor}
\usepackage{MnSymbol}
\graphicspath{ {images/} }

\newcommand{\ket}[1]{{\left| {#1} \right>}}
\newcommand{\bra}[1]{{\left< {#1} \right|}}
\newcommand{\ii}{\mathrm{i}}
\newcommand{\TrAk}{\text{Tr}_{\text{A}_k}}
\newcommand{\TrAl}{\text{Tr}_{\text{A}_\text{l}}}
\newcommand{\Var}{\text{Var}}
\renewcommand{\d}{\mathrm{d}}

\newcommand{\be}{\begin{equation}}
\newcommand{\bel}[1]{\begin{equation}\label{#1}}
\newcommand{\ee}{\end{equation}}

\begin{document}

\title{Open dynamics under rapid repeated interaction}
\author{Daniel Grimmer}
\email{dgrimmer@uwaterloo.ca}
\affiliation{Institute for Quantum Computing, University of Waterloo, Waterloo, ON, N2L 3G1, Canada}
\affiliation{Dept. Physics and Astronomy, University of Waterloo, Waterloo, ON, N2L 3G1, Canada}
\author{David Layden}
\email{dlayden@uwaterloo.ca}
\affiliation{Institute for Quantum Computing, University of Waterloo, Waterloo, ON, N2L 3G1, Canada}
\affiliation{Dept. Applied Math., University of Waterloo, Waterloo, ON, N2L 3G1, Canada}
\author{Robert B. Mann}
\email{rbmann@uwaterloo.ca}
\affiliation{Dept. Physics and Astronomy, University of Waterloo, Waterloo, ON, N2L 3G1, Canada}
\affiliation{Institute for Quantum Computing, University of Waterloo, Waterloo, ON, N2L 3G1, Canada}
\affiliation{Perimeter Institute for Theoretical Physics, Waterloo, ON, N2L 2Y5, Canada}

\author{Eduardo Mart\'{i}n-Mart\'{i}nez}
\email{emartinmartinez@uwaterloo.ca}
\affiliation{Institute for Quantum Computing, University of Waterloo, Waterloo, ON, N2L 3G1, Canada}
\affiliation{Dept. Applied Math., University of Waterloo, Waterloo, ON, N2L 3G1, Canada}
\affiliation{Perimeter Institute for Theoretical Physics, Waterloo, ON, N2L 2Y5, Canada}

\begin{abstract}
We investigate the emergent open dynamics of a quantum system that undergoes rapid repeated unitary interactions with a sequence of ancillary systems. We study in detail how decoherence appears as a subleading effect when a quantum system is ‘bombarded’ by a quick succession of ancillas. In the most general case, these ancillas are a) taken from an ensemble of quantum systems of different dimensions, b) prepared in different states, and c) interacting with the system through different Hamiltonians. We derive an upper bound on decoherence rates in this regime, and show how a rich variety of phenomena in open dynamics (such as projection, thermalization, purification, and dephasing) can emerge out of our general model of repeated interaction. Furthermore, we show a fundamental link between the strength of the leading order dissipation and the intrinsic ``unpredictability'' in the system-ancilla interaction. We also discuss how these results encompass and extend results obtained with other earlier models of repeated interaction.
\end{abstract}

\maketitle
\section{Introduction}
The dynamics of open quantum systems, i.e., of systems interacting with an environment, forms the basis of numerous active areas of research. For instance, understanding and controlling open quantum systems is essential to the development of quantum technologies, such as quantum computers, simulators, and sensors, in which the decohering effects of the environment must be suppressed (or, in some schemes, exploited). Open dynamics are also central to more foundational questions such as the quantum measurement problem: whether the formalism of projective measurements must be postulated or whether it emerges naturally from the complex interplay of a quantum system, a detector, and an environment.

The generic picture of open quantum dynamics involves a quantum system coupled to an environment, which is treated quantum-mechanically. Together, the system and environment are typically assumed to evolve unitarily. Due to our inability to directly characterize the environment, we must average over its possible states in order to describe the dynamics of the system. This averaging typically yields non-unitary system dynamics, which can display a multitude of complex features not found in closed systems; for instance, non-Markovianity, i.e., evolution which is non-local in time due to memory effects in the environment. While the nature of the environment and its coupling to the system are typically unknown in practice, it can be illuminating to examine particular models for environments which couple in a simple manner to the system. 

One such model of open quantum dynamics involves an environment that is composed of numerous independent ancillas which interact with the system one at a time, in succession. In other words, the model consists of a system which couples to a single ancilla for some fixed time, $\delta t$, after which the ancilla is discarded and the system proceeds to interact with a fresh ancilla. This setup, known as a \textit{Repeated Interaction System} or a \textit{Collision Model} \cite{Attal:2006, Attal:2007, Attal:2007b, Vargas:2008, Giovannetti:2012}, has received significant attention of late, as it describes non-trivial open dynamics while remaining relatively amenable to analytical treatment. In particular, it has provided insight into very diverse phenomena, ranging from decoherence and related effects \cite{Scarani:2002, Ziman:2002, Ziman:2005, Ziman:2005b} and quantum thermodynamics \cite{Bruneau:2006, Bruneau:2008, Bruneau:2008b, Karevski:2009, Bruneau:2014, Bruneau:2014b, Hanson:2015} to the measurement problem (due to its close relation with the quantum Zeno effect, see \cite{Layden:2015} and references therein) and even gravitational decoherence \cite{Kafri:2014zsa,Kafri:2015iha, CM}. 

In this paper, we focus on a Repeated Interaction System where the duration of each system-ancilla interaction, $\delta t$, is small as compared to the characteristic timescales of the overall dynamics. Moreover, we allow the coupling strength between the system and each ancilla to be finite and non-vanishing. This regime is of timely relevance, as several authors have recently observed \cite{Zanardi:2014, Zanardi:2015, Layden:2015b} in it a surprising phenomenon: namely, that the reduced dynamics of the system is unitary to leading order in $\delta t$---in sharp contrast to the typical behavior of open quantum systems. What is more, this leading-order unitary system dynamics does not generally commute with the free dynamics of the system. In particular, in \cite{Layden:2015b}, it was shown how this phenomenon could yield rapid quantum control, while Zanardi and Campos Venuti analyzed its emergence in great generality \cite{Zanardi:2014, Zanardi:2015}, and also characterized the subleading-order dynamics of the system in this regime \cite{Zanardi:2016}. In this paper, we build upon the formalism of \cite{Layden:2015b}, generalizing it in three main ways: (i) we allow for different types of ancillas (which may have different dimensions) randomly picked from a known distribution in each cycle, (ii) we consider system-ancilla couplings which can vary randomly from one cycle to the next, and (iii) we describe the resulting system dynamics to all orders in $\delta t$, focusing on the leading order dissipative effects.
In this context, we note concurrent work \cite{ACMZ}
investigating the types of dynamics that can emerge from a model of repeated interactions of a system with a set of ancillas. 
Our results are commensurate with these, albeit from a different perspective.  Whereas the emphasis in Ref.\ \cite{ACMZ} is on the continuum limit and the general considerations under which an effective classical interaction emerges between two subsystems (thinking of gravity-inspired decoherence models), we are  concerned with understanding how thermalization, purification, and dephasing can emerge from finite-frequency repeated interaction effects. 

This paper is divided into five sections: Section \ref{Model} introduces the model and Section \ref{Master_Eq} describes the resulting reduced dynamics of the system. Section \ref{relationship} discusses the relation between the present work and previous results, while Section \ref{error_analysis} contains a detailed error analysis of the approximations employed. Finally, Section \ref{Qubits} presents illustrative examples, in terms of two-level systems and ancillas, of the results developed in previous section.

\section{Model}\label{Model}
We seek to model the effective dynamics of a quantum system which undergoes repeated interactions with a series of ancillary systems. If the ancillas are identical, a very simple example of such a setup could be repeated interaction of the system with a probe which is reset between repetitions. Another example could be the repeated bombardment of a quantum system by a medium of identical constituents.

In a general setup, the ancillas that are repeatedly interacting with the system will not be identical. Therefore, we would like to include the possibility that the target quantum system repeatedly interacts with a variety of ancillary systems allowing the nature and state of the ancillas, as well as the form and strength of the interaction, to change with every repetition.  

With such a general setup in mind, we assume there are different types of ancillas which interact with the target system differently. We assign a distinct label $k$ to every individual type of ancilla. We assume the $k$-th type of ancilla interacts with the target system unitarily for a fixed amount of time $\delta t$.

In \cite{Layden:2015b}, it was assumed that all the ancillas and interactions were identical in every repetition. Here we generalize that setup by assuming that at every repetition of the system-ancilla interaction, with probability $p_k$, the system interacts with an ancilla of the $k$-th type.

In an interaction with a type $k$ ancilla, the system, S, encounters an ancilla, $\text{A}_k$, in the state $\rho_{\text{A}_k}$. The joint system then evolves under a Hamiltonian of the form,
\bel{HamForm0}
H_{k,\delta t}(t)
=H_k(t/\delta t)
\ee
for a fixed time $\delta t$, where we have normalized the time dependence of each cycle for mathematical convenience. Note, we also assume that these Hamiltonians are independent of $\delta t$ except through the ratio $t/\delta t$.

Thus, after this interaction, the joint system has evolved by the unitary operator which is generated by this Hamiltonian:
\bel{Udef}
U_{k,\delta t}(t)
\coloneqq
\mathcal{T}\exp\Big(\int_0^t \d\tau \, H_{k,\delta t}(\tau)\Big).
\ee
Following the interaction, the ancilla is discarded. If the system, S, is hit by a single ancilla of the $k$-th type then S would get updated by the map
\bel{phikdef}
\phi_k(\delta t)[\rho_\text{S}]
\coloneqq
\TrAk\Big(U_{k,\delta t}(\delta t)(\rho_\text{S}\otimes \rho_{\text{A}_k}) U_{k,\delta t}^\dagger(\delta t)\Big).
\ee
However, as stated above, the ancilla which the system interacts with is selected randomly from an ensemble of different types of ancillas. In any interaction, the probability that S interacts with an ancilla of the type $k$ is $p_k$, so the system is effectively updated by the averaged map given by 
\begin{align}\label{phibardef}
\bar{\phi}(\delta t)[\rho_\text{S}]
&\coloneqq\sum_k p_k \, \phi_k(\delta t)[\rho_\text{S}]\\
\nonumber &=\sum_k p_k \, \TrAk\Big(U_{k,\delta t}(\delta t)(\rho_\text{S}\otimes \rho_{\text{A}_k}) U_{k,\delta t}^\dagger(\delta t)\Big).
\end{align}
For example, if all of the ancillas are in the same state (thus, $\text{A}_k=\text{A}$ and $\rho_{\text{A}_k}=\rho_\text{A}$) but there are still several type of Hamiltonians that the joint system could evolve under, then the effective update map for a single interaction is
\begin{align}
\bar{\phi}(\delta t)[\rho_\text{S}]
\!=\!\text{Tr}_\text{A}\!\Big(\!\sum_k p_k \,
U_{k,\delta t}(\delta t)(\rho_\text{S}\otimes \rho_\text{A}) U_{k,\delta t}^\dagger(\delta t)\Big).
\end{align}
This is the partial trace of a mixed-unitary channel acting on the joint state.

Another example is if all of the ancillas live in the same Hilbert space ($\text{A}_k=\text{A}$) and all the Hamiltonians associated with these ancillas are identical (so that $U_{k,\delta t}(\delta t)=U_{\delta t}(\delta t)$) then the effective update map for a single interaction becomes
\begin{align}\label{Stinespring}
\bar{\phi}(\delta t)[\rho_\text{S}]
&=\text{Tr}_\text{A}\Big(
U_{\delta t}(\delta t)(\rho_\text{S}\otimes \sum_k p_k \, \rho_{\text{A}_k}) U_{\delta t}^\dagger(\delta t)\Big)\\
\nonumber
&=\text{Tr}_\text{A}\Big(
U_{\delta t}(\delta t)(\rho_\text{S}\otimes \tilde{\rho}_\text{A}) U_{\delta t}^\dagger(\delta t)\Big).
\end{align}
This is just an interaction with an ancilla known to be in the state $\tilde{\rho}_\text{A}=\sum_k p_k \, \rho_{\text{A}_k}$. By the Stinespring dilation theorem \cite{Stinespring:1955}, any quantum channel acting on the system Hilbert space can be written in this form for suitable choices of $\tilde{\rho}_\text{A}$ and $U_{\delta t}(\delta t)$. Therefore $\bar{\phi}(\delta t)$ can in general be any channel acting on the system.

In our setup, the target system, S, repeatedly interacts with ancillas so that its state is updated over many interactions by repeated application of $\bar{\phi}(\delta t)$. Thus, if the system is initially in the state $\rho_\text{S}(0)$ and undergoes $n$ interactions then the system will be in the state
\begin{align}\label{discreteintmastereqs}
\rho_\text{S}(n \, \delta t)
=\bar{\phi}(\delta t)^n\big[\rho_\text{S}(0)\big] .
\end{align}
Thus, only knowing the discrete state update operator $\bar{\phi}\big(\delta t\big)$ and the initial system state $\rho_\text{S}(0)$, we know the state of the system at the discrete time points $t=n \, \delta t$. For times, $\tau$, between $n \, \delta t$ and $(n+1) \, \delta t$ we can interpolate the state of the system as
\bel{intmastereqs}
\rho_\text{S}(\tau)
=\Omega_{\delta t}(\tau)\big[\rho_\text{S}(0)\big],
\ee
for some interpolation scheme $\Omega_{\delta t}(\tau)$ with
\bel{matchcondition}
\Omega_{\delta t}(n \, \delta t)=
\bar{\phi}(\delta t)^n,
\ee
such that the interpolated dynamics exactly matches the discrete dynamics after each interactions.

In general, there is no unique choice of such an interpolation scheme. 
However, as we will show, if we restrict our attention to Markovian interpolation schemes then we find a unique solution to \eqref{matchcondition}.

The most general form for the time evolution operator of Markovian dynamics is
\bel{Markform}
\Omega_{\delta t}(t)=\exp(t \, \mathcal{L}_{\delta t}), 
\ee
where $\mathcal{L}_{\delta t}$ is a time independent Liouvillian superoperator.

Noting that since $\Omega_{\delta t}(n \, \delta t)
=\big(\Omega_{\delta t}(\delta t)\big)^n$,
any Markovian interpolation scheme which satisfies \eqref{matchcondition} after the first interaction ($n=1$) will automatically satisfy \eqref{matchcondition} after every interaction (i.e., for every $n$). Thus, we need only require
\bel{Markcond}
\Omega_{\delta t}(\delta t)
=\exp(\delta t \, \mathcal{L}_{\delta t})
=\bar{\phi}(\delta t).
\ee
Given $\bar{\phi}(\delta t)$, \eqref{Markcond} has a unique solution for $\mathcal{L}_{\delta t}$ (upon choosing a branch).  We can formally solve \eqref{Markcond} to find this unique (up to choosing a branch cut) effective Liouvillian as
\bel{Ldef}
\mathcal{L}_{\delta t}
\coloneqq\frac{1}{\delta t}\log\big(\bar{\phi}(\delta t)\big).
\ee
This effective Liouvillian plays the role of the generator of time translations for the  interpolation scheme \eqref{Markform}. Thus we have the master equation,
\begin{align}\label{mastereqs}
\frac{\d}{\d t}\rho_\text{S}(t)
=\mathcal{L}_{\delta t}[\rho_\text{S}(t)].
\end{align}

We are particularly interested in repeated interactions in the regime of rapid successive interactions. To this end, in the rest of this section, we expand the master equation \eqref{mastereqs} as a formal series in $\delta t$. 

As discussed above, for each type of interaction, $k$, the corresponding Hamiltonian, $H_k(t/\delta t)$, generates a unitary evolution operator, $U_{k,\delta t}(t)$ by which the joint system evolves during the interaction. Thus, at the end of an interaction the joint system has evolved by the unitary, $U_{k,\delta t}(\delta t)$. For small enough $\delta t$, we can formally take a Dyson expansion of this unitary operator yielding the following power series in $\delta t$:
\begin{align}\label{Ukexpand}
U_{k,\delta t}(\delta t)
\nonumber&=\mathcal{T}\exp\Big(\int_0^{\delta t} \d \tau\, H_{k,\delta t}(\tau)\Big)\\
&=U_{k,0}
+\delta t \ U_{k,1}
+\delta t^2 \ U_{k,2}
+\dots \, ,
\end{align}
where $U_{k,0}=\boldsymbol{1}$ and for $n\geq1$ we have,
\bel{Ukseriesdef}
U_{k,n}\!\coloneqq\!
\frac{1}{(\ii \hbar)^n}\!\!
\int_{0}^{1} \!\! d\xi_1 \!\!
\int_{0}^{\xi_1} \!\! d\xi_2 \!
\dots \!
\int_{0}^{\xi_{n-1}} \!\!\!\! d\xi_n
\prod_{j=1}^{n}
H_k(\xi_j).
\ee

Next, using \eqref{Ukexpand}, we can expand the discrete single interaction update map \eqref{phibardef} as a series in $\delta t$ as:
\begin{align}\label{phibarexpand}
\bar{\phi}(\delta t)[\,\cdot\,]
\nonumber
&=\sum_k  \, p_k \,
\text{Tr}_{\text{A}_k}\Big(
U_{k,\delta t}(\delta t)(\,\cdot\,\otimes \rho_{\text{A}_k}) U_{k,\delta t}(\delta t)^\dagger\Big)\\
&=\bar{\phi}_{0}[\,\cdot\,]
+\delta t \ \bar{\phi}_{1}[\,\cdot\,]
+\delta t^2 \ \bar{\phi}_{2}[\,\cdot\,]
+\dots \, ,
\end{align}
where in Appendix \ref{AppSeries} we show,
\bel{phibarseriesdef}
\bar{\phi}_n[\,\cdot\,]
=\sum_k \, p_k \, \TrAk\Big(
\sum_{m=0}^{n}U_{k,m}(\,\cdot\,\otimes\rho_{\text{A}_k})U_{k,n-m}{}^\dagger\Big).
\ee
To small orders in $\delta t$ we have,
\begin{align}
\bar{\phi}_0[\,\cdot\,]
&=\boldsymbol{1},\\
\bar{\phi}_1[\,\cdot\,]
\nonumber&=
\sum_k \, p_k \, 
\TrAk\Big(U_{k,1}(\,\cdot\,\otimes\rho_{\text{A}_k})
+(\,\cdot\,\otimes\rho_{\text{A}_k})U_{k,1}{}^\dagger\Big),\\
\bar{\phi}_2[\,\cdot\,]
\nonumber
&=\sum_k \, p_k \, 
\TrAk\Big(U_{k,2}(\,\cdot\,\otimes\rho_{\text{A}_k})
+(\,\cdot\,\otimes\rho_{\text{A}_k})U_{k,2}{}^\dagger\\
\nonumber
&\quad+U_{k,1}(\,\cdot\,\otimes\rho_{\text{A}_k})U_{k,1}{}^\dagger\Big).
\end{align}
Finally, using the the expansion \eqref{phibarexpand}, we can expand the effective Liouvillian defined in \eqref{Ldef} as a series in $\delta t$ as,
\begin{align}\label{Lseriesdef}
\mathcal{L}_{\delta t}
\nonumber&=\frac{1}{\delta t}\log\big(\bar{\phi}(\delta t)\big)\\
&=\mathcal{L}_0+\delta t \, \mathcal{L}_1+\delta t^2\mathcal{L}_2+\dots \, ,
\end{align}
where in Appendix \ref{AppSeries} we find a recursive definition for $\mathcal{L}_{M}$ for all $M$. For small orders of $\delta t$ we find
\begin{align}
\label{L0def}
\mathcal{L}_0&=\bar{\phi}_1,\\
\label{L1def}
\mathcal{L}_1&=\bar{\phi}_2-\bar{\phi}_1{}^2/2,\\
\label{L2def}
\mathcal{L}_2&=\bar{\phi}_{3}
-\{\bar{\phi}_1,\bar{\phi}_{2}\}/2
+\bar{\phi}_1{}^3/3.    
\end{align}
Therefore, the master equation \eqref{mastereqs} associated with the Markovian interpolation scheme \eqref{Markform} is expanded as a series in $\delta t$ as,
\begin{align}\label{mastereqsseries}
\frac{\text{d}}{\text{dt}}\rho_\text{S}(t)
&=\mathcal{L}_{\delta t}[\rho_\text{S}(t)]\\
\nonumber&=\mathcal{L}_0[\rho_\text{S}(t)]
+\delta t \, \mathcal{L}_1[\rho_\text{S}(t)]
+\delta t^2\mathcal{L}_2[\rho_\text{S}(t)]
+\dots \, .
\end{align}

\section{Explicit Master Equation}\label{Master_Eq}
The most general system-ancilla Hamiltonian whose free components are time-independent can be written as,
\bel{Hamform}
H_k(\xi)
=H_\text{S}\otimes\boldsymbol{1}
+\boldsymbol{1}\otimes H_{\text{A}_k}
+H_{\text{SA}_k}(\xi);
\ee
that is, as the sum of two local time-independent terms and a possibly time-dependent interaction term.

Note that as stated in \eqref{HamForm0}, we only consider time dependence through the dimensionless ratio $\xi=t/\delta t$. Additionally, we assume that except through this ratio the Hamiltonians are $\delta t$ independent. Taking this form for the system-ancilla Hamiltonians into the series \eqref{Ukseriesdef} and  \eqref{phibarseriesdef} we can find the explicit form for the coefficient maps \eqref{L0def} and \eqref{L1def}. We give the form of these coefficients as well as their interpretations in the following subsections.

\subsection{Zeroth Order Effective Liouvillian}\label{H0section}
As it was first shown in \cite{Layden:2015b}, and as can be seen in Appendix \ref{AppL0}, to zeroth order the evolution is entirely unitary. Namely,
\bel{L0explicit}
\mathcal{L}_0[\,\cdot\,]
=\frac{-\ii}{\hbar}
[H_\text{eff}^{(0)},\,\cdot\,],
\ee
where the effective Hamiltonian $H_\text{eff}^{(0)}$ is given by
\bel{H0effdef}
H_\text{eff}^{(0)}
\coloneqq
H_S+H^{(0)}.
\ee 
That is, the free system Hamiltonian plus a new contribution to the dynamics coming from the repeated interactions. Specifically, $H^{(0)}$ takes the following form:
\bel{H0def}
H^{(0)}
\coloneqq
\sum_k p_k \ 
\Big\langle G_0\big(
H_{\text{SA}_k}(\xi)
\big)\Big\rangle_k,
\ee
where
\bel{G0def} 
G_0(H(\xi))\coloneqq
\int_0^1 H(\xi) \, d\xi \, ,
\ee
and
\bel{langlerangledef}
\langle X\rangle_k
\coloneqq
\TrAk(\rho_{\text{A}_k} X).
\ee
In other words, the contribution to the dynamics from the repeated interaction results from $H_{\text{SA}_k}$ averaged over the duration of the interaction, then traced out over the ancillas' subspaces (weighted by each ancilla's state $\rho_{\text{A}_k}$), and finally taking into account all possible choices for interaction type with their respective weights $p_k$.

This result of unitary evolution is perhaps surprising given that in a general interaction the target system and the ancilla will become entangled and therefore the target system will become mixed once the ancilla is traced out. This may seem at odds with the fact that unitary evolution cannot map pure states to mixed states. However \eqref{L0explicit} shows that the system and ancilla do not become entangled to leading order in $\delta t$: the ancillas do push the system but if $\delta t$ is small enough the ancillas \textit{do not have time} to entangle with it \cite{Layden:2015b}. Of course, this apparent contradiction is resolved at higher orders in $\delta t$ where dissipative effects emerge, as we will see in Section \ref{Dsection}.

\subsection{First Order Effective Louvillian}\label{L1section}
In this paper, we extend the result of \cite{Layden:2015b} to include the dissipative effects coming from the fact that the interaction generates entanglement between the system and the ancillas as well as from the uncertainty about which particular type of ancilla was chosen from the statistical ensemble. To do so we compute the first order contribution in the effective dynamics \eqref{mastereqsseries}. As we show in Appendix \ref{AppL1}, the first order contribution to the effective system dynamics includes both a subleading Hamiltonian correction as well as the leading order dissipative effects:
\bel{L1explicit}
\mathcal{L}_1[\,\cdot\,]
=\frac{-\ii}{\hbar}
[H_\text{eff}^{(1)},\,\cdot\,]
+\frac{1}{2}\mathcal{D}[\,\cdot\,].
\ee
Notice that if the nontrivial part of the zeroth order unitary dynamics, \eqref{H0def}, vanishes (for example if $\TrAk(H_{\text{SA}_k} \rho_{\text{A}_k})=0$ for every $k$), then the first non-trivial contribution to the system's effective generator appears only at first order in $\delta t$. In this case, the leading-order dynamics would contain a dissipative component.

\subsubsection{First Order Hamiltonian Dynamics}\label{H1section}
As we show in Appendix \ref{AppL1}, the effective Hamiltonian coming from the first order dynamics can be written as the sum of three terms, 
\bel{H1effdef}
H_\text{eff}^{(1)}
=H_1^{(1)}+H_2^{(1)}+H_3^{(1)},
\ee
where
\begin{align}
\label{H1def}
H_1^{(1)}
&\coloneqq
\sum_k p_k \ 
\Big\langle G_1\Big(
\frac{-\ii}{\hbar}
[H_{\text{SA}_k}(\xi),H_\text{S}]
\Big)\Big\rangle_k \, ,\\
\label{H2def}
H_2^{(1)}
&\coloneqq
\sum_k p_k \ 
\Big\langle G_2\Big(
\frac{-\ii}{\hbar}
[H_{\text{SA}_k}(\xi),H_{\text{A}_k}]
\Big)\Big\rangle_k \, ,\\
\label{H3def}
H_3^{(1)}
&\coloneqq
\sum_k p_k \ 
\Big\langle G_3\Big(
\frac{-\ii}{\hbar}
[H_{\text{SA}_k}(\xi_1),
H_{\text{SA}_k}(\xi_2)]
\Big)\Big\rangle_k \, ,
\end{align}
with
\begin{align}
\label{G1def}
G_1\big(X(\xi)\big)
&\coloneqq
\int_0^1 \ 
(\xi-1/2) \, X(\xi) \,
d\xi \, ,\\
\label{G2def}
G_2\big(X(\xi)\big)
&\coloneqq
\int_0^1 \ 
\xi \, X(\xi) \, 
d\xi \, ,\\
\label{G3def}
G_3\big(Y(\xi_1,\xi_2)\big)
&\coloneqq
\frac{1}{2}
\int_{0}^1 d\xi_1
\int_{0}^{\xi_1} d\xi_2 \ Y(\xi_1,\xi_2).
\end{align}
As with the zeroth order Hamiltonian \eqref{H0def}, these contributions consist of: a) integrating over time; b) tracing over the ancillas' subspaces weighted by the ancillas' states $\rho_{\text{A}_k}$; and c) averaging over all possible choices of ancillas with weights $p_k$.

One can gain some insight into the physical meaning of $H_1^{(1)}$, $H_2^{(1)}$, and $H_3^{(1)}$ by thinking of the commutator $(\ii\hbar)^{-1}[H_{\text{SA}_k},X]$ as the time evolution of the operator $X$ with respects to the Hamiltonian $H_{\text{SA}_k}$ as in a sort of interaction picture. 
Notice that, in contrast to \cite{Layden:2015b}, we allow for the interaction Hamiltonian to not commute with itself at different times which leads to nontrivial effects in the first order dynamics.

\subsubsection{First Order Dissipative Dynamics}\label{Dsection}
Unlike in zeroth order Liouvillian \eqref{L0explicit}, the first order Liouvillian \eqref{L1explicit} contains  dissipative effects. As we show in Appendix \ref{AppL1}, the dissipative part of \eqref{L1explicit} takes the form,
\begin{align}\label{Ddef}
&\mathcal{D}[\,\cdot\,]
=\frac{1}{\hbar^2}
\big[H^{(0)},[H^{(0)},\,\cdot\,]\big]\\
\nonumber
&-\frac{1}{\hbar^2}\sum_k p_k \, \TrAk \Big(
\big[G_0(H_{\text{SA}_k}),[G_0(H_{\text{SA}_k}),
\,\cdot\,\otimes\rho_{\text{A}_k}]\big]\Big),
\end{align}
where $H^{(0)}$ and $G_0$ are defined in \eqref{H0def} and \eqref{G0def} respectively.

Unlike the unitary part of the dynamics, which preserve the purity of S, the dissipative terms can in general either increase or decrease the system's purity. For any Markovian master equations a necessary and sufficient condition for strictly non-increasing purity is $\mathcal{D}[I]=0$ \cite{Lidar2006, Gorini:1976}. In general, we do not have $\mathcal{D}[I]=0$. Thus rapid repeated interactions with ancillas may generally either increase or decrease the purity of the bombarded system. 


To help in the interpretation of the leading order dissipative term, we define the superoperator $C_k$ which acts on system-ancilla states of the form $\rho_{\text{SA}_k}$ as
\bel{Kdef}
C_k[\rho_{\text{SA}_k}]\coloneqq
\frac{-\ii}{\hbar}[G_0(H_{\text{SA}_k}),\rho_{\text{SA}_k}].
\ee
Namely, it takes as input a system-ancilla density matrix and returns the time-evolved density matrix through the corresponding time averaged interaction Hamltonian $G_0(H_{\text{SA}_k})$. Using \eqref{Kdef} we can rewrite \eqref{Ddef} as
\bel{DdefVar}
\mathcal{D}[\rho_\text{S}]
=\sum_k p_k \, \TrAk \big(\Var(C_k)[\rho_\text{S}\otimes\rho_{\text{A}_k}]\big),
\ee
where the ``variance'' $\Var(C_k)$ is defined as
\be
\Var(C_k)\coloneqq
\llangle C_k{}^2\rrangle-\llangle C_k\rrangle^2.
\ee
We use $\llangle C_k\rrangle$ to denote the application of the superoperator \bel{Avgdef}
\llangle C_k\rrangle[\rho_{\text{SA}_k}]
=\rho_{\text{A}_k}\otimes\sum_l p_l \, \TrAl \big(C_l[\rho_{\text{SA}_l}]\big).
\ee
In other words, the ``average'' $\llangle C_k\rrangle$ takes as input a system-ancilla density matrix, applies the corresponding $C_k$, takes the partial trace over the $\text{A}_k$ ancillary systems, performs a weighted sum over all types of ancillas, and then attaches a fresh ancilla to the resulting reduced state.

We show in Appendix \ref{AppL1} that we can interpret this ``variance'' as how much, on average, a superoperator $C_k$ does not commute with the averaged superoperator $\llangle C_k\rrangle$ since
\begin{align}\label{Varinterpret}
\Var(C_k)
=\Big\llangle\big[
\llangle C_k\rrangle,C_k
\big]\Big\rrangle.
\end{align}
Note that even if there is only one type of ancilla (that is, if $p_k=\delta_{k,1}$) we do not have 
$\Var(C_k)=0$ because even then $C_1\neq\llangle C_1\rrangle$, as can be readily seen from \eqref{Avgdef}. Thus the dissipation includes contributions from our uncertainty about the ancilla state, as well as from which ancilla was chosen from a statistical ensemble, and what type of interaction the system underwent.

In sections \ref{Lindblad},
\ref{Product}, and \ref{Qubits} we see explicit examples of this interpretation of the leading order dissipative effects.

\subsection{First Order Master Equation}\label{FirstOrderMasterEqs}
From \eqref{mastereqs}, \eqref{L0explicit} and \eqref{L1explicit} we can write the interpolated master equation as
\begin{align}\label{truncatedmastereqs}
\frac{\text{d}}{\text{d}t}\rho_\text{S}(t)
\nonumber=&\mathcal{L}_0[\rho_\text{S}(t)]
+\delta t \ \mathcal{L}_1[\rho_\text{S}(t)]\\
=&\frac{-\ii}{\hbar}[
H_\text{eff}^{(0)}
+\delta t \ H_\text{eff}^{(1)}
,\rho_\text{S}(t)]
+ \frac{\delta t}{2}
\mathcal{D}[\rho_\text{S}(t)]\\
\nonumber&
+\mathcal{O}(E^2\delta t^2/\hbar^2),
\end{align}
where $E$ is the largest energy scale relavant to our dynamics. For example if the system and ancillas are qubits then these energy scales would be the free energy of the qubits $\hbar\omega_\text{S}, \hbar\omega_{\text{A}_k}$ and the coupling strengths $\hbar J_k$ so that $E=\hbar \,  \text{max}\{\omega_\text{S},\omega_{\text{A}_k},J_k\}$. For the rest of the paper we truncate the dynamics at this order.  

\subsection{Lindblad Form}\label{Lindblad}

For any Markovian master equation, we can find operators $\mathcal{H},F_n$ and positive numbers $\Gamma_n$ such that:
\bel{LindbladForm}
\frac{\text{d}}{\text{d}t}\rho_\text{S}(t) 
=\frac{-\ii}{\hbar}[\mathcal{H},\rho_\text{S}(t)]
+\sum_n \Gamma_n \,  L(F_n)[\rho_\text{S}(t)],
\ee
where $\mathcal{H}$ is a self-adjoint operator and 
\be
L(X)[\rho]=X\rho X^\dagger-X^\dagger X\rho/2-\rho X^\dagger X/2.
\ee
A master equation written in the form
\eqref{LindbladForm} is said to be in Lindblad form \cite{Lindblad}.
$L(X)[\rho]$ is called the Lindblad superoperator, $\Gamma_n$ are called decoherence rates and $F_n$ are called decoherence modes.

From \eqref{truncatedmastereqs}, we can directly see that
\be 
\mathcal{H}
=H_\text{eff}^{(0)}
+\delta t \ H_\text{eff}^{(1)}
+\mathcal{O}(\delta t^2),
\ee
where the dissipative part of the dynamics \eqref{Ddef} can be written in the form of equation \eqref{LindbladForm} as
\begin{align}\label{DLindblad}
\mathcal{D}[\rho_\text{S}]
\nonumber& =\frac{2}{\hbar^2}
\sum_k\sum_{\alpha_k\neq\beta_k}
q_{(k,\alpha_k)} \ 
L\Big(\bra{\beta_k}G_0(H_{\text{SA}_k})\ket{\alpha_k}\Big)[\rho_\text{S}]\\
& +\frac{2}{\hbar^2}
\sum_{m=1}^M \gamma_m \  L(A_m)[\rho_\text{S}],
\end{align}
as shown in Appendix \ref{AppLindblad}.

In writing \eqref{DLindblad} we have made use of  several definitions. First, we have written the spectral decomposition of each $\rho_{\text{A}_k}$ as
\be
\rho_{\text{A}_k}=\sum_{\alpha_k}\lambda_{\alpha_k}\ket{\alpha_k}\bra{\alpha_k}.
\ee
Second, we defined the probability vector $\bm{q}$ with dimension $N=\sum_k \text{dim}(\text{A}_k)$ and components $q_{(k,\alpha_k)}\coloneqq p_k\lambda_{\alpha_k}$; this is the probability of meeting a type $k$ ancilla and finding it to be in the pure state $\ket{\alpha_k}$.

We then defined the $N\times N$ matrix $\mathrm{\bm{Q}}$ with components 
$Q_{ij}\coloneqq q_i\delta_{ij}-q_i q_j$. We interpret this matrix as the symmetric bilinear form associated with the variance, since associating a vector of outcomes $\bm{X}$ to the the probability vector $\bm{q}$ we see that the variance of $\bm{X}$ is $\text{var}(\bm{X})=\bm{X}^T\mathrm{\bm{Q}}\bm{X}$. This is an example of the fundamental relationship between the leading order dissipation and the uncertainty about the ancillary systems claimed in Section \ref{Dsection}.

Noting that this matrix is real and symmetric, we have denoted its eigenvalues and eigenvectors as $\gamma_m$ and $\bm{v}_m$ respectively, where the components of the $m$-th eigenvector are $v_{m,(k,\alpha_k)}$. In Appendix \ref{AppLindblad}, we show the eigenvalues are real, non-negative, and bounded as $0\leq\gamma_m\leq1$.

Finally, we define
\bel{Amdef} 
A_m
\coloneqq
\sum_{k,\alpha_k} v_{m,(k,\alpha_k)}\bra{\alpha_k}G_0(H_{\text{SA}_k})\ket{\alpha_k}.
\ee
Thus the $A_m$ decoherence modes are linear combinations of the diagonal blocks of all of the interaction Hamiltonians with respects to some eigenvector of $\mathrm{\bm{Q}}$. Note that since the diagonal blocks of a Hermitian matrix are Hermitian, the $A_m$ decoherence modes are Hermitian. Additionally, from \eqref{DLindblad}, we see the decoherence rates of these modes are proportional to their corresponding eigenvalue of $\mathrm{\bm{Q}}$.

Further inspection of  \eqref{DLindblad} indicates the existence of  another type of decoherence mode: each off-diagonal block of every interaction Hamiltonian, $H_{\text{SA}_k}$, appears as a decoherence mode. The decoherence rates of these modes are proportional to the probability that the ancilla is of type $k$ and is found in state $\ket{\alpha_k}$. 

We note that this is not the canonical form of the Lindblad equation. As mentioned above, the decoherence modes in \eqref{DLindblad} are formed from subblocks of the interaction Hamiltonians. In general, these will not be orthogonal to each other or traceless, however, for any particular choice of the $H_{\text{SA}_k}$'s the canonical Lindblad form can be found from \eqref{DLindblad} by diagonalizing the modes against each other and then rediagonalizing the Lindblad form. 

\subsubsection{Decoherence rates}\label{DecoherenceRate}
In this section, we derive an upper bound for all the decoherence rates associated with the leading order dissipation.

The decoherence rate for the $n$-th mode in \eqref{DLindblad}, including the $\delta t/2$ factor outside of $\mathcal{D}$ in \eqref{truncatedmastereqs}, is 
\be
\Gamma_n
=\frac{\delta t}{\hbar^2}P_n E_n{}^2,
\ee
where $E_n$ is the energy scale associated with the $n$-th decoherence mode and $P_n$ is either one of the $q_i$ probabilities or an eigenvalue of $\mathrm{\bm{Q}}$, $\gamma_m\in[0,1]$.

As mentioned above, the decoherence modes in \eqref{DLindblad} are in general not orthogonal, and so we expect them to interfere with each other. We would have the fastest decoherence rate if all of the decoherence modes constructively interfered. The most extreme this interference can be is if all the decoherence modes are identical, in which case the decoherence rate is
\be
\Gamma_\text{max}
\coloneqq\sum_n\Gamma_n
=\frac{\delta t}{\hbar^2}\sum_n P_n \ E_n{}^2,
\ee
which we can upper-bound as
\be
\Gamma_\text{max}
\leq\frac{\delta t \ E^2}{\hbar^2}\sum_n P_n,
\ee
where $E=\text{max}(E_n)$.  We can compute $\sum_n P_n$ by separating the $P_n$'s into those which come from the $q_{(k,\alpha_k)}$ modes and those which come from the $\gamma_m$ modes. Doing this we find
\begin{align}
\sum_n P_n
&=\sum_k\sum_{\alpha_k\neq\beta_k}
q_{(k,\alpha_k)}
+\sum_{m=1}^M \gamma_m\\
&\nonumber
=\sum_k \sum_{\alpha_k}
q_{(k,\alpha_k)}(M_k-1)
+\text{Tr} \ Q\\
&\nonumber
=\sum_k p_k 
\sum_{\alpha_k}\lambda_{\alpha_k}
(M_k-1)
+\sum_{i=1}^M q_i-q_i{}^2\\
&\nonumber
=\sum_k p_k 
(M_k-1)
\sum_{\alpha_k}\lambda_{\alpha_k}
+(1-\vert\bm{q}\vert^2)\\
&\nonumber
=\langle M_k\rangle-1
+(1-\vert\bm{q}\vert^2)\\
&\nonumber
=\langle M_k\rangle-\vert\bm{q}\vert^2,
\end{align}
where $M_k=\text{dim}(\text{A}_k)$ is the dimensionality of the $k$-th type of ancilla and $\langle M_k\rangle=\sum_k p_k M_k$ is the average effective dimension of the ancillas. Thus we have a bound on any decoherence rate, $\Gamma$, as
\bel{sumGammabound}
\Gamma
\leq\Gamma_\text{max}
\leq\frac{\delta t \, E^2}{\hbar^2}
\big(\langle M_k\rangle-\vert\bm{q}\vert^2\big)
+\mathcal{O}(\delta t^2).
\ee
Thus, at leading order, longer, stronger interactions with higher dimensional ancillas can cause the fastest decoherence. On the other hand, rapid, weak interactions with a single type of pure qubit cannot cause fast decoherence.

Noting that the $\vert\bm{q}\vert^2$ factor is bounded by $1$ we see that this upper bound scales as $\delta t  \, E^2  \, \langle M_k\rangle$. We also note that since the bound on the rate of decoherence scales linearly with $\delta t$, the bound on the decoherence due to a single interaction of duration $\delta t$ scales quadratically as $\delta t^2 E^2 \langle M_k\rangle$. Specifically, this means that for small $\delta t$, two interactions of a duration $\delta t/2$ can at most produce half the decoherence that is possible with one interaction of length $\delta t$.

Using \eqref{sumGammabound} we can lower-bound the coherence time $\tau$ associated with any decoherence channel as
\begin{align}
\tau\geq
\tau_\text{min}
=\frac{1}{\Gamma_\text{max}}
&\geq\frac{\hbar^2}{\delta t \ E^2 \ (\langle M_k\rangle-\vert\bm{q}\vert^2)}
+\mathcal{O}(\delta t^0).
\end{align}

\subsection{Example: simple time dependent switching function}\label{Product}
To see an explicit example of \eqref{truncatedmastereqs}, let us consider a single type of interaction with a single type of ancilla. Namely, we choose the interaction Hamiltonian to be $H_{\text{SA}}(\xi)=g(\xi)J_\text{S}\otimes J_\text{A}$. That is, we couple an observable of the system, $J_\text{S}$, with an observable of the ancilla, $J_\text{A}$, via a time dependent switching function. As we show in Appendix \ref{AppProduct}, from \eqref{H0effdef}, \eqref{H1effdef}, and \eqref{Ddef} we then have
\begin{align}
\label{H0effdefproduct}
H_\text{eff}^{(0)}
&=H_\text{S}
+g_0 \, \langle J_\text{A}\rangle J_\text{S} \, ,\\
\label{H1effdefproduct}
H_\text{eff}^{(1)}
&= g_1 \langle J_\text{A}\rangle \, \frac{-\ii}{\hbar}[J_\text{S},H_\text{S}]
+g_2\frac{-\ii}{\hbar}\big\langle[J_\text{A},H_\text{A}]\big\rangle J_\text{S} \, ,\\
\label{Ddefproduct}
\mathcal{D}[\rho_\text{S}]
&=-\frac{g_0{}^2}{\hbar^2}
\Delta_{J_\text{A}}^2
\big[J_\text{S},[J_\text{S},\rho_\text{S}]\big] \, ,
\end{align}
where $H_\text{S}$ and $H_\text{A}$ are the free system and ancilla Hamiltonians respectively as in \eqref{Hamform};
\begin{align}
g_0
&=\int_0^1\ g(\xi) \, d\xi \, ,\\
g_1
&=\int_0^1 \ (\xi-1/2) \, g(\xi) \, d\xi \, ,\\
g_2
&=\int_0^1 \ \xi \, g(\xi) \, d\xi,
\end{align}
are weighted averages of the switching function; $\langle J_\text{A}\rangle$ is the average with respects to the initial ancilla state $\rho_{\text{A}_k}$ as defined in \eqref{langlerangledef}, and $\Delta_{J_\text{A}}^2
=\langle J_\text{A}{}^2\rangle-\langle J_\text{A}\rangle^2$ is the variance of the ancilla observable $J_\text{A}$. Note that the weighted averages of the switching function are not independent since $g_2=g_1+g_0/2$.

We note that the dissipative term \eqref{Ddefproduct} is already written in Lindblad form as defined in \eqref{LindbladForm}. We see that there is only one decoherence mode $J_\text{S}$ with a decoherence rate
\be
\Gamma_{J_\text{S}}=\frac{g_0{}^2}{\hbar^2}\delta t \, \Delta_{J_\text{A}}^2 \, \vert J_\text{S}\vert^2,
\ee
where $\vert J_\text{S}\vert$ is some scale associated with $J_\text{S}$. 

We can again see the connection between dissipation and variance discussed in Section \ref{Dsection}. Here we see that the decoherence rate is proportional to $\Delta_{J_\text{A}}^2$, the variance of the ancilla observable $J_\text{A}$. Thus the rate of dissipation is exactly related to the amount of information we are ignoring by tracking only the system.

We will see specific applications of this form of interaction in Sections \ref{CavesMilburn}, \ref{ZZQubits}, and \ref{XXQubits}.

\section{Relationship with other repeated interaction models}
\label{relationship}

In this section we situate the present results within the existing literature.
\subsection{Caves-Milburn Repeated Measurement Model}\label{CavesMilburn}
Caves and Milburn \cite{CM} considered a system S with one degree of freedom. Let $X_\text{S}$ be its position and $P_\text{S}$ its canonically conjugate momentum. They investigated the dynamics of such a system when its position is repeatedly measured through an interaction with a series of identically prepared probe systems, A, where each individual system-probe Hamiltonian is
\bel{CMHam}
H_{\delta t}(t)
=H_\text{S}\otimes\boldsymbol{1}
+\delta(t-\delta t)X_\text{S}\otimes P_\text{A},
\ee
where $P_\text{A}$ is the probe's linear momentum operator. The probe's position is assumed to be in a stationary pure Gaussian distribution. The system evolves under its free Hamiltonian $H_\text{S}$ except for a series of ``delta kicks'' at times $t=n\delta t$. The probes are assumed to have no free dynamics, so they do not evolve, except at the precise time their respective ``delta kick''. 

The unitary for a single interaction is
\bel{UCM}
U_{\delta t}(\delta t)
=e^{\ii X_S P_A/\hbar} \ 
e^{\ii H_S \delta t/\hbar}.
\ee
The effect of the interaction on each probe's position is roughly a translation in space by $\langle X_\text{S}\rangle$, thus making a rough record of the system position. Since each probe is assumed to be stationary, $\langle P_\text{A} \rangle=0$, the effect of its interaction with the system is roughly to widen the system's momentum distribution $\langle P_\text{S}{}^2\rangle$ by a quantity $\langle P_\text{A}{}^2\rangle=\hbar^2/2\sigma$, where $\sigma$ is the spatial spread of the probe's state.

Critically, in the limit of $\delta t\to0$ the impact of a single interaction, \eqref{UCM}, does not vanish. This is because the average coupling strength in \eqref{CMHam} strengthens as the interaction shortens. Therefore the limit of continuous measurement in this model ($\delta t\to0$), as stated, is singular. In order to overcome this singularity, the authors of \cite{CM} weaken the effect of each interaction by taking $\sigma$ to get larger as $\delta t$ goes to zero. This implies that the positions of the probes become more uncertain as they scan the system more frequently. Therefore, as we approach the limit, the probe states are no longer perfect indicators of the system's position, making the ``measurement'' weak.

In the limits $\delta t\to0$ and $\sigma\to\infty$ keeping their product constant (i.e. $\delta t \ \sigma=\text{const.}$), the master equation for the system becomes \cite{CM}
\bel{CMeqs}
\frac{\text{d}}{\text{dt}}\rho_\text{S}(t)
=\frac{-\ii}{\hbar}[H_\text{S},\rho_\text{S}(t)]
-\frac{1}{4 \, \delta t \ \sigma} \ \big[X_\text{S},[X_\text{S},\rho_\text{S}(t)]\big].
\ee
Comparing \eqref{CMeqs} with \eqref{truncatedmastereqs}, \eqref{H0effdefproduct}, \eqref{H1effdefproduct}, and \eqref{Ddefproduct}, one may wonder, in what way the weak measurement formalism of \cite{CM} is similar to the repeated interaction scheme developed in this paper. At first sight the key difference between the two scenarios is that in our setting the coupling strength is finite, and remains so even in the limit of infinitely frequent interactions. We can nevertheless recover the dynamics in \eqref{CMeqs} as a particular case of our general formalism by taking an interaction with a single type of ancilla of the form described in Section \ref{Product} and introducing a time scale $\tau$ controlling the strength of the interaction in the following way:
\bel{CMHamMod}
H_{\delta t}(t)
=H_\text{S}\otimes\boldsymbol{1}
+\delta(t-\delta t) \, \frac{\delta t}{\tau} X_\text{S}\otimes P_\text{A}.
\ee
The inclusion of the $\delta t/\tau$ factor weakens the interaction as $\delta t\to0$. 

As in \eqref{CMHam}, the system's position $X_\text{S}$ is coupled to the ancillas momentum $P_\text{A}$ through a delta coupling. The ancilla has no free dynamics ($H_\text{A}=0$) and is in a stationary pure spatial Gaussian distribution so that $\langle P_\text{A}\rangle=0$ and $\langle P_\text{A}{}^2\rangle=\hbar^2/2\sigma$ where $\sigma$ is the spatial spread of the probe's state. 

Using these assumptions about the ancilla states we greatly simplify the master equation. Specifically, we see that the zeroth order Hamiltonian  \eqref{H0effdefproduct} reduces to the system Hamiltonian since $\langle P_\text{A}\rangle=0$. Furthermore the first order Hamiltonian contribution to the dynamics \eqref{H1effdefproduct} vanishes since $\langle P_\text{A}\rangle=0$ and $H_\text{A}=0$. Thus all of the nontrivial contributions to the unitary dynamics have disappeared.

Finally the dissipative part \eqref{Ddefproduct} reduces to
\be
-\frac{\delta t}{4\tau^2\sigma}
\big[X_\text{S},[X_\text{S},\rho_\text{S}(t)]\big]
\ee
since $g_0=1$, and $\Delta_{P_\text{A}}^2=\hbar^2/2\sigma$. Again, as discussed in Section \ref{Dsection}, we see that the rate of dissipation is proportional to the uncertainty of the relevant ancilla observable, here momentum.

With all these particularizations considered, the master equation \eqref{truncatedmastereqs} becomes
\bel{OurCM}
\frac{\d}{\d t}\rho_\text{S}(t)
=\frac{-\ii}{\hbar}[H_\text{S},\rho_\text{S}(t)]
-\frac{\delta t}{4\tau^2\sigma}
\big[X_\text{S},[X_\text{S},\rho_\text{S}(t)]\big].
\ee
We note \eqref{OurCM} is of the same form as \eqref{CMeqs}. If we make the interaction strength parameter scale as $\tau\sim\delta t$, mimicking the delta strength in \eqref{CMHam}, then the dissipative prefactor becomes $(4\delta t\sigma)^{-1}$, reproducing exact the result by Caves and Milburn in \cite{CM} as a particular case of the general equation \eqref{truncatedmastereqs}.

\subsection{Zanardi-Campos Venuti dissipation-projected dynamics}

In a series of recent papers \cite{Zanardi:2014, Zanardi:2015, Zanardi:2016}, Zanardi et al.\ examined the effects of adiabatic control in strongly dissipative quantum systems. In particular, they considered a system which evolves according to the Lindblad equation
\begin{equation}
\frac{\mathrm{d}}{\mathrm{d}t} \rho (t)
=
\big( \mathscr{L}_0 + \mathscr{L}_1 \big)
\rho(t),
\label{eq:Zanardi_EOM}
\end{equation}
where $\mathscr{L}_0$ is the system's free Liouvillian and $\mathscr{L}_1$ is a perturbation which plays the role of a control/driving term. The authors examined in particular the dynamics of system states in $\text{ker}[\mathscr{L}_0]$ (i.e., states which remain fixed during evolution by $\mathscr{L}_0$) under \eqref{eq:Zanardi_EOM}. In the special case where $\mathscr{L}_1$ vanishes with time (as $\mathcal{O}(1/t)$ in \cite{Zanardi:2014, Zanardi:2015} and as $\mathcal{O}(1/\sqrt{t})$ in \cite{Zanardi:2016}), they found that the dynamics within $\text{ker}[\mathscr{L}_0]$ were well described by an effective Liouvillian $\mathcal{P}_0 \mathscr{L}_1 \mathcal{P}_0$ up to an error that vanishes as $t \to \infty$, where $\mathcal{P}_0$ is a projector onto $\text{ker}[\mathscr{L}_0]$. Notably, when $\mathscr{L}_1 = -\frac{\mathrm{i}}{\hbar}[H, \, \cdot \;]$ for $H = H^\dagger$, i.e., when the driving is Hamiltonian, this effective dynamics within $\text{ker}[\mathscr{L}_0]$ is unitary \cite{Albert:2015}.

While \cite{Zanardi:2014, Zanardi:2015, Zanardi:2016} are not directly concerned with repeated interaction systems, the highly general results they present can be specialized to describe a quantum system undergoing a series of identical interactions with a succession of identical ancillas. In particular, the authors consider as an example the case where the system is bipartite and composed of a ``small system" S as well as a bath B. They suppose that $\mathscr{L}_0$ acts trivially on S but describes a strongly dissipative dynamics on B with a unique steady-state $\rho_\mathrm{B}$. The overall generator $\mathscr{L}_0 + \mathscr{L}_1$ then describes a dynamics wherein S evolves non-trivially while B feels a strong constant pull towards the state $\rho_\textrm{B}$. Intuitively, this scenario resembles that of a repeated interaction system in which S interacts identically with a sequence of ancillas, each prepared in the state $\rho_\textrm{B}$; the difference being that B is drawn to $\rho_\mathrm{B}$ via dissipation here, rather than being periodically replaced with a fresh copy.

This intuitive connection is formalized in \cite{Zanardi:2015}, where the authors show that the effective dynamics within $\text{ker}[\mathscr{L}_0]$ is well described by $(e^{\frac{t}{n}\mathscr{L}_1 } \mathcal{P}_0)^n$ as $n \to \infty$. In this example, $\text{ker}[\mathscr{L}_0]$ comprises all states of the form $\rho \otimes \rho_\mathrm{B}$, where $\rho$ is an arbitrary state of S. Correspondingly, the projection $\mathcal{P}_0$ is chosen to be $\mathcal{P}_0 = \text{Tr}_\mathrm{B} ( \, \cdot \, ) \otimes \rho_\mathrm{B}$, which in the context of repeated interaction systems, describes the process of replacing a used ancilla with a fresh one in the state $\rho_\mathrm{B}$. Thus, this salient example demonstrates that a repeated interaction system (i) whose total Hamiltonian scales as $\mathcal{O}(1/t)$ (ii) interacting with identical ancillas via (iii) identical interaction Hamiltonians, evolves as $\mathcal{P}_0 \mathscr{L}_1 \mathcal{P}_0$ when the cycles are repeated at an infinite frequency, up to an error that vanishes when $t \to \infty$. Moreover, the effective generator $\mathcal{P}_0 \mathscr{L}_1 \mathcal{P}_0$ has the same form as $H_\text{eff}^{(0)}$ in Eq.~\eqref{H0effdef} when the latter is specialized to the case of identical ancillas and trivial time dependence.

In \cite{Zanardi:2016} the authors extend their analysis to describe the subleading-order effective dynamics within $\text{ker}[\mathscr{L}_0]$ under \eqref{eq:Zanardi_EOM}. In particular, they solve for the effective generator in the case where $\mathcal{P}_0 \mathscr{L}_1 \mathcal{P}_0 = 0$, and find it to scale as $|| \mathscr{L}_1 ||^2$. Whereas their leading-order effective generator describes a unitary dynamics when $\mathscr{L}_1$ is Hamiltonian, their subleading-order generator contains both Hamiltonian and dissipative terms; in agreement with the behavior displayed by $\mathcal{L}_0$ and $\mathcal{L}_1$ [see Eq.~\eqref{Ldef}] in the present work.

The results developed in \cite{Zanardi:2014, Zanardi:2015, Zanardi:2016} are noteworthy both for their great generality and for their timely relevance to a number of areas in the field of quantum information. However, the focus of these papers is rather different than the present one. In particular, Zanardi et al.\ were concerned with very general driven-dissipative systems where $\text{dim}(\text{ker}[\mathscr{L}_0]) > 1$ and $|| \mathscr{L}_1|| \to 0$ at an appropriate rate as $t \to \infty$ (in the spirit of the adiabatic theorem). In contrast, we focus here on a specific physical setting: that of a repeated interaction system. However, we analyze aspects of this setting which have no analogue in \cite{Zanardi:2014, Zanardi:2015, Zanardi:2016}: for instance, the effects of non-identical ancillas and non-identical interaction Hamiltonians, effective dynamics beyond subleading order (i.e., $\mathcal{L}_2, \mathcal{L}_3, \dots$), and mid-cycle non-Markovian effects (see Section \ref{sec:stroboscopic_error}). Moreover, we do not limit ourselves to cases in which the overall system-ancilla Hamiltonian vanishes with large $t$. Given the parallels between the present work and that of Zanardi et al., we believe that our results are complementary to theirs, and that they may provide insights that lead to extensions of \cite{Zanardi:2014, Zanardi:2015, Zanardi:2016}; for instance, in describing the effective dynamics within $\text{ker}[\mathscr{L}_0]$ to higher orders.

\subsection{Other Investigations of Repeated Interaction Systems}

There exist a number of other studies on repeated interaction systems in a variety of contexts. Among them:
\begin{itemize}
\item Refs.\ \cite{Attal:2006, Attal:2007, Attal:2007b, Vargas:2008, Giovannetti:2012} study the mathematical underpinnings of such models, with a focus on ancillas which are prepared in thermal states of the same temperature.

\item Refs.\ \cite{Scarani:2002, Ziman:2002, Ziman:2005, Ziman:2005b} consider a qubit system interacting identically with a series of identical ancillas. They examine various types of system-ancilla couplings and determine which kinds lead to decoherence and which lead to thermalization (or more generally, homogenization).

\item Refs.\ \cite{Bruneau:2006, Bruneau:2008, Bruneau:2008b, Karevski:2009, Bruneau:2014, Bruneau:2014b, Hanson:2015} use repeated interaction systems as a tool to study quantum thermodynamics. In particular, they examine the steady-state towards which the system is asymptotically drawn in the long-time limit.

\item Other authors have used repeated interaction systems to study a wide range of phenomena. For instance, in Ref.\ \cite{ Kafri:2013} the authors use such a model to describe the behavior of force carriers. Their scheme is comprised of cycles whose effect on the system does not vanish as the cycle becomes infinitesimally short---effectively invoking an infinite-energy term in the dynamics, in the spirit of \cite{Giovannetti:2012, CM,ACMZ}.
\end{itemize}

\section{Error Analysis}
\label{error_analysis}

In this section we analyze the degree to which the effective Liouvillian $\mathcal{L}_{\delta t}$ accurately describes the dynamics of a system interacting with a succession of ancillas. There are two distinct sources of discrepancy between the $\mathcal{L}_{\delta t}$-generated effective dynamics and the full open dynamics of the system: (i) ``stroboscopic error", which is important over short timescales [of order $\mathcal{O}(\delta t)$], and (ii) ``truncation error", which is important over long timescales [of order $\mathcal{O}(t)$].

\subsection{Stroboscopic Error}
\label{sec:stroboscopic_error}

The system's effective Liouvillian $\mathcal{L}_{\delta t}$ is derived by coarse-graining the impact of individual cycles to arrive at a smooth ``average" dynamics. Since ancillas are discarded after they interact with the system, information from the system is irreversibly lost to the environment (i.e., the collection of ancillas), yielding an effective system evolution that is Markovian on timescales much longer than a single cycle. The true dynamics of the system, however, is not exactly Markovian. Rather, during each cycle, information from the system can flow to an ancilla and then back into the system, creating non-Markovian dynamics on short timescales of order $\mathcal{O}(\delta t)$. In other words, the duration of each cycle, $\delta t$, is the characteristic timescale of the environment's effective memory.

During each cycle, the system's true dynamics will deviate slightly from the $\mathcal{L}_{\delta t}$-predicted evolution, only to come back into agreement with the latter again at the end of the cycle. We refer to this effect as ``stroboscopic error", due to close analogy with the error that arises from simulating a Hamiltonian stroboscopically using Trotter's formula \cite{Nielsen:2000}. As one might expect, the degree to which non-Markovian effects temporarily cause the system to deviate from the $\mathcal{L}_{\delta t}$-predicted evolution scales with the cycle duration: sufficiently short cycles render the environment effectively memoryless to a good approximation.

We quantify the magnitude of stroboscopic error by extending the approach used in \cite{Layden:2015b}, taking each interaction and each ancilla to be identical for simplicity. In particular, we consider the state of S at the end of the $n$\textsuperscript{th} cycle, $\rho_\mathrm{S}(t_n)$, where $t_n = n \delta t$. We then compute the system's state during cycle $n+1$, at a time $t_n + \tau$ (where $0 < \tau < \delta t$), in two different ways. First, using the full S-A evolution and tracing out A to obtain the exact system state $\rho_\mathrm{S}^{\text{(ex)}}(t_n + \tau)$. Then, evolving the system by $\mathcal{L}_{\delta t}$, effectively ignoring non-Markovian effects, and denoting the result as $\rho_\mathrm{S}^{\text{(eff)}} (t_n + \tau)$. To quantify magnitude of stroboscopic error, we examine the difference $\rho_\mathrm{S}^{\text{(ex)}} (t_n + \tau) - \rho_\mathrm{S}^{\text{(eff)}} (t_n + \tau)$, which describes the deviation between the exact and effective system dynamics. Expressing the result in powers of $\delta t$, we have
\begin{multline}
||\rho_\mathrm{S}^{\text{(ex)}}(t_n + \tau) - \rho_\mathrm{S}^{\text{(eff)}} (t_n + \tau)||
\le
c_1 \, \delta t
+
c_2 \, \delta t^2 
+ \mathcal{O}(\delta t^3),
\label{eq:stroboscopic}
\end{multline}
with 
\begin{align}
c_1 &= \frac{4}{\hbar} ||H_\mathrm{SA}||_\text{max},\\
c_2 &= \frac{1}{\hbar^2} ||H_\mathrm{SA}||_\text{max} 
\Big( 
17 || H_\mathrm{S}|| + 16 || H_\mathrm{A}||
+ \frac{17}{2} || H_\mathrm{SA}||_\text{max}
\Big),
\end{align}
where we have maximized over $\tau$. Here, $||\,\cdot\,|| = \text{Tr} \, |\,\cdot\,|$ denotes the trace norm (which we have assumed to take on finite values), and $||H_\mathrm{SA}||_\text{max} \coloneqq \max_{0 < \xi < 1} || H_\mathrm{SA} (\xi)||$. The calculation is straightforward but lengthy, and is presented in Appendix \ref{Stroboscopic}. Observe that to leading order in $\delta t$, the magnitude of stroboscopic error depends linearly on the S-A coupling strength, as found in \cite{Layden:2015b}. To $\mathcal{O}(\delta t)$, however, stroboscopic error depends not only on the interaction Hamiltonian, but it also scales non-trivially with the free Hamiltonians of S and A. As the previous intuition suggested, we find stroboscopic error to vanish as the cycle duration $\delta t$ becomes short.

\subsection{Truncation Error}
As discussed above, the interpolation described by Eq.~\eqref{mastereqs} ensures agreement between the exact and $\mathcal{L}_{\delta t}$-generated dynamics of the system at the end of each cycle when using the full expression for $\mathcal{L}_{\delta t}$ [see Eq.~\eqref{Ldef}]. However, if the system-ancilla interactions occur in sufficiently rapid succession, i.e., if $\delta t$ is sufficiently short, it is reasonable (and highly convenient) to truncate $\mathcal{L}_{\delta t}$ to a finite order in $\delta t$. In particular, the master equation in \eqref{LindbladForm} arises from approximating $\mathcal{L}_{\delta t}$ as
\be
\label{L_approx}
\mathcal{L}_{\delta t} \approx \mathcal{L}_0+\delta t \ \mathcal{L}_1.
\ee
We refer to the deviation between the exact system dynamics under repeated interactions and that generated by \eqref{L_approx} at cycle endpoints as ``truncation error". In particular, truncating $\mathcal{L}_{\delta t}$ to $\mathcal{O}(\delta t)$ as above produces after one cycle a truncation error that goes as
\be
\exp\big(\delta t 
\big(\mathcal{L}_0+\delta t \ \mathcal{L}_1
\big)\big)-\bar{\phi}(\delta t)=\mathcal{O}\big(E^3\delta t^3/\hbar^3\big),
\ee
where $E$ is the largest relevant energy scale of the system-ancilla dynamics, as defined in \ref{FirstOrderMasterEqs}. In other words, the effect on the system of a single cycle is well-described by the approximation in Eq.~\eqref{L_approx} provided $\delta t$ is not unreasonably large. 

In contrast with the stroboscopic error, however, truncation error accumulates with $t$ instead of vanishing at the end of each cycle. Concretely, after $n$ cycles, the deviation between the exact system dynamics and that generated by \eqref{L_approx} scales as $\mathcal{O}(n \, E^3\delta t^3/\hbar^3) = \mathcal{O}(t \, E^3 \delta t^2/\hbar^3)$.

\section{General Repeated Qubit-Qubit Interaction}\label{Qubits}
In this section, we illustrate the results of this paper with three simple examples of physical interest \cite{Mason:2012, Ha:1996}. We consider the case where the system and ancillas are all qubits. The most general Hamiltonian of the form \eqref{Hamform} for two interacting qubits is 
\be
H_k(\xi)
=\hbar\bm{\omega}_\text{S}
\cdot\bm{\sigma}_\text{S}
+\hbar\bm{\sigma}_{\text{A}_k}
\cdot\bm{\omega}_{\text{A}_k}
+\hbar\bm{\sigma}_{\text{A}_k}
\bm{\mathrm{J}}_{k}(\xi)
\bm{\sigma}_\text{S},
\ee
where $\bm{\sigma}_\text{S}$ and $\bm{\sigma}_{\text{A}_k}$ are the system and ancillas Pauli vectors, $\bm{\omega}_\text{S}, \ \bm{\omega}_{\text{A}_k}\in\mathbb{R}^3$ and $\bm{\mathrm{J}}_k(\xi)$ is a $3\times3$ matrix.

For notational convenience let us introduce the Bloch vectors $\bm{R}_k$ corresponding to the ancilla states $\rho_{\text{A}_k}$ defined as
\be
\bm{R}_k=\TrAk\big(\rho_{\text{A}_k}\bm{\sigma}_{\text{A}_k}\big).
\ee
We will use Einstein's summation notation with row vectors having subscripts and column vectors having superscripts. The indices $k$ and $l$ are reserved for the indices of the ancilla types and are always explicitly summed over. Greek indices will be taken to run from 1 to 3. Additionally, where necessary, we introduce $\{\,\cdot\,\}$ around objects to separate the Hilbert space/interaction type labels from the vector labels. 

We show in Appendix \ref{AppQubit} that directly from equations \eqref{truncatedmastereqs},  \eqref{H0def}, \eqref{H1def}, and \eqref{Ddef}, one finds the following master equation for the system's effective dynamics under repeated interactions:
\begin{align}\label{QubitRhoEqs}
\frac{\text{d}}{\text{dt}}\rho_\text{S}(t) 
&=-\ii \, \big[
\big(\bm{\omega}_\text{eff}^{(0)}
+\delta t \ \bm{\omega}_\text{eff}^{(1)}\big)
\bm{\sigma}_\text{S},
\rho_\text{S}(t)\big]\\
\nonumber
&+\delta t \, \bm{D}_0\bm{\sigma}_\text{S}
-\frac{\delta t}{2}
\tensor{D_{1}{}}{_\mu_\nu} \, 
\big[\sigma_\text{S}{}^\mu,
[\sigma_\text{S}{}^\nu,
\rho_\text{S}]\big],
\end{align}
where we define the objects in \eqref{QubitRhoEqs} below.

The leading order unitary contribution to the dynamics is described by 
$\bm{\omega}_\text{eff}^{(0)}
=\bm{\omega}_\text{S}
+\bm{\omega}^{(0)}$
where the components of $\bm{\omega}^{(0)}$ are given by
\bel{omega0}
\tensor{\{\omega^{(0)}\}}{_\beta}
=\sum_k p_k \, 
\tensor{\{R_k\}}{_\alpha} \, 
G_0\Big(\tensor{\{J_k(\xi)\}}{^\alpha_\beta}\Big),
\ee
where $\tensor{\{R_k\}}{_\alpha}$ represent the $\alpha$-th component of the row vector $\bm{R}_k$, $\tensor{\{J_k(\xi)\}}{^\alpha_\beta}$ represents the $(\alpha,\beta)$ entry of the $3\times3$ matrix $\bm{\mathrm{J}}_k(\xi)$, and where $G_0$ is the unweighted time average, defined in \eqref{G0def}.

The first (subleading) order correction to the unitary dynamics is 
$\bm{\omega}_\text{eff}^{(1)}
=\bm{\omega}_1^{(1)}
+\bm{\omega}_2^{(1)}
+\bm{\omega}_3^{(1)}$
where the components  of each vector are given by
\begin{align}
\label{omega1}
\tensor{\{\omega_1^{(1)}\}}{_\mu}
&\!=2\!\sum_k p_k \, 
\tensor{\varepsilon}{_\mu^\beta^\gamma} \,
\tensor{\{\omega_\text{S}\}}{_\beta} \, 
\tensor{\{R_k\}}{_\alpha}
G_1\Big(\tensor{\{J_k(\xi)\}}{^\alpha_\gamma}\Big),\\ 
\label{omega2}
\tensor{\{\omega_2^{(1)}\}}{_\mu}
&\!=2\!\sum_k p_k
\tensor{\{R_k\}}{_\gamma}
\tensor{\varepsilon}{^\gamma_{\!\alpha}_\beta}
\tensor{\{\omega_{\text{A}_k}\}}{^\alpha}
G_2\Big(\tensor{\{J_k(\xi)\}}{^\beta_\mu}\Big),\\
\label{omega3}
\tensor{\{\omega_3^{(1)}\}}{_\mu}
&\!=\!\sum_k p_k \, 
\tensor{\varepsilon}{_\mu^{\!\gamma}^\nu}
\tensor{\delta}{_\alpha_\beta}
G_3\Big(\tensor{\{J_k(\xi_2)\}}{^\alpha_\gamma}
\tensor{\{J_k(\xi_1)\}}{^\beta_\nu}\Big),
\end{align}
where $\tensor{\{\omega_{\text{A}_k}\}}{^\alpha}$ represents the $\alpha$-th component of the column vector $\bm{\omega}_{\text{A}_k}$ and $G_1$, $G_2$, and $G_3$ are the time averages defined in \eqref{G1def}, \eqref{G2def}, and \eqref{G3def}, respectively.

The leading order dissipative effects are given by the terms
\bel{qubitD0}
\tensor{\bm{D}_0{}}{_\eta}
\!=\!\!\sum_k p_k \, 
\tensor{\varepsilon}{^\mu^\nu_\eta}\!
\tensor{\{R_k\}}{_\gamma}
\tensor{\varepsilon}{^\gamma_\alpha_\beta}
G_0\Big(\!\tensor{\{J_k(\xi)\}}{^\alpha_\mu}\Big)
G_0\Big(\!\tensor{\{J_k(\xi)\}}{^\beta_\nu}\Big)
\ee
and
\begin{align}\label{qubitD1}
\tensor{\bm{\mathrm{D}}_{1}{}}{_\mu_\nu}
=\sum_{k,l} \Big(&p_k \, 
\tensor{\delta}{_k_l} \, 
\tensor{\delta}{_\alpha_\beta} \, 
-p_k \, p_l \, 
\tensor{\{R_k\}}{_\alpha} \, 
\tensor{\{R_l\}}{_\beta}\Big)\\
\nonumber
&\times G_0\Big(\tensor{\{J_k(\xi)\}}{^\alpha_\mu}\Big)
G_0\Big(\tensor{\{J_l(\xi)\}}{^\beta_\nu}\Big).
\end{align}
It is convenient to represent the state of the system with its Bloch vector:
\be
\bm{a}(t)=\text{Tr}_\text{S}\big(\rho_\text{S}(t)\bm{\sigma}_\text{S}\big).
\ee
The dynamics \eqref{QubitRhoEqs} can be equivalently recast in terms of the system's Bloch vector as
\bel{BlochDynamics}
\bm{a}'(t)
=-2\big(\bm{\omega}_\text{eff}^{(0)}
+\delta t \, \bm{\omega}_\text{eff}^{(1)}
\big)\times\bm{a}(t)
-2\delta t \, \bm{\mathrm{B}} \, \bm{a}(t)
+2\delta t \, \bm{b}.
\ee
In Eq.~\eqref{BlochDynamics} the entries of the matrix $\bm{\mathrm{B}}$ are given by 
\mbox{$\tensor{B}{^\beta_\gamma}
=\tensor{\varepsilon}{^\beta^\mu_\alpha}
\tensor{\varepsilon}{^\nu^\alpha_\gamma}
\tensor{D_1{}}{_\mu_\nu} =\tensor{\delta}{^\beta_\gamma}
\tensor{\delta}{^\nu^\mu}
\tensor{D_1{}}{_\nu_\mu} - \tensor{\delta}{^\alpha^\beta}
\tensor{D_1{}}{_\alpha_\gamma}$},
the vector $\bm{b}=\bm{D}_0{}^T$, and the symbol `$\times$' is the regular vector cross product.

In view of Eq.~\eqref{BlochDynamics}, we can understand the unitary part of the dynamics as a rotation of the system's Bloch vector about the axis given by $\bm{\omega}_\text{eff}^{(0)}+\delta t \, \bm{\omega}_\text{eff}^{(1)}$. Notice that we can recast the cross product as $\bm{\omega}\times\bm{a}=\bm{\mathrm{\Omega}} \, \bm{a}$ where $\bm{\mathrm{\Omega}}$ is a $3\times3$ antisymmetric matrix. Thus the unitary part of the dynamics can be thought of as a linear antisymmetric transformation of the system's Bloch vector.

Additionally, in \eqref{BlochDynamics}, we have the dissipative terms $\bm{\mathrm{B}}$ and $\bm{b}$. We note that $\bm{\mathrm{B}}$ is a symmetric matrix, hence it does not contribute to the unitary dynamics. Finally, notice that dissipative part of the dynamics also gives rise to an affine contribution to the dynamics on the Bloch sphere. Note, however, that $\bm{\mathrm{B}}$ and $\bm{b}$ are not independent of each other.

We now invoke a result from \cite{Lidar2006} which states that dissipative dynamics with $\mathcal{D}[I]=0$, where $I$ describes the maximally mixed state, cannot increase the purity of a quantum system. The Bloch vector of the maximally mixed state is $\bm{a}=0$, thus any linear transformation on $\bm{a}$ will not increase its purity. Therefore, we can interpret $B$ as the part of the dissipation which decreases purity and the affine term $\bm{b}$ as that which can potentially increase the purity.

Next we present a few relevant concrete examples of the kinds of dynamics that \eqref{BlochDynamics} can generate. We see several types of behaviour relating to different fundamental phenomena. These include projection, evolution toward the maximally mixed state, as well as attraction to a fixed point of arbitrary purity.

\begin{figure*}
\includegraphics[width=0.24\textwidth]{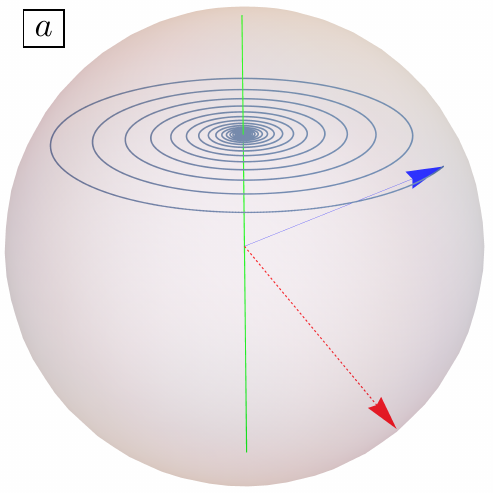}
\includegraphics[width=0.24\textwidth]{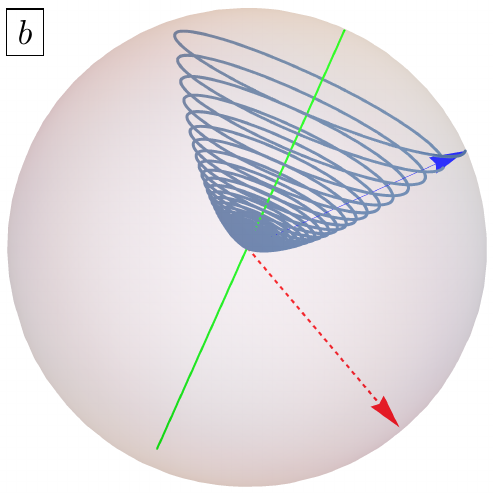}
\includegraphics[width=0.24\textwidth]{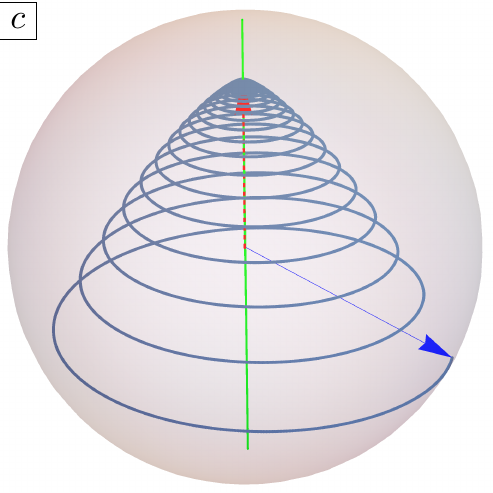}
\includegraphics[width=0.24\textwidth]{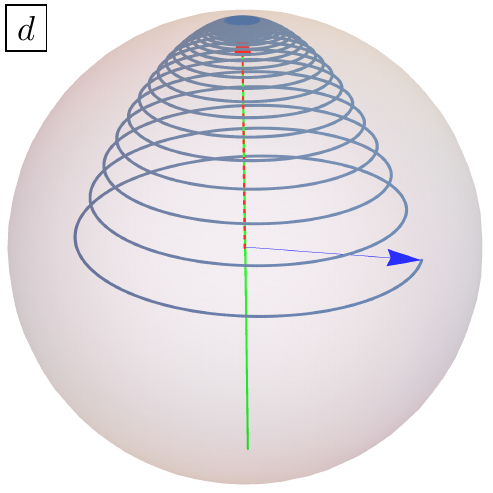}
\caption{
(Color online.) A range of phenomena in rapid repeated interactions between qubits. In all figures the system's initial Bloch vector is the blue (solid) vector and follows the path beginning there. The ancillas' Bloch vectors are the red (dashed) vector. The green (solid) line marks the $\omega_\text{eff}$ axis. In subfigure a) qubits interact via a ZZ coupling. The result is that system qubit is projected onto the $\omega_\text{eff}$ axis; b) qubits interact via an XX coupling. The result is that the system qubit is completely dephased; c) qubits interact via a $\bm{\sigma}\cdot\bm{\sigma}$ coupling. The result is that system qubit is thermalized to the same purity as the ancillas; d) qubits interact via a $\bm{\sigma}\cdot\bm{\sigma}$ coupling. The system qubit begins mixed and is fully purified by the interaction.}
\label{FigQubit}
\end{figure*}

\subsection{ZZ coupling}\label{ZZQubits}
Let us consider a qubit, S, repeatedly interacting with a series of identically prepared qubits, A, via a time dependent $ZZ$ coupling. The Hamiltonian for this interaction is
\bel{ZZHam}
H(t)=\hbar\omega_\text{S} \, \sigma_\text{S,z}
+\hbar\omega_{\text{A}} \, \sigma_{\text{A},z}
+\hbar J(t/\delta t) \, 
\sigma_\text{S,z}
\sigma_{\text{A},z}.
\ee
From \eqref{QubitRhoEqs}, we find that
the system's density matrix evolves as
\be
\frac{\text{d}}{\text{dt}}\rho_\text{S}(t) 
=-\ii \, \omega \, [\sigma_{\text{S},\text{z}},\rho_\text{S}(t)]
-\Gamma 
\big[\sigma_{\text{S},\text{z}},
[\sigma_{\text{S},\text{z}},
\rho_\text{S}(t)]\big]/4
\ee
where
$\omega=\omega_\text{S}+J_0\langle\sigma_{\text{A},\text{z}}\rangle$
and
$\Gamma
=2 \, \delta t \, \big(1-\langle\sigma_{\text{A},z}\rangle^2\big) \, J_0{}^2$,
and where $J_0=G_0(J(\xi))=\int_0^1 d\xi J(\xi)$. Note that all  $\bm{\omega}_\text{eff}^{(1)}$ terms
vanish because  $J(t/\delta t)$ only has non-vanishing $zz$ components.

Equivalently, from \eqref{BlochDynamics}, the trajectory of the system's Bloch vector is given by
\begin{align}
a_\text{x}'(t)
&=2\omega \, a_\text{y}(t)
-\Gamma a_\text{x}(t),\\
a_\text{y}'(t)
&=-2\omega \, a_\text{x}(t)
-\Gamma a_\text{y}(t),\\
a_\text{z}'(t)&=0.
\end{align}
We can solve these equations, with initial conditions $\bm{a}(0)=(a_\text{x0},a_\text{y0},a_\text{z0})^T$ to find
\begin{align}
a_\text{x}(t)
&=e^{-\Gamma t}[a_\text{x0}\cos(2\omega t)+a_\text{y0}\sin(2\omega t)],\\
a_\text{y}(t)
&=e^{-\Gamma t}[a_\text{y0}\cos(2\omega t)-a_\text{x0}\sin(2\omega t)],\\
a_\text{z}(t)
&=a_\text{z0}.
\end{align}
Thus the Bloch vector maintains a constant $z$ component and the $x$ and $y$ components rotate around the $z$ axis at a rate $2\omega$ while decaying exponentially at a rate $\Gamma$, as shown in Fig. \ref{FigQubit} a).

The decoherence rate for this projection is 
\be
\Gamma
=2 \, \delta t \, \big(1-\langle\sigma_{\text{A},z}\rangle^2\big) \, J_0{}^2.
\ee 

As discussed in Section \ref{Dsection}, the dissipative terms can be understood in terms of uncertainty. The decoherence rate is proportional to the uncertainty of the ancillas' $z$ component, $\Delta_{\sigma_{\text{A},z}}^2\!\!\!\!\!\!=1-\langle\sigma_{\text{A},z}\rangle^2$. In other words, if we are certain about the $z$ polarization of the ancilla then there is no decoherence. If we are maximally uncertain about the $z$ polarization of the ancilla (say the ancilla is $x$-polarized or maximally mixed) the decoherence is strongest. 

In sum, the decoherence of the system S introduced by the interaction \eqref{ZZHam} can be understood as a projection onto the $z$ axis of the Bloch sphere. The dephasing time of this decoherence channel is
\be
T_2=\frac{1}{\Gamma}=\frac{1}{2 \, \delta t \, \big(1-\langle\sigma_{\text{A},z}\rangle^2\big)
 \, J^2}
 +O(\delta t^0).
\ee

\subsection{XX coupling}\label{XXQubits}
Next, let us consider a qubit, S, repeatedly interacting with a series of identically prepared qubits, A, via a time dependent $XX$ coupling. The Hamiltonian for this interaction is
\bel{XXHam}
H(t)=\hbar\omega_\text{S}\sigma_\text{S,z}
+\hbar\omega_{\text{A}}\sigma_{\text{A},z}
+\hbar J(t/\delta t)
\sigma_\text{S,x}
\sigma_{\text{A},x}.
\ee
Using \eqref{QubitRhoEqs}, we then find that
the system's density matrix evolves as
\begin{align}
\frac{\d}{\d t}\rho_\text{S}(t) =&-\ii \, \omega_\text{S}[\sigma_\text{S,z},\rho_\text{S}(t)]\\
\nonumber
&-\ii \, \big(\omega_x^{(0)}+\omega_x^{(1)}\big)[\sigma_\text{S,x},\rho_\text{S}(t)]\\
\nonumber
&-\Gamma\big[\sigma_{\text{S},\text{x}},
[\sigma_{\text{S},\text{x}},
\rho_\text{S}(t)]\big]/4,
\end{align}
where 
$\omega_x^{(0)}
=J_0 \, \langle\sigma_{\text{A},\text{x}}\rangle$, 
$\omega_x^{(1)}
=2 J_2 \, \delta t \, \omega_A \, \langle\sigma_{A,y}\rangle$,
and $\Gamma
=2 \, \delta t \, J_0{}^2  \, \big(1-\langle\sigma_{\text{A},\text{x}}\rangle^2\big)$, where $J_0=G_0(J(\xi))=\int_0^1 J(\xi)d\xi$ and $J_2=G_2(J(\xi))=\int_0^1\xi J(\xi)d\xi$.

Equivalently, from \eqref{BlochDynamics}, we can rewrite the dynamics of the system in terms of its Bloch vector as
\begin{align}
a_\text{x}'(t)
=& 2\omega_S \, a_\text{y}(t),\\
a_\text{y}'(t)
=&-2\omega_S \, a_\text{x}(t)
+2\omega_x \, a_\text{z}(t)
-\Gamma \, a_\text{y}(t),\\
a_\text{z}'(t)
=&-2\omega_x \, a_\text{y}(t)
-\Gamma \, a_\text{z}(t)   
\end{align}
where $\omega_x=\omega_x^{(0)}+\omega_x^{(1)}$.

The $\omega_S$ terms are the system's free evolution which rotates the Bloch vector around the $z$ axis. The $\omega_x$ terms are the corrections to the unitary dynamics which cause some rotation about the $x$ axis. Finally the $\Gamma$ terms capture the decoherence caused by the repeated interaction which serve to exponentially suppress the $y$ and $z$ components. 

We note that both the leading (zeroth) order and subleading corrections to the unitary dynamics are rotations around the $x$ direction. Roughly speaking, the leading order correction comes from the ancillas Bloch vector having some $x$ component which the system senses through the $XX$ coupling. The subleading correction comes from the ancillas $y$ component rotating into the $x$ direction during the interaction due to the ancillas' local dynamics.

If the ancillas are not fully polarized in the $x$ direction, together these effects cause any initial system state to become maximally mixed as shown in Fig. \ref{FigQubit} b). The $y$ and $z$ components are suppressed directly by the $\Gamma$ term, while the $x$ component rotates into the $y$ and $z$ directions and is thereby also suppressed.

Again we see that, as discussed in Section \ref{Dsection}, dissipation in this setting can be understood in terms of uncertainty. Concretely, the decoherence rate for this example, $\Gamma$, is proportional to the uncertainty of the ancillas $x$ component, $\Delta_{\sigma_{\text{A},x}}^2\!\!\!\!\!\!=1-\langle\sigma_{\text{A},x}\rangle^2$. Therefore, if we are certain about the $x$ polarization of the ancilla then there is no decoherence. If we are maximally uncertain about the $x$ polarization of the ancillas (say the ancillas are $z$-polarized or maximally mixed) the decoherence is the fastest.

\subsection{Isotropic spin coupling (\texorpdfstring{$\bm{\sigma}\cdot\bm{\sigma}$}{text})}
Finally, let us consider a qubit, S, repeatedly interacting with a series of identically prepared qubits, A, via a time dependent $\bm{\sigma}\cdot \bm{\sigma}$ coupling. The Hamiltonian for this interaction is
\bel{SSHam}
H(t)=\hbar\omega_\text{S}\sigma_\text{S,z}
+\hbar\omega_{\text{A}}\sigma_{\text{A},z}
+\hbar J(t/\delta t)
\bm{\sigma}_\text{S}
\cdot\bm{\sigma}_{\text{A}}.
\ee
We recall from \eqref{QubitRhoEqs} that the system's density matrix evolves as
\begin{align}
\frac{\text{d}}{\text{dt}}\rho_\text{S}(t) 
&=-\ii \, \big[
\big(\bm{\omega}_\text{eff}^{(0)}
+\delta t \ \bm{\omega}_\text{eff}^{(1)}\big)
\bm{\sigma}_\text{S},
\rho_\text{S}(t)\big]\\
\nonumber
&+\delta t \ \bm{D}_0\bm{\sigma}_\text{S}
-\frac{\delta t}{2}
\tensor{D_{1}{}}{_\mu_\nu} \, 
\big[\sigma_\text{S}{}^\mu,
[\sigma_\text{S}{}^\nu,
\rho_\text{S}]\big].
\end{align}

For the specific interaction $\eqref{SSHam}$, the leading order unitary dynamics is described by 
\bel{SSomega0}
\bm{\omega}_\text{eff}^{(0)}
=\omega_\text{S} \, \bm{\hat{z}}
+J_0 \, \bm{R},
\ee 
where $J_0=G_0(J(\xi))=\int_0^1 J(\xi)d\xi$. That is, it goes as the free evolution of the system plus a contribution from the repeated interactions, which points in the direction of each ancillas initial Bloch vector.

The first (subleading) order correction to the unitary dynamics is 
\bel{SSomega1}
\bm{\omega}_\text{eff}^{(1)}
=2 J_1(\omega_\text{S}-\omega_\text{A}) \ \bm{\hat{z}}\times\bm{R}
+J_0 \ \omega_\text{A} \ \bm{\hat{z}}\times\bm{R},
\ee
where
$J_1=G_1(J(\xi))=\int_0^1(\xi-1/2) J(\xi)d\xi$. 

The leading order dissipative effects are given by 
\be
\bm{D}_0=2J_0{}^2 \bm{R},
\ee
and
\be
\tensor{\bm{\mathrm{D}}_{1}{}}{_\mu_\nu}
=J_0{}^2\big(
\tensor{\delta}{_\mu_\nu} \, 
-\tensor{R}{_\mu} \tensor{R}{_\nu}\big).
\ee
Using \eqref{BlochDynamics}, we can write the system's dynamics in terms of its Bloch vector. Doing this, we find
\bel{SSBlochDynamics}
\bm{a}'(t)
=-2\big(\bm{\omega}_\text{eff}^{(0)}
+\delta t \, \bm{\omega}_\text{eff}^{(1)}
\big)\times\bm{a}(t)
-2\delta t \, \bm{\mathrm{B}}\bm{a}(t)
+2\delta t \, \bm{b},
\ee
where $\bm{\omega}_\text{eff}^{(0)}$ and $\bm{\omega}_\text{eff}^{(1)}$ are defined in \eqref{SSomega0} and \eqref{SSomega1} respectively and where
\be
\bm{b}=2 J_0{}^2 \bm{R}^T,
\ee
and
\be
\bm{\mathrm{B}}
=J_0{}^2\big(
(2-\vert \bm{R}\vert^2)\bm{\mathrm{I}}
+\bm{R}\bm{R}^T\big).
\ee
As discussed above, the first term in \eqref{SSBlochDynamics} corresponds to the effective unitary evolution of the system, whereas the effect of the $\bm{\mathrm{B}}$ matrix is to reduce system's purity. These two effects combine with the affine term $\bm{b}$ to produce a fixed point that could be anywhere near the $\bm{\omega}_\text{eff}^{(0)}$ axis, depending on the initial state of the ancilla.

For example, if we take the ancillas Bloch vector to be
$\bm{R}=R \, \hat{z}$, then we find the Bloch vector dynamics to be
\begin{align}
a_\text{x}'(t)
&= \ \ 2\omega \, a_\text{y}(t)
-(\Gamma_1+\Gamma_2) \, a_\text{x}(t),\\
a_\text{y}'(t)
&=-2\omega \, a_\text{x}(t)
-(\Gamma_1+\Gamma_2) \, a_\text{y}(t),\\
a_\text{z}'(t)
&=-\Gamma_1 (a_\text{z}(t)-R).   
\end{align}
where $\omega=\omega_S+J_0 R$,
\be
\Gamma_1 
=2\delta t J_0^2,
\ee 
and
\be
\Gamma_2 
=\delta t J_0^2(1-R^2).
\ee

The $\omega$ terms captures the unitary part of the dynamics and serves to rotate the system's Bloch vector around the $z$ axis. The $\Gamma$ terms exponentially suppress the $x$ and $y$ components to 0, and drive the $z$ component towards $a_z=R$. Note that there is some base decoherence rate $\Gamma_1$ suppressing all of the components of the Bloch vector as well as an additional decoherence rate $\Gamma_2$, which suppresses the the $x$ and $y$ components and is proportional to the variance of the ancillas $z$ polarization, $\Delta_{\sigma_{\text{A},z}}^2=1-R^2$. This additional decoherence in the $x$ and $y$ components comes from uncertainty about the ancillas' $z$ polarization. 

We solve these equations with initial conditions $\bm{a}(0)=(a_\text{x0},a_\text{y0},a_\text{z0})^T$ to find
\begin{align}
a_\text{x}(t)
&=e^{-(\Gamma_1+\Gamma_2) t}[a_\text{x0}\cos(2\omega t)+a_\text{y0}\sin(2\omega t)],\\
a_\text{y}(t)
&=e^{-(\Gamma_1+\Gamma_2) t}[a_\text{y0}\cos(2\omega t)-a_\text{x0}\sin(2\omega t)],\\
a_\text{z}(t)
&=R+(a_\text{z0}-R)e^{-\Gamma_1 t}.
\end{align}

The end result of this dynamics is that the system is ``thermalized'' with the ancilla, i.e., the system's purity at the fixed point of this dynamics becomes the same as the ancillas'. We see this behavior in Fig. \ref{FigQubit} c). Note that if the ancillas are eigenstates of $\sigma_{A,z}$, the repeated interaction serves to drive the system to a fixed point which is pure as in Fig. \ref{FigQubit} d).

Note that these examples of qubit-qubit interactions are just particular cases of our general formalism, which can be applied to a wide variety of different systems. For example, it was shown in \cite{Layden:2015b} that this kind of formalism can be applied to infinite dimensional quantum systems such as a harmonic oscillator repeatedly coupled to qubits. Furthermore, this formalism is particularly well suited to analyze more complex relevant situations such as, for instance, the light-matter interaction \cite{Scully:1997}, entanglement farming \cite{Martin:2013}, and gravitational decoherence \cite{CM,Pikovski:2015, ACMZ}.

\section{Conclusion}
In this paper, we studied the emergent open dynamics of a quantum system which undergoes repeated unitary interactions (of duration $\delta t$) with a sequence of ancillary systems. We extended previous results (see \cite{Layden:2015b}) in the following ways:
\begin{itemize}
\item We fully determined the leading order (in $\delta t$) dissipative effects. We also characterized the first subleading corrections to unitary dynamics that arise when we deviate from the limit $\delta t\to 0$, which formed the basis of the scheme presented in \cite{Layden:2015b}.
\item We found the effective master equation (in Lindblad form) describing the emergent dynamics of the system incorporating the effects induced by the repeated interactions with the ancillas. We characterize the decoherence rates of this emergent open dynamics. We also found an upper bound to the decoherence rates and studied how it scales with the effective dimension of the ancillary systems.
\item We relaxed the restriction in \cite{Layden:2015b} that the ancillas repeatedly interacting with the system be identical. Instead, we allow, in general, the ancillas to be taken from an ensemble of different types of ancillas. Namely, these ancillas can now be quantum systems of different dimensions, can be prepared in different states, and can interact with the system through different Hamiltonians.
\end{itemize}

Remarkably, we found that the strength of the dissipative part of the dynamics is intrinsically linked with the fundamental ``unpredictability'' in the system-ancilla interaction. This unpredictability can come from a) quantum uncertainty in the observable through which the ancilla is coupled to the system b)  classical uncertainty (mixedness) in the state of the ancilla or, in the general case where the ancillas are randomly chosen from a statistical ensemble, c)  classical uncertainty as to which type of ancilla are chosen at every particular iteration of the repeated interaction. In the particular case of qubit-qubit interactions, we showed how the strength of the dissipation is proportional to the variance $\Delta_{\sigma_i}^2$ in the ancilla state, where $\sigma_i$ is the observable of the ancilla that we couple to the system.

We compared our findings with existing results in the literature. In particular, (i) we showed how our formalism could be adapted to include, as a special case, the results of \cite{CM}, (ii) we discussed the relation between the present work and \cite{Zanardi:2014, Zanardi:2015, Zanardi:2016}, extending a special case of their results to higher orders in $\delta t$, and (iii) we situated our results within the existing literature on repeated interaction systems \cite{Attal:2006, Attal:2007, Attal:2007b, Vargas:2008, Giovannetti:2012, Scarani:2002, Ziman:2002, Ziman:2005, Ziman:2005b, Bruneau:2006, Bruneau:2008, Bruneau:2008b, Karevski:2009, Bruneau:2014, Bruneau:2014b, Hanson:2015, Layden:2015}. 

We analyzed several examples wherein the system and the ancillas are qubits.  We showed the rich variety of phenomena that can arise from different types of couplings even between low-dimensional systems. In particular, we considered transverse, longitudinal and isotropic system-ancilla couplings, and observed that the system's dynamics gives rise to effective projection, depolarization, themalization of the system with the ancillas, and purification of the system through the repeated interactions. The characterization of more general conditions under which the repeated interaction can purify is an interesting topic in its own right and will be reported elsewhere.

The results presented in this paper have potental application in numerous settings. Most immediately, they could be used to describe the effective dynamics of a particle scattering through media, which, in turn, could enable one to characterize the medium in question through its effective impact on a probe system. Similarly, our results could provide insight into the process of decoherence, complementing existing results involving repeated interaction systems in other regimes \cite{Bruneau:2006, Attal:2007, Vargas:2008, Bruneau:2008b, Giovannetti:2012, Bruneau:2014, Bruneau:2014b, Hanson:2015}, and in the field of quantum thermodynamics \cite{Goold:2016}.

This work significantly extends the formalism that was developed in \cite{Layden:2015b} for the purpose of coherent quantum control. The present extensions allow not only for a better understanding of errors that may arise in such control schemes, but paves the way for non-unitary control, which could be used for dissipative state preparation and for Lindblad simulation. Finally, our results may provide insight into the quantum measurement problem \cite{Layden:2015}, and even into the quantum nature of gravity through gravitational decoherence phenomena  \cite{Kafri:2014zsa,
Kafri:2015iha, CM, ACMZ}.

\appendix
\section{Derivation of Formal Power Series}\label{AppSeries}
In this appendix, we derive the series expansions \eqref{phibarexpand} and \eqref{Lseriesdef} from Section \ref{Model}. We expand $\bar{\phi}(\delta t)$, defined in \eqref{phibardef}, as a formal series in powers of $\delta t$. Using the expansion \eqref{Ukexpand} for $U_{\delta t,k}(\delta t)$, we have
\begin{align}
\bar{\phi}(\delta t)[ \,\cdot \,]
&=\sum_k p_k \TrAk\!\big(
U_{\delta t,k}(\delta t)
(\cdot\otimes \rho_{\text{A}_k})
U_{\delta t,k}(\delta t)^\dagger\big)\\
\nonumber
&=\sum_k p_k \TrAk\!\Big(
\sum_{i=0}^\infty \delta t^i U_{k,i} 
(\cdot\otimes \rho_{\text{A}_k})
\sum_{j=0}^\infty \delta t^j  U_{k,j}^\dagger\Big)\\
\nonumber
&=\sum_k p_k\sum_{i,j=0}^\infty
\delta t^{i+j}
\TrAk\!\Big(U_{k,i}
(\cdot\otimes \rho_{\text{A}_k})
U_{k,j}^\dagger\Big)\\
\nonumber
&=\sum_{n=0}^\infty
\delta t^{n}
\sum_k p_k
\sum_{m=0}^n
\TrAk\!\Big(U_{k,m}
(\cdot\otimes\rho_{\text{A}_k}\!)
U_{k,n-m}^\dagger\Big)\\
\nonumber
&=\sum_{n=0}^\infty
\delta t^{n} \, 
\phi_{n}[\,\cdot\,]
\end{align}
as claimed in $\eqref{phibarexpand}$, where
\be
\phi_{n}[\,\cdot\,]
\coloneqq
\sum_k p_k\sum_{m=0}^n
\TrAk\!\Big(U_{k,m}
(\cdot\otimes \rho_{\text{A}_k})
U_{k,n-m}^\dagger\Big),
\ee
as claimed in $\eqref{phibarseriesdef}$.

Next, we find a recursive definition of the coefficients for the expansion of $\mathcal{L}_{\delta t}$ claimed in \eqref{Lseriesdef}. We note that the problem of solving for $\mathcal{L}_j$ is closely related to that of converting between Dyson and Magnus series \cite{Argeri:2014}. Recall, in Section \ref{Model}, we defined $\mathcal{L}_{\delta t}$ as the unique (up to choosing a branch cut) Liouvillian superoperator satisfying \eqref{Markcond}, namely $\exp(\delta t \, \mathcal{L}_{\delta t})=\bar{\phi}(\delta t)$. 

To summarize our calculation, we will first formally expand $\mathcal{L}_{\delta t}$ as a series in $\delta t$ as $\mathcal{L}_{\delta t}=\mathcal{L}_0+\delta t \, \mathcal{L}_1+\delta t^2 \, \mathcal{L}_2+\dots$. Then, using this expression, we will expand \eqref{Markcond} as a series in $\delta t$ and require it to match order by order with the expansion for $\bar{\phi}(\delta t)$ given by \eqref{phibarexpand}. This gives a recursive definition for the coefficients of $\mathcal{L}_{\delta t}$.

First, we expand $\Omega_{\delta t}(t)$, defined in \eqref{Markform}, as a series in both $\delta t$ and $t$ as
\bel{Lderivation1}
\Omega_{\delta t}(t)
=\exp\Big(t \sum_{j=0}^\infty \delta t^j \ \mathcal{L}_j\Big)\\
=\sum_{n=0}^\infty \frac{t^n}{n!} \Big(\sum_{j=0}^\infty \delta t^j \, \mathcal{L}_j\Big)^n.
\ee
We can expand this further using a multinomial expansion. Keeping in mind that the $\mathcal{L}_j$ operators do not commute, we get
\bel{Lderivation2}
\Omega_{\delta t}(t)
=\boldsymbol{1}+\sum_{n=1}^\infty \frac{t^n}{n!}\sum_{\bm{\beta}\in\mathbb{Z}_{\geq 0}^n} \prod_{i=1}^{n} \mathcal{L}_{\beta_i}\delta t^{\beta_i},
\ee
where the set $\mathbb{Z}_{\geq 0}^n$ contains the non-negative integer vectors of dimension $n$. In \eqref{Lderivation2}, each $\bm{\beta}$ in the sum corresponds to a way of picking one term from each of the $n$ sums in \eqref{Lderivation1}. For example, $n=4$ and $\bm{\beta}=(0,1,0,3)$ corresponds to the term 
\be
\big(\mathcal{L}_{0}\delta t^0\big)
\big(\mathcal{L}_{1}\delta t^1\big)
\big(\mathcal{L}_{0}\delta t^0\big)
\big(\mathcal{L}_{3}\delta t^3\big)
=\mathcal{L}_0 \mathcal{L}_1 \mathcal{L}_0 \mathcal{L}_3
\ \delta t^4.
\ee
In order to collect terms with the same powers of $\delta t$ in \eqref{Lderivation2}, we define $J_{\bm{\beta}}=\sum_{i=1}^n{\beta_i}$ and regroup the sum over all $\bm{\beta}$'s into a sum over $J$ and a sum over $\bm{\beta}$'s with $J_{\bm{\beta}}=J$ as
\begin{align}\label{Lderivation4}
\Omega_{\delta t}(t)
&=\boldsymbol{1}
+\sum_{n=1}^\infty \frac{t^n}{n!}
\sum_{J=0}^\infty\delta t^J
\sum_{\bm{\beta}\in\mathbb{Z}_{\geq 0}^n\vert J_{\bm{\beta}}=J} \prod_{i=1}^{n} \mathcal{L}_{\beta_i}.
\end{align}
The set of $\bm{\beta}$'s with $J_{\bm{\beta}}=J$ are exactly the weak compositions of $J$ of length $n$ which we denote by $\text{C}_\text{w}(J,n)$, that is, the ordered lists of nonnegative integers of length $n$ which sum to $J$. For example, $\text{C}_\text{w}(3,2)$ are all the ways of writing $3$ as the sum of two non-negative integers:
\be\text{C}_\text{w}(3,2)
=\{
(3,0),
(0,3),
(2,1),
(1,2)
\}.
\ee
And $\text{C}_\text{w}(2,3)$ are all the ways of writing $2$ as the sum of three non-negative integers:
\begin{align}
\text{C}_\text{w}(2,3)
=\{
&(2,0,0),
(0,2,0),
(0,0,2),\\
&\nonumber
(1,1,0),
(1,0,1),
(0,1,1)
\}.
\end{align}
Adopting this notation and evaluating \eqref{Lderivation4} at $t=\delta t$ we find
\begin{align}
\Omega_{\delta t}(\delta t)
=& \boldsymbol{1}
+\sum_{J=0}^\infty \sum_{n=1}^\infty \frac{\delta t^{J+n}}{n!}\sum_{\bm{\beta}\in\text{C}_\text{w}(J,n)} \prod_{i=1}^{n} \mathcal{L}_{\beta_i}\\
\nonumber
=& \boldsymbol{1}
+\sum_{M=1}^\infty \delta t^{M}
\sum_{n=1}^{M}\frac{1}{n!}
\sum_{\bm{\beta}\in\text{C}_\text{w}(M-n,n)}\prod_{i=1}^{n} \mathcal{L}_{\beta_i}.
\end{align}
Thus we have expanded the left hand side of \eqref{Markcond} as a series in $\delta t$. Now we require this to match order by order with the expansion of $\bar{\phi}(\delta t)$ given in \eqref{phibarexpand}. They automatically match at zeroth order, and for $M\geq1$ we require
\be
\bar{\phi}_M
=\sum_{n=1}^{M}\frac{1}{n!}
\sum_{\bm{\beta}\in\text{C}_\text{w}(M-n,n)}\prod_{i=1}^{n} \mathcal{L}_{\beta_i}.
\ee
Next, we turn this relationship into a recursive definition for $\mathcal{L}_M$. Separating the $n=1$ term from the sum isolates an $\mathcal{L}_{M-1}$ term since $\text{C}_\text{w}(M-1,1)=\{(M-1)\}$. Explicitly, we have
\begin{align}
\bar{\phi}_M
&=\sum_{\bm{\beta}\in\text{C}_\text{w}(M-1,1)}
\mathcal{L}_{\beta_i}
+\sum_{n=2}^{M}\frac{1}{n!}
\sum_{\bm{\beta}\in\text{C}_\text{w}(M-n,n)} \prod_{i=1}^{n} \mathcal{L}_{\beta_i}\\
\nonumber
&=\mathcal{L}_{M-1}
+\sum_{n=2}^{M}\frac{1}{n!}
\sum_{\bm{\beta}\in\text{C}_\text{w}(M-n,n)} \prod_{i=1}^{n} \mathcal{L}_{\beta_i}.
\end{align}
Solving this for $\mathcal{L}_{M-1}$ we find
\be
\mathcal{L}_{M-1} 
= \bar{\phi}_M
-\sum_{n=2}^{M}\frac{1}{n!}
\sum_{\bm{\beta}\in\text{C}_\text{w}(M-n,n)} \prod_{i=1}^{n} \mathcal{L}_{\beta_i}.
\ee
Finally, to simplify, we change $M\to M+1$ and shift the index $n$ by one, yielding
\bel{Lderivation3}
\mathcal{L}_{M} 
=\bar{\phi}_{M+1}
-\sum_{n=1}^{M}\!\frac{1}{(n+1)!}
\sum_{\bm{\beta}\in\text{C}_\text{w}(M-n,n+1)}\prod_{i=1}^{n+1} \mathcal{L}_{\beta_i}.
\ee
This is the recursive definition for $\mathcal{L}_M$ claimed in Eq.~\eqref{Lseriesdef}. It is easily seen to be recursive, since $n\geq1$ implies all $\bm{\beta}\in\text{C}_\text{w}(M-n,n+1)$ have $\beta_i\leq M-n\leq M-1$. Thus right hand side of \eqref{Lderivation3} only contains $\mathcal{L}_m$'s with $m\leq M-1$.

Next, we compute $\mathcal{L}_m$ for small $m$. Taking $M=0$ in \eqref{Lderivation3} we get an empty sum over $n$ and thus $\mathcal{L}_0=\bar{\phi}_1$ as claimed in Eq.~\eqref{L0def}. Taking $M=1$ in \eqref{Lderivation3} we get
\begin{align}
\mathcal{L}_{1} 
&= \bar{\phi}_{2}
-\sum_{n=1}^{1}\frac{1}{(n+1)!}
\sum_{\bm{\beta}\in\text{C}_\text{w}(1-n,n+1)} \prod_{i=1}^{n+1} \mathcal{L}_{\beta_i}\\
\nonumber
&= \bar{\phi}_{2}
-\frac{1}{2}
\sum_{\bm{\beta}\in\text{C}_\text{w}(0,2)} \prod_{i=1}^{2} \mathcal{L}_{\beta_i}\\
\nonumber
&= \bar{\phi}_{2}
-\mathcal{L}_0{}^2/2\\
\nonumber
& =\bar{\phi}_{2}-\bar{\phi}_1{}^2/2,
\end{align}
as claimed in Eq.~\eqref{L1def}. Finally, taking $M=2$ in \eqref{Lderivation3} we get
\begin{align}
\mathcal{L}_{2} 
&= \bar{\phi}_{3}
-\sum_{n=1}^{2}\frac{1}{(n+1)!}
\sum_{\bm{\beta}\in\text{C}_\text{w}(2-n,n+1)} \prod_{i=1}^{n+1} \mathcal{L}_{\beta_i}\\
\nonumber
&=\bar{\phi}_{3}
-\frac{1}{2!}\!
\sum_{\bm{\beta}\in\text{C}_\text{w}(1,2)} \prod_{i=1}^{2} \mathcal{L}_{\beta_i}
-\frac{1}{3!}\!
\sum_{\bm{\beta}\in\text{C}_\text{w}(0,3)} \prod_{i=1}^{3} \mathcal{L}_{\beta_i}\\
\nonumber
&=\bar{\phi}_{3}
-\frac{1}{2}(\mathcal{L}_0\mathcal{L}_1+\mathcal{L}_1\mathcal{L}_0)
-\frac{1}{6}\mathcal{L}_0{}^3\\
\nonumber
& = \bar{\phi}_{3}
-(\bar{\phi}_1\bar{\phi}_{2}+\bar{\phi}_{2}\bar{\phi}_1)/2
+\bar{\phi}_1{}^3/3,
\end{align}
as claimed in Eq.~\eqref{L2def}.

\section{Derivation of \texorpdfstring{$\mathcal{L}_0$}{text}}\label{AppL0}
In this appendix, we calculate $\mathcal{L}_0$ using the general form of Hamiltonians given in \eqref{Hamform}.  Using \eqref{L0def} and \eqref{phibarseriesdef} we calculate
\begin{align}
\mathcal{L}_0[\,\cdot\,]
&=\bar{\phi}_1[\,\cdot\,]\\
\nonumber
&=\sum_k \, p_k \, \TrAk\! \Big(U_{k,1}(\cdot\otimes\rho_{\text{A}_k})
+(\cdot\otimes\rho_{\text{A}_k}) U_{k,1}{}^\dagger\Big)\\
\nonumber
&=\sum_k \, p_k \, \TrAk\! \Big(U_{k,1}(\cdot\otimes\rho_{\text{A}_k})
-(\cdot\otimes\rho_{\text{A}_k}) U_{k,1}\Big)\\
\nonumber
&=\sum_k \, p_k \, \TrAk\! \Big([U_{k,1},(\cdot\otimes\rho_{\text{A}_k})]\Big)\\
\nonumber
&=\sum_k \, p_k \, \big[\TrAk\! \big(U_{k,1}\rho_{\text{A}_k}\big),\,\cdot\,\big]\\
\nonumber
&=\Big[\sum_k \, p_k \, \TrAk\! \big(U_{k,1}\rho_{\text{A}_k}\big),\,\cdot\,\Big]\\
\nonumber
&=\frac{-\ii}{\hbar}[H_\text{eff}^{(0)},\,\cdot\,],
\end{align}
where we have used that $U_{k,1}$, defined in \eqref{Ukseriesdef}, is antihermitian. We have also defined 
\be 
H_\text{eff}^{(0)}
\coloneqq\ii\hbar\sum_k \, p_k \, \TrAk\! \big(\rho_{\text{A}_k} U_{k,1}\big).
\ee 
Thus,we have confirmed that $\mathcal{L}_0$ is of the form claimed in \eqref{L0explicit}, meaning that to zeroth order the evolution is completely unitary.

Now we convert $H_\text{eff}^{(0)}$ into the form claimed in \eqref{H0effdef}. First, we calculate $U_{k,1}$ defined in \eqref{Ukseriesdef} as
\begin{align}\label{Uk1explicit}
U_{k,1}
&=\frac{-\ii}{\hbar}\int_{0}^1 H_{S}\otimes\boldsymbol{1}
+\boldsymbol{1}\otimes H_{\text{A}_k}
+H_{SA_k}(\xi) d\xi\\
\nonumber
&=\frac{-\ii}{\hbar}\Big(
H_{S}\otimes\boldsymbol{1}
+\boldsymbol{1}\otimes H_{\text{A}_k}
+G_0(H_{SA_k}(\xi))\Big),
\end{align}
where $G_0(X)=\int_{0}^1 X(\xi) \, d\xi$, as defined in \eqref{G0def}. Using this we can simplify $H_\text{eff}^{(0)}$ as
\begin{align}
&H_\text{eff}^{(0)}
=(\ii\hbar)\sum_k \, p_k \, \TrAk\! \big(\rho_{\text{A}_k} U_{k,1}\big)\\
\nonumber
&=\sum_k p_k \TrAk\!\Big(\rho_{\text{A}_k} \big(
H_{S}\otimes\boldsymbol{1}
+\boldsymbol{1}\otimes H_{\text{A}_k}
+G_0(H_{SA_k})\big)\Big)\\
\nonumber
&=H_{S}
+\boldsymbol{1}\sum_k p_k \langle H_{\text{A}_k} \rangle_k
+\sum_k p_k
\big\langle G_0(H_{SA_k})\big\rangle_k\\
\nonumber
&=H_{S}+H^{(0)},
\end{align}
where $H^{(0)}=\sum_k \, p_k \, 
\big\langle G_0(H_{SA_k})\big\rangle_k$ as in \eqref{H0def}. Note that we have dropped the term proportional to the identity, and used $\langle X \rangle_k\coloneqq\TrAk\!\big(\rho_{\text{A}_k} \, X\big)$ as defined in \eqref{langlerangledef}.

Thus, we have confirmed that the effective Hamiltonian at leading order is the free system Hamiltonian, $H_\text{S}$, plus a contribution from the repeated interactions, $H^{(0)}$, as claimed in \eqref{H0effdef}.

\section{Derivation of \texorpdfstring{$\mathcal{L}_1$}{text}}\label{AppL1}
In this appendix, we calculate $\mathcal{L}_1$, as defined in \eqref{L1def}. 

The calculation proceeds as follows:

1. First, we separate $U_{k,2}$, defined in \eqref{Ukseriesdef}, into $U_{k,1}{}^2/2$ plus a remainder. We then use this decomposition of $U_{k,2}$ to calculate $\bar{\phi}_2$, defined in \eqref{phibarseriesdef}. 

2. Then, recalling that $\mathcal{L}_1=\bar{\phi}_2-\bar{\phi}_1{}^2/2$, we separate this expression for $\bar{\phi}_2$ into $\bar{\phi}_1{}^2/2$ plus additional terms, which we identify as the Hamiltonian and dissipative terms claimed in \eqref{L1explicit}.

3. Finally, we rewrite these Hamiltonian and dissipative parts in the forms claimed in \eqref{H1effdef} and \eqref{Ddef}.

\subsection{Derivation of \texorpdfstring{$\bar{\phi}_2$}{phi bar 2}}
As mentioned above, we first seek to rewrite $U_{k,2}$, defined in \eqref{Ukseriesdef}, as
\bel{Uk2form}
U_{k,2}=U_{k,1}{}^2/2+u_{k,2}.
\ee
We compute
\begin{widetext}
\begin{align}
2(\ii \hbar)^2 U_{k,2}
&=2\int_{0}^1 d\xi_1
\int_{0}^{\xi_1} d\xi_2
H_k(\xi_1)H_k(\xi_2)\\
\nonumber
&=\int_{0}^1 d\xi_1
\int_{0}^{\xi_1} d\xi_2
H_k(\xi_1)H_k(\xi_2)
+\int_{0}^1 d\xi_2
\int_{0}^{\xi_2} d\xi_1
H_k(\xi_2)H_k(\xi_1)\\
\nonumber
&=\int_{0}^1 d\xi_1
\int_{0}^{\xi_1} d\xi_2
H_k(\xi_1)H_k(\xi_2)
+\int_{0}^1 d\xi_2
\int_{0}^{\xi_2} d\xi_1
H_k(\xi_1)H_k(\xi_2)
+\int_{0}^1 d\xi_2
\int_{0}^{\xi_2} d\xi_1
[H_k(\xi_2),H_k(\xi_1)]\\
\nonumber
&=\int_{0}^1 d\xi_1
\int_{0}^{\xi_1} d\xi_2
H_k(\xi_1)H_k(\xi_2)
+\int_{0}^1 d\xi_1
\int_{\xi_1}^{1} d\xi_2
H_k(\xi_1)H_k(\xi_2)
+\int_{0}^1 d\xi_2
\int_{0}^{\xi_2} d\xi_1
[H_k(\xi_2),H_k(\xi_1)]\\
\nonumber
&=\int_{0}^1 d\xi_1
\int_{0}^{1} d\xi_2
H_k(\xi_1)H_k(\xi_2)
+\int_{0}^1 d\xi_2
\int_{0}^{\xi_2} d\xi_1
[H_k(\xi_2),H_k(\xi_1)]\\
\nonumber
&=\Big(\int_{0}^1 H_k(\xi_1) d\xi_1\Big)
\Big(\int_{0}^{1} H_k(\xi_2) d\xi_2\Big)
+\int_{0}^1 d\xi_2
\int_{0}^{\xi_2} d\xi_1
[H_k(\xi_2),H_k(\xi_1)]\\
\nonumber
&=(\ii\hbar)^2 U_{k,1}{}^2
+\int_{0}^1 d\xi_1
\int_{0}^{\xi_1} d\xi_2
[H_k(\xi_1),H_k(\xi_2)],
\end{align}
\end{widetext}
where we recall from \eqref{Ukseriesdef} that \mbox{$(\ii\hbar) U_{k,1}=\int_{0}^1 H_k(\xi) d\xi$}.

To summarize the lengthy calculation in words: we have symmetrized over the dummy variables $\xi_1$ and $\xi_2$ and then manipulated the resultant terms to match each other in both integrand and order of integration. Specifically, in one term we switched the order of the operators $H_k(\xi_1)$ and $H_k(\xi_2)$ at the cost of adding a commutator term, we then changed the order of integration to match the other term. Combining these two integrals, we recognize the result as the product $U_{k,1}^2$.

Thus we have rewritten $U_{k,2}$ in the form \eqref{Uk2form}, where
\bel{uk2form}
u_{k,2}=\frac{1}{2(\ii\hbar)^2}\int_{0}^1 d\xi_1
\int_{0}^{\xi_1} d\xi_2
[H_k(\xi_1),H_k(\xi_2)].
\ee
We note that $U_{k,1}$ is antihermitian such that $U_{k,1}{}^2$ is Hermitian, whereas $u_{k,2}$ is antihermitian. Thus, we have
\be
U_{k,2}{}^\dagger=U_{k,1}{}^2/2-u_{k,2}.
\ee

Using this form for $U_{k,2}$, we can find $\bar{\phi}_{2}$, defined in \eqref{phibarseriesdef}, as
\begin{align}
2\bar{\phi}_2[\,\cdot\,]
&=2\!\sum_k p_k \TrAk\! 
\Big(U_{k,1}(\cdot\otimes\rho_{\text{A}_k})U_{k,1}{}^\dagger\\
&\nonumber 
+U_{k,2}(\cdot\otimes\rho_{\text{A}_k})
+(\cdot\otimes\rho_{\text{A}_k})U_{k,2}{}^\dagger\Big)\\
&\nonumber
=\sum_k p_k \TrAk\! 
\Big(-2 \, U_{k,1}(\cdot\otimes\rho_{\text{A}_k})U_{k,1}{}\\
&\nonumber
+(U_{k,1}{}^2+2u_{k,2})(\cdot\otimes\rho_{\text{A}_k})
+(\cdot\otimes\rho_{\text{A}_k})(U_{k,1}{}^2-2u_{k,2})\Big)\\
&\nonumber
=\sum_k p_k \TrAk\!\Big(
2[u_{k,2},\,\cdot\otimes\rho_{\text{A}_k}]
+\big[U_{k,1},[U_{k,1},\,\cdot\otimes\rho_{\text{A}_k}]\big]\Big)\\
&\nonumber
=2\sum_k p_k \TrAk\!\Big(
[u_{k,2},\,\cdot\otimes\rho_{\text{A}_k}]\Big)\\
&\nonumber
+\sum_k p_k \TrAk\!\Big(\big[U_{k,1},[U_{k,1},\cdot\otimes\rho_{\text{A}_k}]\big]\Big)\\
&\nonumber
=2\Big[
\sum_k p_k \TrAk\!\big(u_{k,2} \ \rho_{\text{A}_k}\big),\,\cdot\,\Big]\\
&\nonumber
+\sum_k p_k \TrAk\!\Big(\big[U_{k,1},[U_{k,1},\cdot\otimes\rho_{\text{A}_k}]\big]\Big),
\end{align}
so that
\begin{align}\label{phibar2end}
\bar{\phi}_2[\,\cdot\,]
& =\Big[
\sum_k p_k \TrAk\!\big(u_{k,2} \ \rho_{\text{A}_k}\big),\,\cdot\,\Big]\\
&\nonumber
+\frac{1}{2}\sum_k p_k \TrAk\!\Big(\big[U_{k,1},[U_{k,1},\cdot\otimes\rho_{\text{A}_k}]\big]\Big).
\end{align}
\subsection{Separation of 
\texorpdfstring{$\mathcal{L}_1$}{L 1}}\label{AppL1form}
Next, we seek to find an expression for $\mathcal{L}_1
=\bar{\phi}_2-\bar{\phi}_1{}^2/2$ separated into Hamiltonian and dissipative parts as claimed in \eqref{L1explicit}. In order to do this we expand the double $U_{k,1}$ commutator in $\bar{\phi}_2$ given by \eqref{phibar2end}. We then collect the terms which are manifestly Hamiltonian as well as those which make up $\bar{\phi}_1{}^2/2$.

To begin, we recall that from \eqref{Uk1explicit} we have
\be
(\ii\hbar)U_{k,1}
=H_\text{S}\otimes\boldsymbol{1}
+\boldsymbol{1}\otimes H_{\text{A}_k}
+G_0(H_{\text{SA}_k}).
\ee
Thus by linearity the double commutator in \eqref{phibar2end} involves picking pairs of these three terms. We analyze each of these possible combinations in turn, beginning with all those involving $H_{\text{A}_k}$.

If we pick $\boldsymbol{1}\otimes H_{\text{A}_k}$ in the outer commutator then the double commutator vanishes, since
\be
\TrAk\!\Big(
\big[\boldsymbol{1}\otimes H_{\text{A}_k},
[U_{k,1},
\,\cdot\,\otimes\rho_{\text{A}_k}]\big]\Big)
=0
\ee
due to cyclic property of partial trace, that is, the fact that $\TrAk \big( (\boldsymbol{1} \otimes X) Y \big) = \TrAk \big( Y (\boldsymbol{1} \otimes X) \big)$ for arbitrary operators $X$ and $Y$ on $\mathrm{A}_k$ and S-$\mathrm{A}_k$ respectively. Likewise, the terms with $H_\text{S}\otimes\boldsymbol{1}$ in the outer commutator and $\boldsymbol{1}\otimes H_{\text{A}_k}$ in the inner commutator vanish  \begin{align}
&\TrAk\!\Big(
\big[H_\text{S}\otimes\boldsymbol{1},
[\boldsymbol{1}\otimes H_{\text{A}_k},
\,\cdot\,\otimes\rho_{\text{A}_k}]\big]\Big)\\
&\nonumber
=\Big[H_\text{S},\TrAk\!\big(
[\boldsymbol{1}\otimes H_{\text{A}_k},
\,\cdot\,\otimes\rho_{\text{A}_k}]\big)\Big]\\
&\nonumber
=0
\end{align}
since the trace can freely move through the $H_\text{S}$ commutator and vanishes at the $H_{\text{A}_k}$ commutator by the aforementioned cyclic property of partial trace. 

However, the term with $G_0(H_{\text{SA}_k})$ in the outer commutator and $\boldsymbol{1}\otimes H_{\text{A}_k}$ in the inner commutator does not vanish. We find,
\begin{align}
&\sum_k p_k\TrAk\!\Big(
\big[G_0(H_{\text{SA}_k}),
[\boldsymbol{1}\otimes H_{\text{A}_k},
\,\cdot\,\otimes\rho_{\text{A}_k}]\big]\Big)\\
&\nonumber
=\sum_k p_k\TrAk\!\Big(
\big[G_0(H_{\text{SA}_k}),
\,\cdot\,\otimes[H_{\text{A}_k},
\rho_{\text{A}_k}]\big]\Big)\\
&\nonumber
=\sum_k p_k\Big[
\TrAk\!\big(G_0(H_{\text{SA}_k}) \ [H_{\text{A}_k},\rho_{\text{A}_k}]\big),
\,\cdot\,\Big]\\
&\nonumber
=\Big[\sum_k p_k
\TrAk\!\big(G_0(H_{\text{SA}_k}) \ [H_{\text{A}_k},\rho_{\text{A}_k}]\big),
\,\cdot\,\Big].
\end{align}
Thus, we have examined all terms involving $H_{\text{A}_k}$ and found that all but one vanish. 

Next, we handle all the remaining terms involving $H_\text{S}$. If we pick the terms with $H_\text{S}\otimes\boldsymbol{1}$ in both the inner and outer commutators, we find
\begin{align}
&\sum_k p_k\TrAk\!\Big(
\big[H_\text{S}\otimes\boldsymbol{1},
[H_\text{S}\otimes\boldsymbol{1},
\,\cdot\,\otimes\rho_{\text{A}_k}]\big]\Big)\\
&\nonumber
=\Big[H_\text{S},
\sum_k p_k\TrAk\!\big(
[H_\text{S}\otimes\boldsymbol{1},
\,\cdot\,\otimes\rho_{\text{A}_k}]\big)\Big]\\
&\nonumber
=\big[H_\text{S},[H_\text{S},\,\cdot\,]\big] \ 
\sum_k p_k\TrAk\!\big(\rho_{\text{A}_k}\big)\\
&\nonumber
=\big[H_\text{S},[H_\text{S},\,\cdot\,]\big],
\end{align}
since $\rho_{\text{A}_k}$ has unit trace and $\sum_k p_k=1$. On the other hand, if we pick the term with $H_\text{S}\otimes\boldsymbol{1}$ in the outer commutator and $G_0(H_{\text{SA}_k})$ in the inner commutator, we find
\begin{align}
&\sum_k p_k\TrAk\!\Big(
\big[H_\text{S}\otimes\boldsymbol{1},
[G_0(H_{\text{SA}_k}),
\,\cdot\,\otimes\rho_{\text{A}_k}]\big]\Big)\\
&\nonumber
=\sum_k p_k\Big[H_\text{S},
\TrAk\!\big([G_0(H_{\text{SA}_k}),
\,\cdot\,\otimes\rho_{A_k}]\big)\Big]\\
&\nonumber
=\sum_k p_k\Big[H_\text{S},
\big[\TrAk\!(G_0(H_{\text{SA}_k})\rho_{\text{A}_k}),
\,\cdot\,\big]\Big]\\
&\nonumber
=\Big[H_\text{S},
\big[\sum_k p_k\TrAk\!(G_0(H_{\text{SA}_k})\rho_{\text{A}_k}),
\,\cdot\,\big]\Big]\\
&\nonumber
=\big[H_\text{S},[H^{(0)},\,\cdot\,]\big],
\end{align}
where we recall $H^{(0)}=\sum_k p_k\TrAk\!(G_0(H_{\text{SA}_k})\rho_{\text{A}_k})$ from \eqref{H0def}. Finally, if we pick the term with $G_0(H_{\text{SA}_k})$ in the outer commutator and $H_\text{S}\otimes\boldsymbol{1}$ in the inner commutator, we find
\begin{align}
&\sum_k p_k\TrAk\!\Big(
\big[G_0(H_{\text{SA}_k}),
[H_\text{S}\otimes\boldsymbol{1},
\,\cdot\,\otimes\rho_{\text{A}_k}]\big]\Big)\\
&\nonumber
=\sum_k p_k\TrAk\!\Big(
\big[G_0(H_{\text{SA}_k}),
[H_\text{S},
\,\cdot\,\big]\otimes\rho_{\text{A}_k}]\Big)\\
&\nonumber
=\sum_k p_k\Big[\TrAk\!\big(G_0(H_{\text{SA}_k})\rho_{\text{A}_k}\big),
[H_\text{S},\,\cdot\,]\Big]\\
&\nonumber
=\Big[\sum_k p_k\TrAk\!\big(G_0(H_{\text{SA}_k})\rho_{\text{A}_k}\big),
[H_\text{S},\,\cdot\,]\Big]\\
&\nonumber
=\big[H^{(0)},
[H_\text{S},\,\cdot\,]\big],
\end{align}
where we have again recalled from \eqref{H0def} that $H^{(0)}=\sum_k p_k\TrAk\!(G_0(H_{\text{SA}_k})\rho_{\text{A}_k})$. Thus, in full we have
\begin{align}
&\bar{\phi}_2[\,\cdot\,]
=\Big[
\sum_k p_k \TrAk\!\big(u_{k,2} \ \rho_{\text{A}_k}\big),\,\cdot\,\Big]\\
&\nonumber
+\frac{1}{2(\ii\hbar)^2}
\Big[\sum_k p_k
\TrAk\!\big(G_0(H_{\text{SA}_k}) \ [H_{\text{A}_k},\rho_{\text{A}_k}]\big),
\,\cdot\,\Big]\\
&\nonumber
+\frac{1}{2(\ii\hbar)^2}
\Big(\big[H_\text{S},[H_\text{S},\,\cdot\,]\big]
+\big[H_\text{S},[H^{(0)}\!\!,\,\cdot\,]\big]
+\big[H^{(0)}\!\!,[H_\text{S},\,\cdot\,]\big]\Big)\\
&\nonumber
+\frac{1}{2(\ii\hbar)^2}
\sum_k p_k\TrAk\!\Big(
\big[G_0(H_{\text{SA}_k}),
[G_0(H_{\text{SA}_k}),
\,\cdot\,\otimes\rho_{\text{A}_k}]\big]\Big).
\end{align}
The first two terms are both single commutators and so they can be combined together into a Hamiltonian term. Additionally, from \eqref{L0def} and \eqref{L0explicit} we see that the next three terms almost have the form of
\begin{align}
\frac{1}{2}\bar{\phi}_1{}^2[\,\cdot\,]
&=\frac{1}{2(\ii\hbar)^2}
\big[H_\text{eff}^{(0)},[H_\text{eff}^{(0)},\,\cdot\,]\big]\\
&\nonumber
=\frac{1}{2(\ii\hbar)^2}
\big[H_\text{S}+H^{(0)},[H_\text{S}+H^{(0)},\,\cdot\,]\big];
\end{align}
they are missing only the term $(\ii\hbar)^{-2}\big[H^{(0)},[H^{(0)},\,\cdot\,]\big]/2$. Using these observations, we have
\begin{align}
&\bar{\phi}_2[\,\cdot\,]
=\frac{-\ii}{\hbar}
\Big[H_\text{eff}^{(1)},\,\cdot\,\Big]\\
&\nonumber
+\frac{1}{2}\bar{\phi}_1{}^2[\,\cdot\,]
-\frac{1}{2(\ii\hbar)^2}
\big[H^{(0)},[H^{(0)},\,\cdot\,]\big]\\
&\nonumber
+\frac{1}{2(\ii\hbar)^2}
\sum_k p_k\TrAk\!\Big(
\big[G_0(H_{\text{SA}_k}),
[G_0(H_{\text{SA}_k}),
\,\cdot\,\otimes\rho_{\text{A}_k}]\big]\Big),
\end{align}
where
\begin{align}\label{H1effdefApp}
H_\text{eff}^{(1)}
&=(\ii\hbar)\sum_k p_k \TrAk\!\big(u_{k,2} \ \rho_{\text{A}_k}\big)\\
&\nonumber
+\frac{1}{2(\ii\hbar)}\sum_k p_k
\TrAk\!\big(G_0(H_{\text{SA}_k}) \ [H_{\text{A}_k},\rho_{\text{A}_k}]\big).
\end{align}
Finally, recalling from \eqref{L1def} that $\mathcal{L}_1
=\bar{\phi}_2-\bar{\phi}_1{}^2/2$,
we have
\be
\mathcal{L}_1[\,\cdot\,]
=\frac{-\ii}{\hbar}
\big[H_\text{eff}^{(1)},\,\cdot\,\big]
+\frac{1}{2}\mathcal{D}[\,\cdot\,],
\ee
where 
\begin{align}\label{AppDdef}
&\mathcal{D}[\,\cdot\,]
\coloneqq\frac{-1}{(\ii\hbar)^2}
\big[H^{(0)},[H^{(0)},\,\cdot\,]\big]\\
&\nonumber
+\frac{1}{(\ii\hbar)^2}
\sum_k p_k\TrAk\!\Big(
\big[G_0(H_{\text{SA}_k}),
[G_0(H_{\text{SA}_k}),
\,\cdot\,\otimes\rho_{\text{A}_k}]\big]\Big).
\end{align}
Thus we have confirmed the form of $\mathcal{L}_1$ claimed in Eq.~\eqref{L1explicit}.

\subsection{Simplifying \texorpdfstring{$H_\text{eff}^{(1)}$}{H eff 1}}
In Appendix \ref{AppL1form}, we showed that the unitary part of \eqref{L1explicit} takes the form \eqref{H1effdefApp}, namely
\begin{align}\label{H1effdefApp2}
H_\text{eff}^{(1)}
&\coloneqq(\ii\hbar)\sum_k p_k \TrAk\!\big(u_{k,2} \ \rho_{\text{A}_k}\big)\\
&\nonumber
+\frac{1}{2(\ii\hbar)}\sum_k p_k
\TrAk\!\big(G_0(H_{\text{SA}_k}) \ [H_{\text{A}_k},\rho_{\text{A}_k}]\big),
\end{align}
where from \eqref{uk2form} we have
\be
u_{k,2}=\frac{1}{2(\ii\hbar)^2}\int_{0}^1 d\xi_1
\int_{0}^{\xi_1} d\xi_2
[H_k(\xi_1),H_k(\xi_2)].
\ee

Recall that our Hamiltonian is given by \eqref{Hamform} as
\be 
H_k(\xi)
=H_\text{S}\otimes\boldsymbol{1}+\boldsymbol{1}\otimes H_{\text{A}_k}+H_{\text{SA}_k}(\xi).
\ee
In this subsection we will drop the $\boldsymbol{1}$'s for convenience and define an aggregate free Hamiltonian  $H_{0,k}=H_\text{S}+H_{\text{A}_k}$ such that
\begin{align}
H_k(\xi)
&=H_\text{S}+H_{\text{A}_k}+H_{\text{SA}_k}(\xi)\\
&\nonumber
=H_{0,k}+H_{\text{SA}_k}(\xi).
\end{align} 
Thus we can expand $u_{k,2}$ as
\begin{align}
2(\ii\hbar)^2 u_{k,2}
&=\int_{0}^1 d\xi_1
\int_{0}^{\xi_1} d\xi_2
[H_k(\xi_1),H_k(\xi_2)]\\
&\nonumber
=\int_{0}^1 d\xi_1
\int_{0}^{\xi_1} d\xi_2
\underbrace{[H_{0,k},H_{0,k}]}_{=0}\\[-2mm]
&\nonumber
+\int_{0}^1 d\xi_1
\int_{0}^{\xi_1} d\xi_2
[H_{0,k},H_{SA_k}(\xi_2)]\\
&\nonumber
+\int_{0}^1 d\xi_1
\int_{0}^{\xi_1} d\xi_2
[H_{SA_k}(\xi_1),H_{0,k}]\\
&\nonumber
+\int_{0}^1 d\xi_1
\int_{0}^{\xi_1} d\xi_2
[H_{SA_k}(\xi_1),H_{SA_k}(\xi_2)].
\end{align}
We can simplify the first two terms as
\begin{align}
&\int_{0}^1 d\xi_1
\int_{0}^{\xi_1} d\xi_2
[H_{\text{SA}_k}(\xi_1),H_{0,k}]\\
&\nonumber
+\int_{0}^1 d\xi_1
\int_{0}^{\xi_1} d\xi_2
[H_{0,k},H_{\text{SA}_k}(\xi_2)]\\
&\nonumber
=\int_{0}^1 d\xi_1
\int_{0}^{\xi_1} d\xi_2
[H_{\text{SA}_k}(\xi_1),H_{0,k}]\\
&\nonumber
-\int_{0}^1 d\xi_1
\int_{0}^{\xi_1} d\xi_2
[H_{\text{SA}_k}(\xi_2),H_{0,k}]\\
&\nonumber
=\int_{0}^1 d\xi_1
\int_{0}^{\xi_1} d\xi_2
[H_{\text{SA}_k}(\xi_1),H_{0,k}]\\
&\nonumber
-\int_{0}^1 d\xi_2
\int_{0}^{\xi_2} d\xi_1
[H_{\text{SA}_k}(\xi_1),H_{0,k}]\\
&\nonumber
=\int_{0}^1 d\xi_1
\int_{0}^{\xi_1} d\xi_2
[H_{\text{SA}_k}(\xi_1),H_{0,k}]\\
&\nonumber
-\int_{0}^1 d\xi_1
\int_{\xi_1}^{1} d\xi_2
[H_{\text{SA}_k}(\xi_1),H_{0,k}]\\
&\nonumber
=\int_{0}^1 d\xi_1
\xi_1 \ 
[H_{\text{SA}_k}(\xi_1),H_{0,k}]\\
&\nonumber
-\int_{0}^1 d\xi_1
(1-\xi_1) \ 
[H_{\text{SA}_k}(\xi_1),H_{0,k}]\\
&\nonumber
=\int_{0}^1 d\xi_1
(2\xi_1-1) \ 
[H_{\text{SA}_k}(\xi_1),H_{0,k}]\\
&\nonumber
=2 \, G_1\big([H_{\text{SA}_k}(\xi),H_{0,k}]\big)\\
&\nonumber
=2 \, G_1\big([H_{\text{SA}_k}(\xi),H_\text{S}]\big)
+2 \, G_1\big([H_{\text{SA}_k}(\xi),H_{\text{A}_k}]\big),
\end{align}
where $G_1(X(\xi)):=\int_{0}^1 \ (\xi-1/2)X(\xi) \ d\xi$ as in \eqref{G1def}. Additionally, defining $G_3(X(\xi_1,\xi_2))=
\frac{1}{2}\int_{0}^1 d\xi_1
\int_{0}^{\xi_1} d\xi_2 X(\xi_1,\xi_2)$ as in \eqref{G3def}, we have
\begin{align}
(\ii\hbar)^2 u_{k,2}
&=
G_1\big([H_{\text{SA}_k}(\xi),H_\text{S}]\big)\\
&\nonumber
+G_1\big([H_{\text{SA}_k}(\xi),H_{\text{A}_k}]\big)\\
&\nonumber
+G_3\big(
[H_{\text{SA}_k}(\xi_1),H_{\text{SA}_k}(\xi_2)]\big).
\end{align}
The $u_{k,2}$ term in $H_\text{eff}^{(1)}$ is thus
\begin{align}\label{uk2part}
&(\ii\hbar)\sum_k p_k \TrAk\!\big(u_{k,2} \ \rho_{\text{A}_k}\big)\\
&\nonumber
=\frac{1}{\ii\hbar}\sum_k p_k \TrAk\!\Big(G_1\big([H_{\text{SA}_k}(\xi),H_\text{S}]\big) \ \rho_{\text{A}_k}\Big)\\
&\nonumber
+\frac{1}{\ii\hbar}\sum_k p_k \TrAk\!\Big(G_1\big([H_{\text{SA}_k}(\xi),H_{\text{A}_k}]\big) \ \rho_{\text{A}_k}\Big)\\
&\nonumber
+\frac{1}{\ii\hbar}\sum_k p_k \TrAk\!\Big(G_3\big(
[H_{\text{SA}_k}(\xi_1),H_{\text{SA}_k}(\xi_2)]\big) \ \rho_{\text{A}_k}\Big)\\
&\nonumber
=\frac{1}{\ii\hbar}\sum_k p_k \Big\langle G_1\big([H_{\text{SA}_k}(\xi),H_\text{S}]\big)\Big\rangle\\
&\nonumber
+\frac{1}{\ii\hbar}\sum_k p_k \Big\langle G_1\big([H_{\text{SA}_k}(\xi),H_{\text{A}_k}]\big)\Big\rangle\\
&\nonumber
+\frac{1}{\ii\hbar}\sum_k p_k \Big\langle G_3\big(
[H_{\text{SA}_k}(\xi_1),H_{\text{SA}_k}(\xi_2)]\big)\Big\rangle.
\end{align}
The other term in $H_\text{eff}^{(1)}$ can be rewritten as
\begin{align}\label{otherpart}
&\frac{1}{2(\ii\hbar)}\sum_k p_k
\TrAk\!\big(G_0(H_{\text{SA}_k}) \, [H_{\text{A}_k},\rho_{\text{A}_k}]\big)\\
&\nonumber
=\frac{1}{2(\ii\hbar)}\sum_k p_k
\TrAk\!\big([G_0(H_{\text{SA}_k}), H_{\text{A}_k}]\, \rho_{\text{A}_k}\big).
\end{align}
This follows quickly from the aforementioned cyclic property of partial trace and is analogous to the identity for the full trace, $\text{Tr}(X \, [Y,Z])=\text{Tr}([X,Y] \, Z)$. 

Finally, combining \eqref{uk2part} and \eqref{otherpart}, we get 
\begin{align}
H_\text{eff}^{(1)}
&=\frac{1}{\ii\hbar}\sum_k p_k \Big\langle G_1\big([H_{\text{SA}_k},H_\text{S}]\big)\Big\rangle\\
&\nonumber
+\frac{1}{\ii\hbar}\sum_k p_k \Big\langle G_2\big([H_{\text{SA}_k},H_{\text{A}_k}]\big)\Big\rangle\\
&\nonumber
+\frac{1}{\ii\hbar}\sum_k p_k \Big\langle G_3\big(
[H_{\text{SA}_k}(\xi_1),H_{\text{SA}_k}(\xi_2)]\big)\Big\rangle,
\end{align}
where $G_2=G_1+\frac{1}{2}G_0$. Thus we have $H_\text{eff}^{(1)}=H_1^{(1)}+H_2^{(1)}+H_3^{(1)}$ as claimed in \eqref{H1effdef}, where
\begin{align}
H_1^{(1)}
&\coloneqq
\sum_k p_k \ 
\Big\langle G_1\Big(
\frac{-\ii}{\hbar}
[H_{\text{SA}_k}(\xi),H_\text{S}]
\Big)\Big\rangle_k,\\
H_2^{(1)}
&\coloneqq
\sum_k p_k \ 
\Big\langle G_2\Big(
\frac{-\ii}{\hbar}
[H_{\text{SA}_k}(\xi),H_{\text{A}_k}]
\Big)\Big\rangle_k,\\
H_3^{(1)}
&\coloneqq
\sum_k p_k \ 
\Big\langle G_3\Big(
\frac{-\ii}{\hbar}
[H_{\text{SA}_k}(\xi_1),
H_{\text{SA}_k}(\xi_2)]
\Big)\Big\rangle_k,
\end{align}
as claimed in Eqs.~\eqref{H1def},
\eqref{H2def}, and
\eqref{H3def}.

\subsection{Simplifying \texorpdfstring{$\mathcal{D}[\,\cdot\,]$}{D}}
In Appendix \ref{AppL1form}, we showed that the dissipative part of \eqref{L1explicit} takes the form \eqref{AppDdef}, namely
\begin{align}\label{AppDdef2}
&\mathcal{D}[\,\cdot\,]
=\frac{-1}{(\ii\hbar)^2}
\big[H^{(0)},[H^{(0)},\,\cdot\,]\big]\\
&\nonumber
+\frac{1}{(\ii\hbar)^2}
\sum_k p_k\TrAk\!\Big(
\big[G_0(H_{\text{SA}_k}),
[G_0(H_{\text{SA}_k}),
\,\cdot\,\otimes\rho_{A_k}]\big]\Big),
\end{align}
as claimed in \eqref{Ddef}, where $H^{(0)}$ and $G_0$ are defined in \eqref{H0def} and \eqref{G0def} respectively.

Here, we seek to rewrite $\mathcal{D}$ in the form claimed in \eqref{DdefVar}. In order to do this we define 
\bel{KdefApp}
C_k[\rho_{\text{SA}_k}]\coloneqq
\frac{-\ii}{\hbar}[G_0(H_{\text{SA}_k}),\rho_{\text{SA}_k}]
\ee
as in \eqref{Kdef}, and
\bel{AvgdefApp}
\llangle C_k\rrangle[\rho_{\text{SA}_k}]
=\rho_{\text{A}_k}\otimes\sum_l p_l \, \TrAl \big(C_l[\rho_{\text{SA}_l}]\big)
\ee
as in \eqref{Avgdef}. We can immediately rewrite the second term in \eqref{AppDdef2} as
\begin{align}\label{varpart2}
&\nonumber
\frac{1}{(\ii\hbar)^2}
\sum_k p_k\TrAk\!\Big(
\big[G_0(H_{\text{SA}_k}),
[G_0(H_{\text{SA}_k}),
\,\cdot\,\otimes\rho_{\text{A}_k}]\big]\Big)\\
&=\sum_k p_k \, \TrAk\!\Big(
C_k\big[C_k[\,\cdot\,\otimes\rho_{\text{A}_k}]\big]\Big)\\
&\nonumber
=\sum_l p_l \, \TrAl\Big(\rho_{\text{A}_l}\otimes\sum_k p_k \, \TrAk\!\big(
C_k\big[C_k[\,\cdot\,\otimes\rho_{\text{A}_k}]\big]\big)\Big)\\
&\nonumber
=\sum_l p_l \, \TrAl\Big(\llangle C_k{}^2\rrangle[\,\cdot\,\otimes\rho_{\text{A}_k}]\Big).
\end{align}
Similarly, we can rewrite each $H^{(0)}$ commutator in the first term of \eqref{AppDdef2} as
\begin{align}
\frac{1}{\ii\hbar}[H^{(0)},\,\cdot\,]
&=\frac{-\ii}{\hbar}\Big[\sum_k p_k\TrAk\!\big(G_0(H_{\text{SA}_k})\rho_{\text{A}_k}\big),
\,\cdot\,\Big]\\
&\nonumber
=\sum_k p_k\TrAk\!\Big(\frac{-\ii}{\hbar}[G_0(H_{\text{SA}_k}),
\,\cdot\,\otimes\rho_{\text{A}_k}]\Big)\\
&\nonumber
=\sum_k p_k\TrAk\!\big(C_k[\,\cdot\,\otimes\rho_{\text{A}_k}]\big)\\
&\nonumber
=\sum_l p_l\TrAl\Big(\rho_{\text{A}_l}\otimes\sum_k p_k\TrAk\!\big(C_k[\,\cdot\,\otimes\rho_{\text{A}_k}]\big)\Big)\\
&\nonumber
=\sum_l p_l\TrAl\big(\llangle C_k\rrangle [\,\cdot\,\otimes\rho_{\text{A}_k}]\big).
\end{align}
Thus we can express the double commutator as
\begin{align}\label{varpart1}
&\frac{1}{(\ii\hbar)^2}
\big[H^{(0)},[H^{(0)},\,\cdot\,]\big]\\
&\nonumber
=\sum_m p_m\text{Tr}_{\text{A}_m}\Big(\llangle C_n\rrangle \big[\rho_{\text{A}_n}\otimes
\sum_l p_l\TrAl\big(\llangle C_k\rrangle [\rho_{\text{A}_k}\otimes\,\cdot\,]\big)\big]\Big)\\
&\nonumber
=\sum_m p_m\text{Tr}_{\text{A}_m}\Big(\llangle C_n\rrangle \big[\llangle\llangle C_k\rrangle\rrangle [\rho_{\text{A}_k}\otimes\,\cdot\,]\big)\big]\Big)\\
&\nonumber
=\sum_m p_m\text{Tr}_{\text{A}_m}\big(\llangle C_n\rrangle \big[\llangle C_k\rrangle [\rho_{\text{A}_k}\otimes\,\cdot\,]\big)\big]\Big)\\
&\nonumber
=\sum_m p_m\text{Tr}_{\text{A}_m}\big(\llangle C_k\rrangle^2 [\,\cdot\,\otimes\rho_{\text{A}_k}]\big).
\end{align}
In the above calculation, we temporarily switched the order of the system and the ancilla in the tensor product for convenience. We also used the result $\llangle\llangle C_k\rrangle\rrangle=\llangle C_k\rrangle$ which is proven later in this section. 

Thus, combining \eqref{varpart1} and \eqref{varpart2}, we arrive at
\bel{DdefVarApp}
\mathcal{D}[\rho_\text{S}]
=\sum_k p_k \, \TrAk \big(\Var(C_k)[\rho_\text{S}\otimes\rho_{\text{A}_k}]\big),
\ee
where $\Var(C_k)=\llangle C_k\rrangle^2-\llangle C_k{}^2\rrangle$.

In order to interpret this ``variance'' we must first prove some properties of the average $\llangle\,\cdot\,\rrangle$. Firstly, as claimed above,
\begin{align}
&\llangle\llangle C_k \rrangle\rrangle[\rho_{\text{SA}_k}]\\
&\nonumber
=\rho_{\text{A}_n}\otimes\sum_k p_k \, \TrAk \big( \rho_{\text{A}_k}\otimes\sum_l p_l \, \TrAl \big(C_l [\rho_{\text{SA}_l}]\big)\big)\\
&\nonumber
=\rho_{\text{A}_n}\otimes\sum_k p_k \, \sum_l p_l \, \TrAl \big(C_l [\rho_{\text{SA}_l}]\big)\big)\\
&\nonumber
=\rho_{\text{A}_n}\otimes\sum_l p_l \, \TrAl \big(C_l [\rho_{\text{SA}_l}]\big)\big)\\
&\nonumber
=\llangle C_k\rrangle[\rho_{\text{SA}_k}].
\end{align}
Secondly, we show
\begin{align}
\llangle C_k \, X\rrangle[\rho_{\text{SA}_k}]
&=\rho_{\text{A}_k}\otimes\sum_l p_l \, \TrAl \big(C_l \, X[\rho_{\text{SA}_l}]\big)\\
&\nonumber
=\rho_{\text{A}_k}\otimes\sum_l p_l \, \TrAl \big(C_l \big[X[\rho_{\text{SA}_l}]\big]\big)\\
&\nonumber
=\llangle C_k\rrangle\big[X[\rho_{\text{SA}_k}]\big]\\
&\nonumber
=\llangle C_k\rrangle X[\rho_{\text{SA}_k}],
\end{align}
so that $\llangle C_k \, X\rrangle =\llangle C_k\rrangle X$. Specifically, this means 
\be
\llangle C_k\llangle C_l\rrangle\rrangle 
=\llangle C_k\rrangle \llangle C_l\rrangle
=\llangle C_k\rrangle^2.
\ee
Note, however, that $\llangle X \, C_k\rrangle \neq X\llangle C_k\rrangle$. Finally, we show,
\begin{align}
&\llangle \llangle C_k\rrangle C_l\rrangle
[\rho_{\text{SA}_k}]\\
&\nonumber
=\rho_{\text{A}_n}\otimes\sum_k p_k \, \TrAk \big( \rho_{\text{A}_k}\otimes\sum_l p_l \, \TrAl \big(C_l\big[C_l [\rho_{\text{SA}_l}]\big]\big)\big)\\
&\nonumber
=\rho_{\text{A}_n}\otimes\sum_k p_k \, \sum_l p_l \, \TrAl \big(C_l\big[C_l[ [\rho_{\text{SA}_l}]\big]\big)\big)\\
&\nonumber
=\rho_{\text{A}_n}\otimes\sum_l p_l \, \TrAl \big(C_l\big[C_l[\rho_{\text{SA}_l}]\big]\big)\big)\\
&\nonumber
=\llangle C_k{}^2\rrangle.
\end{align}

Thus, we can interpret 
\begin{align}
\Var(C_k)&=\llangle C_k{}^2\rrangle
-\llangle C_k\rrangle^2\\
&\nonumber
=\llangle\llangle C_k\rrangle C_l\rrangle
-\llangle C_l\llangle C_k\rrangle\rrangle\\
&\nonumber
=\llangle\llangle C_k\rrangle C_l
-C_l\llangle C_k\rrangle\rrangle\\
&\nonumber
=\Big\llangle\big[\llangle C_k\rrangle, C_l\big]\Big\rrangle.
\end{align}
As an aside, naively, we might think to interpret the variance as
\be
\Big\llangle\big(C_l-\llangle C_k\rrangle\big)^2\Big\rrangle.
\ee
However, a simple computation shows this expression to vanish:
\begin{align}
&\Big\llangle\big(C_l-\llangle C_k\rrangle\big)^2\Big\rrangle\\
&\nonumber
=\big\llangle
C_l{}^2
-\llangle C_k\rrangle C_l
-C_l\llangle C_k\rrangle
+\llangle C_k\rrangle^2
\big\rrangle\\
&\nonumber
=\llangle C_l{}^2\rrangle
-\big\llangle\llangle C_k\rrangle C_l\big\rrangle
-\big\llangle C_l\llangle C_k\rrangle\big\rrangle
+\big\llangle\llangle C_k\rrangle^2
\big\rrangle\\
&\nonumber
=\llangle C_l{}^2\rrangle
-\llangle C_k{}^2\rrangle
-\llangle C_l\rrangle^2
+\llangle C_k\rrangle^2\\
&\nonumber
=0.
\end{align}

\section{Derivation of Lindblad Form}\label{AppLindblad}
In this section, we convert the dissipative part of the dynamics \eqref{truncatedmastereqs} to the Lindblad form shown in \eqref{LindbladForm}. From \eqref{truncatedmastereqs} we see that
\be 
\mathcal{H}=
H_\text{eff}^{(0)}
+\delta t \ H_\text{eff}^{(1)}
+\mathcal{O}(\delta t^2).
\ee
The dissipative part of \eqref{truncatedmastereqs} takes the form \eqref{Ddef}, namely
\begin{align}\label{Ddef2}
&\mathcal{D}[\rho_\text{S}]
=\frac{1}{\hbar^2}
\big[H^{(0)},[H^{(0)},\rho_\text{S}]\big]\\
\nonumber
&-\frac{1}{\hbar^2}\sum_k p_k \, \TrAk \Big(
\big[G_0(H_{\text{SA}_k}),[G_0(H_{\text{SA}_k}),
\rho_\text{S}\otimes\rho_{\text{A}_k}]\big]\Big),
\end{align}
where $H^{(0)}$ and $G_0$ are defined in \eqref{H0def} and \eqref{G0def} respectively. To put this into Lindblad form, we begin by attempting to rewrite \eqref{Ddef2} in the form
\bel{LindbladDerivation0}
\mathcal{D}[\rho_\text{S}]=\{G,\rho_\text{S}\}+\sum_n c_n \ F_n \rho_\text{S} F_n^\dagger
\ee
for some operators $F_n$ and $G$ and some positive numbers $c_n$. Once we have our dissipative part in this form \eqref{LindbladDerivation0}, we can convert it into Lindblad form \eqref{LindbladForm} by the following argument: 

First, note that, since the dynamics are trace preserving, we have $\text{Tr}_{\text{S}}\big(\mathcal{D}[\rho_\text{S}]\big)=0$ for every $\rho_\text{S}$. Thus
\begin{align}
0
& =\text{Tr}_{\text{S}}\big(\{G,\rho_\text{S}\}
+\sum_n c_n \, 
F_n \rho_\text{S} F_n^\dagger\big)\\
&\nonumber
=\text{Tr}_{\text{S}}\big(2G\rho_\text{S}
+\sum_n c_n \, 
F_n^\dagger F_n\rho_\text{S} \big)\\
&\nonumber
=\text{Tr}_{\text{S}}\Big(\big(2G
+\sum_n c_n \, 
F_n^\dagger F_n\big)\rho_\text{S}\Big).
\end{align}
Since the holds for every $\rho_\text{S}$, we must have 
\be 
G=-\frac{1}{2}\sum_n c_n \ 
F_n^\dagger F_n.
\ee
Thus, we can rewrite
\begin{align}\label{LindbladDerivation1}
\mathcal{D}[\rho_\text{S}]
&=\{G,\rho_\text{S}\}
+\sum_n c_n \ 
F_n \rho_\text{S} F_n^\dagger\\
&\nonumber
=\sum_n c_n\Big(
F_n \rho_\text{S} F_n^\dagger
-\frac{1}{2}F_n^\dagger F_n \, \rho_\text{S}
-\frac{1}{2}\rho_\text{S} \, F_n^\dagger F_n\Big)\\
&\nonumber
=\sum_n c_n \ 
L(F_n)[\rho_\text{S}],
\end{align}
where $L(F_n)[\rho_\text{S}]=
F_n \rho_\text{S} F_n^\dagger
-F_n^\dagger F_n \, \rho_\text{S}/2
-\frac{1}{2}\rho_\text{S} \, F_n^\dagger F_n/2$ is the Lindblad superoperator. From \eqref{LindbladDerivation0} we can read off the decoherence modes $F_n$ and, including the $\delta t/2$ factor in front of $\mathcal{D}$ in \eqref{truncatedmastereqs}, the decoherence rates $\Gamma_n=\delta t \, c_n/2$. Thus, we seek to write each of the two terms in \eqref{Ddef2} in the form \eqref{LindbladDerivation0}. 

First, we rewrite the second term in \eqref{Ddef2} as
\begin{align}\label{LindbladDerivation2}
&\sum_k p_k \TrAk\!\Big(
\big[G_0(H_{SA_k}),[G_0(H_{SA_k}),\rho_\text{S}\otimes\rho_{\text{A}_k}]\big]\Big)\\
&\nonumber
=\{G',\rho_\text{S}\}-\!2\!\sum_k p_k \TrAk\!\big(G_0(H_{SA_k})(\rho_\text{S}\otimes\rho_A) G_0(H_{SA_k})\big),
\end{align}
where we have used the identity
\bel{Identity0}
\big[H,[H,\rho]\big]=\{H^2,\rho\}-2H\rho H,
\ee
the linearity of partial trace, and defined $G'=\TrAk(G_0(H_{SA_k})^2\rho_{\text{A}_k})$.

 Next, we decompose $\rho_{\text{A}_k}$ as  
\be
\rho_{\text{A}_k}=\sum_{\alpha_k}\lambda_{\alpha_k}\ket{\alpha_k}\bra{\alpha_k},
\ee
and use this to rewrite
\begin{align}\label{LindbladDerivation3}
&\TrAk\!\Big(H(\rho_\text{S}\otimes \rho_{\text{A}_k})H\Big)\\
&\nonumber
= \TrAk\!\Big(H\big(\rho_\text{S}\otimes 
\sum_{\alpha_k}\lambda_{\alpha_k}\ket{\alpha_k}\bra{\alpha_k}\big)H\Big)\\
&\nonumber
= \sum_{\alpha_k,\beta_k}\lambda_{\alpha_k}\bra{\beta_k}\big( H(\rho_\text{S}\otimes 
\ket{\alpha_k}\bra{\alpha_k})H\big)\ket{\beta_k}\\
&\nonumber
= \sum_{\alpha_k,\beta_k}\lambda_{\alpha_k}\bra{\beta_k}H\ket{\alpha_k}\rho_\text{S} \bra{\alpha_k}H\ket{\beta_k}\\
&\nonumber
= \sum_{\alpha_k,\beta_k}\lambda_{\alpha_k}
\bra{\beta_k}H\ket{\alpha_k}\rho_\text{S} \big(\bra{\beta_k}H\ket{\alpha_k}\big)^\dagger,
\end{align}
for any Hermitian $H$.

Finally, using \eqref{LindbladDerivation3} we can rewrite \eqref{LindbladDerivation2} as
\begin{align}\label{LindbladDerivation4}
&\sum_k p_k \TrAk\!\Big(
\big[G_0(H_{SA_k}),[G_0(H_{SA_k}),\rho_\text{S}\otimes\rho_{\text{A}_k}]\big]\Big)\\
&\nonumber
=\{G',\rho_\text{S}\}\\
&\nonumber
-2\!\!\!\!\!\!\sum_{k,\alpha_k,\beta_k}\!\!\!\! p_k \, \lambda_{\alpha_k}
\bra{\beta_k}G_0(H_{SA_k})\ket{\alpha_k}\rho_\text{S} \big(\bra{\beta_k}G_0(H_{SA_k})\ket{\alpha_k}\big)^\dagger\\
&\nonumber
=\{G',\rho_\text{S}\}\\
&\nonumber
-2\!\!\!\!\!\!\sum_{k,\alpha_k,\beta_k}\!\!\!\! q_{(k,\alpha_k)}
\bra{\beta_k}G_0(H_{SA_k})\ket{\alpha_k}\rho_\text{S} \big(\bra{\beta_k}G_0(H_{SA_k})\ket{\alpha_k}\big)^\dagger,
\end{align}
where we have defined the probability vector $\bm{q}$ with dimension $N=\sum_k\text{dim}(\text{A}_k)$ and components $q_{(k,\alpha_k)}=p_k \lambda_{\alpha_k}$. Thus we have written the second term in \eqref{Ddef2} in the form \eqref{LindbladDerivation0}.

Next, we rewrite the first term in \eqref{Ddef2} in the form \eqref{LindbladDerivation0}. Using the identity \eqref{Identity0}, we have 
\bel{LindbladDerivation5}
\big[H^{(0)},[H^{(0)},\rho_\text{S}]\big]
=\{G'',\rho_\text{S}\}-2H^{(0)}\rho_\text{S} H^{(0)}{}^\dagger
\ee
for $G''=(H^{(0)})^2$ since $H^{(0)}$ is Hermitian.  

From equations \eqref{Ddef2} and \eqref{LindbladDerivation5} we naively read off a decoherence mode as $H^{(0)}$ with decoherence rate $-\delta t/\hbar^2\vert H^{(0)}\vert^2$, where $\vert H^{(0)}\vert$ is some energy scale associated with $H^{(0)}$. However, this is incorrect as decoherence rates must be positive. Thus there must be interference between the terms in \eqref{LindbladDerivation4} and \eqref{LindbladDerivation5}. To identify this interference, we rewrite $H^{(0)}$ as
\begin{align}
H^{(0)}
&=\sum_k p_k \, \TrAk\!(G_0(H_{SA_k})\rho_{\text{A}_k})\\
&\nonumber
=\sum_k p_k \, \sum_{\alpha_k}\lambda_{\alpha_k}\bra{\alpha_k}G_0(H_{SA_k})\ket{\alpha_k}\\
&\nonumber
=\sum_{k,\alpha_k}
q_{(k,\alpha_k)} \,  \bra{\alpha_k}G_0(H_{SA_k})\ket{\alpha_k}.
\end{align}
Such that,
\begin{align}
H_0\rho_\text{S} H_0{}^\dagger
=&\Big(\sum_{k,\alpha_k}
q_{(k,\alpha_k)} \,  \bra{\alpha_k}G_0(H_{SA_k})\ket{\alpha_k}\Big)\\
\nonumber
&\nonumber\rho_\text{S}
\Big(\sum_{l,\beta_l}
q_{(l,\beta_l)} \,
\bra{\beta_l}G_0(H_{SA_l})\ket{\beta_l}\Big)^\dagger\\
\label{LindbladDerivation6}
=&\nonumber
\sum_{k,\alpha_k}
\sum_{l,\beta_l}\!
q_{(k,\alpha_k)}
q_{(l,\beta_l)}\\
\nonumber
&\bra{\alpha_k}G_0(H_{SA_k})\ket{\alpha_k}\rho_\text{S}
\big(\bra{\beta_l}G_0(H_{SA_l})\ket{\beta_l}\big)^\dagger.
\end{align}

Finally, combining \eqref{Ddef2}, \eqref{LindbladDerivation4}, \eqref{LindbladDerivation5} and, \eqref{LindbladDerivation6} we find
\begin{align}
\mathcal{D}[\rho_\text{S}]
&=\{G''',\rho_\text{S}\}\\
\nonumber
&+\frac{2}{\hbar^2}\!\!\sum_{k,\alpha_k,\beta_k}\!\!\!\! q_{(k,\alpha_k)}\\
\nonumber
&\bra{\beta_k}G_0(H_{SA_k})\ket{\alpha_k}\rho_\text{S} \big(\bra{\beta_k}G_0(H_{SA_k})\ket{\alpha_k}\big)^\dagger\\
\nonumber
&-\frac{2}{\hbar^2}\sum_{k,\alpha_k}\sum_{l,\beta_l}
q_{(k,\alpha_k)} \ 
q_{(l,\beta_l)}\\
\nonumber
&\bra{\alpha_k}G_0(H_{SA_k})\ket{\alpha_k}\rho_\text{S}\big(\bra{\beta_l}G_0(H_{SA_l})\ket{\beta_l}\big)^\dagger,
\end{align}
for $G'''=G'+G''$. To manifest the interference discussed earlier, we transfer the terms with $\alpha_k=\beta_k$ from the first sum to the second sum. This yields,
\begin{align}\label{LindbladDerivation7}
\mathcal{D}[\rho_\text{S}]
&=\{G''',\rho_\text{S}\}\\
\nonumber
&+\frac{2}{\hbar^2}
\sum_k\sum_{\alpha_k\neq\beta_k}
q_{(k,\alpha_k)}\\
\nonumber
&\bra{\beta_k}G_0(H_{SA_k})\ket{\alpha_k}\rho_\text{S} \big(\bra{\beta_k}G_0(H_{SA_k})\ket{\alpha_k}\big)^\dagger\\
\nonumber
&+\frac{2}{\hbar^2}\sum_{k,\alpha_k}\sum_{l,\beta_l}
\Big(q_{(k,\alpha_k)}\delta_{(k,\alpha_k),(l,\beta_l)}-q_{(k,\alpha_k)} \ 
q_{(l,\beta_l)}\Big)\\
\nonumber
&\bra{\alpha_k}G_0(H_{SA_k})\ket{\alpha_k}\rho_\text{S}\big(\bra{\beta_l}G_0(H_{SA_l})\ket{\beta_l}\big)^\dagger.
\end{align}
In \eqref{LindbladDerivation7}, the first term is written in the form required by \eqref{LindbladDerivation0}, however the second term needs to be diagonalized. Defining the $N\times N$ matrix $\mathrm{\bm{Q}}$ with components $Q_{ij}=q_i\delta_{ij}-q_i q_j$ and the operators $L_{(k,\alpha_k)}=\bra{\alpha_k}G_0(H_{SA_k})\ket{\alpha_k}$ the second sum becomes
\bel{LindbladDerivation8}
\frac{2}{\hbar^2}\sum_{ij}
Q_{ij} \ L_i\rho_\text{S}L_j.
\ee
In order to cast this in the form  \eqref{LindbladDerivation0} we need to find $\mathrm{\bm{Q}}$'s eigenvalues and eigenvectors. 

We now prove several properties of $\mathrm{\bm{Q}}$ by associating it with a statistical variance. If we associate a vector of outcomes $\bm{X}$ to the the probability vector $\bm{q}$, we can compute the variance of $\bm{X}$ as
\begin{align}
\Var(\bm{X})
&=\sum_i q_i x_i{}^2
-\big(\sum_i q_i x_i\big)
\big(\sum_j q_j x_j\big)\\
&\nonumber
=\sum_{ij} x_i(q_i \delta_{ij}
- q_i q_j)x_j\\
&\nonumber
=\bm{X}^T\mathrm{\bm{Q}}\bm{X}.
\end{align}
Since the variance is always nonnegative, $\mathrm{\bm{Q}}$ is positive semidefinite. Moreover, noting that $\text{Tr}(\mathrm{\bm{Q}})=\sum_i q_i-q_i{}^2=1-\vert\bm{q}\vert^2\leq1$, we see that $\mathrm{\bm{Q}}$'s eigenvalue are bounded by 1 for every $\bm{q}$.

We proceed to diagonalize \eqref{LindbladDerivation8}. We denote the eigenvectors of $\mathrm{\bm{Q}}$ as $\bm{v}_m$ with components $v_{m,i}$ and the eigenvalues as $\gamma_m$. Using these we can rewrite $Q=U\Gamma U^\dagger$ where $\Gamma=\text{Diag}(\gamma_m)$ and $U$ is the unitary matrix constructed from taking the  eigenvectors as its columns. In components this is written as $Q_{ij}=\sum_m v_{m,i}\gamma_m v_{m,j}{}^*$. Using this we diagonalize \eqref{LindbladDerivation8} as
\begin{align}
\sum_{ij} Q_{ij} \ L_i\rho L_j{}^\dagger
& =\sum_{ij} \sum_m v_{m,i}\gamma_m v_{m,j}{}^* L_i\rho L_j{}^\dagger\\
&\nonumber
=\!\sum_m\!\gamma_m \big(\sum_{i}\!v_{m,i}L_i\big)  \rho \big(\sum_{j}\!v_{m,j}{}^* L_j{}^\dagger\big)\\
&\nonumber
=\sum_m \gamma_m \big(\sum_{i}v_{m,i}L_i\big)  \rho \big(\sum_{j} v_{m,j} L_j\big)^\dagger\\
&\nonumber
=\sum_m \gamma_m A_m \rho A_m{}^\dagger,
\end{align}
where 
\begin{align}
A_m
&\!=\!\sum_{i} v_{i,m} L_i
=\sum_{k,\alpha_k}\!\!\!v_{m,(k,\alpha_k)}\bra{\alpha_k}G_0(H_{\text{SA}_k})\ket{\alpha_k}.
\end{align}

Thus having the second term in \eqref{LindbladDerivation7} written in the form of \eqref{LindbladDerivation0}, we can put \eqref{Ddef2} in Lindblad form as described above. Doing so, we find
\begin{align}
\mathcal{D}[\rho_\text{S}]
\nonumber& =\frac{2}{\hbar^2}
\sum_k\sum_{\alpha_k\neq\beta_k}
q_{(k,\alpha_k)} \ 
L\Big(\bra{\beta_k}G_0(H_{\text{SA}_k})\ket{\alpha_k}\Big)[\rho_\text{S}]\\
&+\frac{2}{\hbar^2}
\sum_{m=0}^M \gamma_m \  L(A_m)[\rho_\text{S}],
\end{align}
as claimed in \eqref{DLindblad}.

\section{Derivation of Master equation for Product Interaction}\label{AppProduct}
In this section, we find the explicit form for \eqref{truncatedmastereqs}, assuming all ancillas to be identical for simplicity (thus $p_k=\delta_{k,0}$, and so we can drop the sum over $k$). We assume the interaction Hamiltonian to have the factorable form
\bel{ProdHint}
H_\text{SA}
=g(\xi)J_\text{S}\otimes J_\text{A}.
\ee 
To describe the effective system dynamics, we first need to compute \eqref{H0effdef}, \eqref{H1effdef}, and \eqref{Ddef}.

Using \eqref{H0def} and \eqref{ProdHint}, and recalling \eqref{langlerangledef}, namely, that $\langle X\rangle=
\text{Tr}_\text{A}(\rho_{\text{A}} X)$, we have
\begin{align}\label{H0defProd}
H^{(0)}
&=\Big\langle G_0\big(
H_\text{SA}
\big)\Big\rangle\\
&\nonumber
=\Big\langle G_0\big(
g(\xi)J_\text{S}\otimes J_\text{A}
\big)\Big\rangle\\
&\nonumber
=G_0\big(g(\xi)\big) \, 
\big\langle J_\text{A}\big\rangle \, 
J_\text{S}\\
&\nonumber
=g_0 \, \big\langle J_\text{A}\big\rangle \, 
J_\text{S},
\end{align}
where $g_0=G_0\big(g(\xi)\big)=\int_0^1 g(\xi)d\xi$. Thus, using \eqref{H0effdef}, we arrive at \eqref{H0effdefproduct}:
\begin{align}
H_\text{eff}^{(0)}
&=H_S+H^{(0)}\\
&\nonumber
=H_S+g_0 \, \big\langle J_\text{A}\big\rangle \, 
J_\text{S}.
\end{align}

To compute $H_\text{eff}^{(1)}$ we recall \eqref{H1effdef}, namely:
\be
H_\text{eff}^{(1)}
=H_1^{(1)}+H_2^{(1)}+H_3^{(1)}.
\ee

Using \eqref{H1def} and \eqref{ProdHint}, we have
\begin{align}\label{H1defProd}
H_1^{(1)}
&=\Big\langle G_1\Big(
\frac{-\ii}{\hbar}
[H_{\text{SA}}(\xi),H_\text{S}]
\Big)\Big\rangle_k\\
&\nonumber
=\Big\langle G_1\Big(
\frac{-\ii}{\hbar}
[g(\xi)J_\text{S}\otimes J_\text{A},H_\text{S}]
\Big)\Big\rangle_k\\
&\nonumber
=G_1\big(g(\xi)\big)
\big\langle J_\text{A}\big\rangle
\frac{-\ii}{\hbar}
[J_\text{S},H_\text{S}]\\
&\nonumber
=g_1
\big\langle J_\text{A}\big\rangle
\frac{-\ii}{\hbar}
[J_\text{S},H_\text{S}],
\end{align}
where $g_1=G_1\big(g(\xi)\big)=\int_0^1 (\xi-1/2)g(\xi)d\xi$.

Using \eqref{H2def}, and \eqref{ProdHint} we have
\begin{align}\label{H2defProd}
H_2^{(1)}
&=\Big\langle G_2\Big(
\frac{-\ii}{\hbar}
[H_{\text{SA}}(\xi),H_{\text{A}_k}]
\Big)\Big\rangle\\
&\nonumber
=\Big\langle G_2\Big(
\frac{-\ii}{\hbar}
[g(\xi)J_\text{S}\otimes J_\text{A},H_{\text{A}_k}]
\Big)\Big\rangle\\
&\nonumber
=G_2\big(g(\xi)\big)
\frac{-\ii}{\hbar}\big\langle
[J_\text{A},H_{\text{A}_k}]\big\rangle
J_\text{S},
\end{align}
where $g_2=G_2\big(g(\xi)\big)=\int_0^1 \xi g(\xi)d\xi$.

Using \eqref{H3def}, and \eqref{ProdHint} we have
\begin{align}\label{H3defProd}
H_3^{(1)}
&=\Big\langle G_3\Big(
\frac{-\ii}{\hbar}
[H_{\text{SA}}(\xi_1),
H_{\text{SA}}(\xi_2)]
\Big)\Big\rangle\\
&\nonumber
=\Big\langle G_3\Big(
\frac{-\ii}{\hbar}
[g(\xi_1)J_\text{S}\otimes J_\text{A},
g(\xi_2)J_\text{S}\otimes J_\text{A}]
\Big)\Big\rangle\\
&\nonumber
=G_3\big(g(\xi_1)g(\xi_2)\big)
\Big\langle \frac{-\ii}{\hbar}
\underbrace{[J_\text{S}\otimes J_\text{A},
J_\text{S}\otimes J_\text{A}]}_{=0}
\Big\rangle\\[-3mm]
&\nonumber
=0;
\end{align}
in other words, the interaction Hamiltonian now commutes with itself at all times. 

Combining \eqref{H1effdef}, \eqref{H1defProd}, \eqref{H2defProd} and \eqref{H3defProd}, we find
\be
H_\text{eff}^{(1)}
= g_1 \langle J_\text{A}\rangle \, \frac{-\ii}{\hbar}[J_\text{S},H_\text{S}]
+g_2\frac{-\ii}{\hbar}\big\langle[J_\text{A},H_\text{A}]\big\rangle J_\text{S},
\ee
as claimed in \eqref{H1effdefproduct}.

Finally we compute the dissipative part. Using \eqref{H0defProd} we have as the first term of \eqref{Ddef}
\begin{align}
\big[H^{(0)},[H^{(0)},\,\cdot\,]\big]
&\nonumber
=\big[g_0 \, \big\langle J_\text{A}\big\rangle \, 
J_\text{S},[g_0 \, \big\langle J_\text{A}\big\rangle \, 
J_\text{S},\,\cdot\,]\big]\\
&=g_0{}^2
\big\langle J_\text{A}\big\rangle^2 \, 
\big[J_\text{S},[J_\text{S},\,\cdot\,]\big].
\end{align}
The second term of \eqref{Ddef} is
\begin{align}
&\TrAk \Big(
\big[G_0(H_{\text{SA}}),[G_0(H_{\text{SA}}),
\,\cdot\,\otimes\rho_{\text{A}}]\big]\Big)\\
&\nonumber
=\TrAk \Big(
\big[G_0(g(\xi)J_\text{S}\otimes J_\text{A}),[G_0(g(\xi)J_\text{S}\otimes J_\text{A}),
\,\cdot\,\otimes\rho_{\text{A}}]\big]\Big)\\
&\nonumber
=g_0{}^2\TrAk\Big(
\big[J_\text{S}\otimes J_\text{A},[J_\text{S}\otimes J_\text{A},
\,\cdot\,\otimes\rho_{\text{A}}]\big]\Big)\\
&\nonumber
=g_0{}^2
\text{Tr}_\text{A}\big(J_\text{A}{}^2 \rho_\text{A}\big) \, 
\big[J_\text{S},[J_\text{S},\,\cdot\,]\big]\\
&\nonumber
=g_0{}^2
\big\langle J_\text{A}{}^2\big\rangle \, 
\big[J_\text{S},[J_\text{S},\,\cdot\,]\big].
\end{align}
Putting these together, we get
\begin{align}
\mathcal{D}[\rho_\text{S}]
&=\frac{g_0{}^2}{\hbar^2}
\big(\big\langle J_\text{A}\big\rangle^2-\big\langle J_\text{A}{}^2\big\rangle\big)
\big[J_\text{S},[J_\text{S},\rho_\text{S}]\big]\\
&\nonumber
=-\frac{g_0{}^2}{\hbar^2}
\Delta_{J_\text{A}}^2
\big[J_\text{S},[J_\text{S},\rho_\text{S}]\big],
\end{align}
where 
$\Delta_{J_\text{A}}^2
=\big\langle J_\text{A}{}^2\big\rangle
-\big\langle J_\text{A}\big\rangle^2$. This comfirms the claim in \eqref{Ddefproduct}.

\section{Derivation of Stroboscopic Error Magnitude}\label{Stroboscopic}

In this section we present a derivation of Eq.~\eqref{eq:stroboscopic}. To begin, consider a superoperator-valued function $A(t)$. Using the trace norm and the operator norm it induces throughout, we have for $0 < x < 1$ that
\begin{align}
|| \int_0^1 A(t) dt - & \frac{1}{x} \int_0^x A(t) d t ||\\
&=
|| \big( 1 - \frac{1}{x} \big) 
\int_0^1 A(t) dt + \int_x^1 A(t) dt || \nonumber \\
&\le 
\int_0^1 || A(t) || dt +
\big( \frac{1}{x} - 2 \big)
\int_0^x ||A(t)|| dt \nonumber
\end{align}
using the triangle inequality. Defining the superoperator-valued function
\begin{equation}
G(x) := 
\left(2 - \frac{1}{x} \right)
\int_0^x ||A(t)|| dt
\label{eq:G_def}
\end{equation}
and substituting \eqref{eq:G_def} into the previous inequality yields
\begin{align}
|| \int_0^1 A(t) dt - \frac{1}{x} & \int_0^x A(t) dt ||\\
&\le 
G(1) - G(x)
= G'(\xi) (1 - x),  \nonumber
\end{align}
for some $\xi \in (x, 1)$, where we have assumed $A(t)$ to be continuous in order to use the Mean Value Theorem for $||A(t)||$. Examining the derivative of $G$, we have
\begin{align}
G'(\xi) 
&=
\frac{1}{\xi^2} \int_0^\xi ||A(t)|| dt + 
\left( 2 - \frac{1}{\xi} \right) ||A(\xi)|| \nonumber \\
&\le 
\frac{2}{\xi} \max_{0 < t < 1} ||A (t)||.
\end{align}
It follows that 
\begin{equation}
|| \int_0^1 A(t) dt - \frac{1}{x} \int_0^x A(t) dt ||
\le 
\frac{2(1 - x)}{x}  \max_{0 < t < 1} ||A (t)||.
\label{eq:A_bound}
\end{equation}
To simplify later notation, we define an operator $\Gamma$ which acts on superoperator-valued maps as
\begin{equation}
(\Gamma A)(x) \coloneqq
\int_0^1 A(t) dt - \frac{1}{x} \int_0^x A(t) dt.
\end{equation}
The left-hand side of Eq.~\eqref{eq:A_bound} can now be written in the compact form $|| (\Gamma A)(x)||$.
Let us make some more preliminary observations:
\begin{itemize}
\item It is simple to show that if $||\mathcal{L}|| = -\frac{i}{\hbar} [H, \, \cdot \;]$ then $||\mathcal{L}|| \le \frac{2}{\hbar} ||H||$.
\item Similarly, if $\mathcal{L}$ is a superoperator on S-A and $\mathcal{L}_\text{red} := \text{Tr}_\text{A} [ \mathcal{L} ( \; \cdot \, \otimes \rho_\text{A}) ]$ is a superoperator on S, then $|| \mathcal{L}_\text{red}|| \le || \mathcal{L}||$. This follows immediately from the contractivity of the partial trace under the trace norm \cite{Nielsen:2000}.
\item The operator norm induced by the trace norm is submultiplicative, i.e., $|| \mathcal{L}_1 \mathcal{L}_2 || \le ||\mathcal{L}_1|| \, || \mathcal{L}_2||$ for superoperators $\mathcal{L}_1$ and $\mathcal{L}_2$. This is a general property of induced norms.
\end{itemize}
We now derive the main result concerning stroboscopic error. Using the notation employed in the main text, we have that
\begin{multline}
\rho_\mathrm{S}^{\text{(ex)}} (t_n + \tau)
=
\text{Tr}_\mathrm{A} \left \{
\mathcal{T} e^{\int_0^\tau \mathcal{L}(t') dt'}
\big[ \rho_\mathrm{S}(t_n) \otimes \rho_\mathrm{A}
\big]
\right \},
\label{eq:rho_exact}
\end{multline}
where we've used $\mathcal{L}(t') = -\frac{i}{\hbar} [H_{\delta t} (t'), \, \cdot \;]$, the Liouvillian corresponding to the total S-A Hamiltonian. We decompose this Liouvillian superoperator into free parts on S and A, as well as an interaction term, as $\mathcal{L} (t') = \mathcal{L}_\mathrm{S} + \mathcal{L}_\mathrm{A} + \mathcal{L}_\mathrm{SA} ( t' )$, following Eq.~\eqref{Hamform}. Expanding \eqref{eq:rho_exact} in powers of $\tau$ yields
\begin{multline}
\rho_\mathrm{S}^{\text{(ex)}} (t_n + \tau)
=
\rho_\mathrm{S} (t_n)\\
+ 
\underbrace{
\text{Tr}_\mathrm{A} \left \{
\int_0^\tau dt' \mathcal{L} (t') 
\big[ \rho_\mathrm{S}(t_n) \otimes \rho_\mathrm{A}
\big]
\right \}
}_{\Xi_1 \, \rho_\mathrm{S}(t_n)}
\\
+
\underbrace{
\text{Tr}_\mathrm{A} \left \{
\int_0^\tau dt' \int_0^{t'} dt'' 
\mathcal{L}(t') \mathcal{L}(t'')
\big[ \rho_\mathrm{S}(t_n) \otimes \rho_\mathrm{A}
\big]
\right \}
}_{\Xi_2 \, \rho_\mathrm{S}(t_n)}
+ \mathcal{O}(\tau^3),
\label{eq:rho_ex_series}
\end{multline}
where we define superoperators $\Xi_j$ as indicated in the expression above. On the other hand, the $\mathcal{L}_{\delta t}$-generated dynamics gives
\begin{align} \label{eq:rho_eff_series}
\rho_\mathrm{S}^{\text{(eff)}} (t_n + \tau)
&=
e^{ \mathcal{L}_{\delta t} \tau} \rho_\mathrm{S}(t_n)\\
&=
\Big[ I  + \tau \mathcal{L}_{\delta t}  + \frac{\tau^2}{2} \mathcal{L}_{\delta t}^2 + \mathcal{O}(\tau^3) \Big] \, \rho_\mathrm{S}(t_n). \nonumber
\end{align}.

To characterize the magnitude of stroboscopic error, we subtract Eqs.~\eqref{eq:rho_ex_series} and \eqref{eq:rho_eff_series} and collect powers of $\delta t$ and $\tau$. In particular, we extract the $\tau$ and $\delta t$ -dependence from all terms in the form of a prefactor $\delta t^\alpha \tau^\beta$, and collect terms having the same value of $\alpha + \beta$. The physical justification for this approach comes from the fact that we seek here to describe dynamics on timescales of order $\mathcal{O}(\delta t)$, and so $\tau$ and $\delta t$ must be comparable in magnitude. Concretely, we have that
\begin{multline}
\rho_\mathrm{S}^{\text{(ex)}} (t_n + \tau)
-
\rho_\mathrm{S}^{\text{(eff)}} (t_n + \tau)\\
=
\Big \{ 
\big[ I + \Xi_1 + \Xi_2 + \dots \big] 
-
\big[ I + \tau \mathcal{L}_0 + (\tau \delta t \mathcal{L}_1 + \frac{\tau^2}{2} \mathcal{L}_0^2) + \dots \big] 
\Big \}
\rho_\mathrm{S} (t_n).
\label{eq:stroboscopic_diff}
\end{multline}
We characterize here the stroboscobic error to order $\mathcal{O}(\delta t^2)$. While lengthy, it is straightforward to extend this procedure to higher orders.

Inserting into the expression above the decomposition of $\mathcal{L}$ into system, ancilla and interaction terms, one finds that the leading-order component of \eqref{eq:stroboscopic_diff} goes as
\begin{equation}
\Xi_1 - \tau \mathcal{L}_0
=
-\tau \, \text{Tr}_\mathrm{A} 
\bigg \{
\Big[ (\Gamma \mathcal{L}_\mathrm{SA})(\tau/\delta t)
\Big]
 ( \; \cdot \, \otimes \rho_\mathrm{A} )
\bigg \}.
\label{eq:LO_stroboscopic_prop}
\end{equation}
Using Eq.~\eqref{eq:A_bound} and our other preliminary observations, it follows immediately that the leading-order term, Eq.~\eqref{eq:LO_stroboscopic_prop}, scales as
\begin{equation}
||\Xi_1 - \tau \mathcal{L}_0||
\le 
\frac{4 (\delta t - \tau)}{\hbar} || H_\mathrm{SA}||_\text{max},
\label{eq:leading_strob}
\end{equation}
where $|| H_\mathrm{SA}||_\text{max} := \max_{0 < \tau < \delta t} ||H_\mathrm{SA} (\tau)||$. Similarly, the subleading order terms in \eqref{eq:stroboscopic_diff} go as
\begin{widetext}
\begin{multline}
\label{eq:NLO_stroboscopic_prop}
\Xi_2 - \tau \delta t \mathcal{L}_1 - \frac{\tau^2}{2} \mathcal{L}_0^2
= \frac{\tau}{2} (\delta t - \tau)
\bigg[
\int_0^1 \Big\{ \mathcal{L}_\mathrm{S},  \,  \text{Tr}_\mathrm{A}  \big[ \mathcal{L}_\mathrm{SA}  (\zeta)  ( \; \cdot \, \otimes \rho_\mathrm{A} ) \big]  \Big \} d \zeta 
 +
\Big(\int_0^1  \text{Tr}_\mathrm{A}  \big[ \mathcal{L}_\mathrm{SA}  (\zeta) ( \; \cdot \, \otimes \rho_\mathrm{A} ) \big] d\zeta \Big)^2
\bigg] \\
-
\tau \delta t \, \text{Tr}_\mathrm{A} 
\bigg \{
\Big[ 
(\Gamma A)(\tau / \delta t)
\Big] ( \; \cdot \, \otimes \rho_\mathrm{A} )
\bigg \}, 
\end{multline}
\end{widetext}
where
\begin{multline}
A(\zeta) = \zeta \mathcal{L}_\mathrm{SA}(\zeta) (\mathcal{L}_\mathrm{S} + \mathcal{L}_\mathrm{A} )
+\\
\big[ 
\mathcal{L}_\mathrm{S} + \mathcal{L}_\mathrm{A}
+ \mathcal{L}_\mathrm{SA}(\zeta) 
\big]
\int_0^{\zeta} \mathcal{L}_\mathrm{SA}(\zeta') d\zeta'
\end{multline}
and $\{ \, \cdot\, , \, \cdot \, \}$ denotes an anti-commutator. Thus, subleading order contribution to stroboscopic error is bounded in norm as
\begin{multline}
||\Xi_2 - \tau \delta t \mathcal{L}_1 + \frac{\tau^2}{2} \mathcal{L}_0^2||
\le \frac{ 2 (\delta t - \tau)}{\hbar^2} ||H _\mathrm{SA}||_\text{max}\\
 \times \bigg \{
\tau \Big(
2 ||H_\mathrm{S}|| + ||H _\mathrm{SA}||_\text{max}
\Big)\\
+ 4 \delta t \Big( 2||H_\mathrm{S} || + 2||H_\mathrm{A}|| + ||H _\mathrm{SA}||_\text{max} \Big)
\bigg \}.
\label{eq:subleading_strob}
\end{multline}
One arrives immediately at Eq.~\eqref{eq:stroboscopic} by combining \eqref{eq:leading_strob} and \eqref{eq:subleading_strob}, and maximizing over $\tau \in (0, \delta t)$.

\section{Derivation of Qubit Master Equation}\label{AppQubit}
In this section, we derive the master equation \eqref{QubitRhoEqs} for arbitrary rapid repeated interactions between qubits. Specifically, we consider interactions under the Hamiltonian
\begin{align}\label{QubitHamApp}
H_k(\xi)
&=\hbar\bm{\omega}_\text{S}
\cdot\bm{\sigma}_\text{S}
+\hbar\bm{\sigma}_{\text{A}_k}
\cdot\bm{\omega}_{\text{A}_k}
+\hbar\bm{\sigma}_{\text{A}_k}
\bm{\mathrm{J}}_{k}(\xi)
\bm{\sigma}_\text{S}\\
&\label{QubitHamAppcomp}
=\hbar
\tensor{\{\omega_\text{S}\}}{_\alpha}
\tensor{\{\sigma_\text{S}\}}{^\alpha}
+\hbar
\tensor{\{\sigma_{\text{A}_k}\}}{_\beta}
\tensor{\{\omega_{\text{A}_k}\}}{^\beta}\\
&\nonumber
+\hbar
\tensor{\{\sigma_{\text{A}_k}\}}{_\mu}
\tensor{\{J_k(\xi)\}}{^\mu_\nu}
\tensor{\{\sigma_\text{S}\}}{^\nu}.
\end{align}
In \eqref{QubitHamApp}, the Hamiltonian is written in vector notation where $\bm{\sigma}_\text{S}$ and $\bm{\sigma}_{\text{A}_k}$ are the system and ancillas Pauli vectors respectively, $\bm{\omega}_\text{S}, \ \bm{\omega}_{\text{A}_k}\in\mathbb{R}^3$ are vectors which set the system and ancillas' free Hamiltonians, and $\bm{\mathrm{J}}_k(\xi)$ is a $3\times3$ matrix.

In \eqref{QubitHamAppcomp}, the same Hamiltonian is written in terms of the components of those vectors. The vector labels are written outside the braces using Einstein's summation notation with row vectors having subscripts and column vectors having superscripts. The indices $k$ and $l$ are reserved for the ancillas and are always explicitly summed over. Greek indices are taken to run from 1 to 3.

The ancillas are initial in the state $\rho_{\text{A}_k}$, which we can write in terms of the ancillas Bloch vector, $\bm{R}_k$, as
\be
\rho_{\text{A}_k}
=(\boldsymbol{1}
+\bm{R}_k\cdot\bm{\sigma}_{\text{A}_k})/2
=(\boldsymbol{1}
+\tensor{\delta}{^\alpha^\beta}
\tensor{\{\bm{R}_k\}}{_\alpha}
\tensor{\{\sigma_{\text{A}_k}\}}{_\beta})/2.
\ee 
Note that 
\bel{sigmatoR}
\langle
\tensor{\{\sigma_{\text{A}_k}\}}{_\alpha}
\rangle_k
=\TrAk\!\big(
\tensor{\{\sigma_{\text{A}_k}\}}{_\alpha}  \, \rho_{\text{A}_k}\big)
=\tensor{\{R_k\}}{_\alpha}.
\ee
From \eqref{H0effdef}, the leading order Hamiltonian dynamics is given by $H_\text{eff}^{(0)}
=H_\text{S}+H^{(0)}$.
Trivially, we have $H_\text{S}=\hbar \, \bm{\omega}_\text{S}
\cdot\bm{\sigma}_\text{S}$, and it is straightforward to compute
\begin{align}
H^{(0)}
&=\sum_k p_k \, 
\Big\langle G_0\big(
H_{\text{SA}_k}(\xi)
\big)\Big\rangle_k\\
&\nonumber
=\sum_k p_k \, 
\Big\langle G_0\big(
\hbar
\tensor{\{\sigma_{\text{A}_k}\}}{_\mu}
\tensor{\{J_k(\xi)\}}{^\mu_\nu}
\tensor{\{\sigma_\text{S}\}}{^\nu}
\big)\Big\rangle_k\\
&\nonumber
=\hbar\sum_k p_k \, 
\big\langle \tensor{\{\sigma_{\text{A}_k}\}}{_\mu}\big\rangle_k
G_0\big(\tensor{\{J_k(\xi)\}}{^\mu_\nu}\big)
\tensor{\{\sigma_\text{S}\}}{^\nu}\\
&\nonumber
=\hbar\sum_k p_k \, 
\tensor{\{R_k\}}{_\mu} \, 
G_0\big(\tensor{\{J_k(\xi)\}}{^\mu_\nu}\big)
\tensor{\{\sigma_\text{S}\}}{^\nu}\\
&\nonumber
=\hbar
\tensor{\{\omega^{(0)}\}}{_\nu}
\tensor{\{\sigma_\text{S}\}}{^\nu},
\end{align}
where we have used \eqref{sigmatoR} and defined
\be
\tensor{\{\omega^{(0)}\}}{_\beta}
=\sum_k p_k \, 
\tensor{\{R_k\}}{_\alpha} \, 
G_0\big(\tensor{\{J_k(\xi)\}}{^\alpha_\beta}\big)
\ee
as in \eqref{omega0}. Thus in vector notation we have 
\begin{align}
H_\text{eff}^{(0)}
&\nonumber
=H_\text{S}+H^{(0)}\\
&\nonumber
=\hbar\big(
\bm{\omega}_\text{S}
+\bm{\omega}^{(0)}
\big)\cdot\bm{\sigma}_\text{S}\\
&\nonumber
=\hbar\bm{\omega}_\text{eff}^{(0)}
\cdot\bm{\sigma}_\text{S},
\end{align}
where $\bm{\omega}_\text{eff}^{(0)}
=\bm{\omega}_\text{S}
+\bm{\omega}^{(0)}$.

Next, we compute the subleading Hamiltonian correction, given by \eqref{H1effdef} as $H_\text{eff}^{(1)}
=H_1^{(1)}+H_2^{(1)}+H_3^{(1)}$. Computing this term by term for the qubit-qubit interaction, we find
\begin{align}
\label{H1defApp}
&H_1^{(1)}
=\sum_k p_k \, 
\Big\langle G_1\Big(
\frac{-\ii}{\hbar}
[H_{\text{SA}_k}(\xi),H_\text{S}]
\Big)\Big\rangle_k\\
&\nonumber
=\sum_k p_k \, 
\Big\langle G_1\Big(
-\ii\hbar
\big[
\tensor{\{\sigma_{\text{A}_k}\}}{_\mu}
\tensor{\{J_k(\xi)\}}{^\mu_\nu}
\tensor{\{\sigma_\text{S}\}}{^\nu},
\tensor{\{\omega_\text{S}\}}{_\alpha}
\tensor{\{\sigma_\text{S}\}}{^\alpha}\big]
\Big)\Big\rangle_k\\
&\nonumber
=-\ii\hbar\sum_k p_k \, 
\big\langle
\tensor{\{\sigma_{\text{A}_k}\}}{_\mu}
\big\rangle_k
G_1\big(\tensor{\{J_k(\xi)\}}{^\mu_\nu}\big)
\tensor{\{\omega_\text{S}\}}{_\alpha}
\big[\tensor{\{\sigma_\text{S}\}}{^\nu},
\tensor{\{\sigma_\text{S}\}}{^\alpha}\big]\\
&\nonumber
=2\hbar\sum_k p_k \, 
\tensor{\{R_k\}}{_\mu}
G_1\big(\tensor{\{J_k(\xi)\}}{^\mu_\nu}\big)
\tensor{\{\omega_\text{S}\}}{_\alpha}
\tensor{\varepsilon}{^\nu^\alpha_\beta}
\tensor{\{\sigma_\text{S}\}}{^\beta}\\
&\nonumber
=\hbar
\tensor{\{\omega_1^{(1)}\}}{_\beta}
\tensor{\{\sigma_\text{S}\}}{^\beta},
\end{align}
where we have used $-\ii[\tensor{\{\sigma_\text{S}\}}{^\nu},
\tensor{\{\sigma_\text{S}\}}{^\alpha}]
=2\tensor{\varepsilon}{^\nu^\alpha_\beta}
\tensor{\{\sigma_\text{S}\}}{^\beta}$ and \eqref{sigmatoR}, and defined
\be
\tensor{\{\omega_1^{(1)}\}}{_\beta}
=2\!\sum_k p_k \, 
\tensor{\{R_k\}}{_\mu}
G_1\big(\tensor{\{J_k(\xi)\}}{^\mu_\nu}\big)
\tensor{\{\omega_\text{S}\}}{_\alpha}
\tensor{\varepsilon}{^\nu^\alpha_\beta}
\ee
as in \eqref{omega1}. Next, we compute $H_2^{(1)}$ as
\begin{align}
\label{H2defApp}
&H_2^{(1)}
=\sum_k p_k \ 
\Big\langle G_2\Big(
\frac{-\ii}{\hbar}
[H_{\text{SA}_k}(\xi),H_{\text{A}_k}]
\Big)\Big\rangle_k\\
&\nonumber
=\sum_k p_k
\Big\langle G_2\Big(\!\!
-\!\ii\hbar
\big[
\tensor{\{\sigma_{\text{A}_k}\}}{_\mu}
\tensor{\{J_k(\xi)\}}{^\mu_\nu}
\tensor{\{\sigma_\text{S}\}}{^\nu}\!\!,
\tensor{\{\sigma_{\text{A}_k}\}}{_\beta}
\tensor{\{\omega_{\text{A}_k}\}}{^\beta}
\big]\Big)\Big\rangle_k\\
&\nonumber
=-\ii\hbar\!\sum_k p_k
\Big\langle\big[
\tensor{\{\sigma_{\text{A}_k}\}}{_\mu},
\tensor{\{\sigma_{\text{A}_k}\}}{_\beta}
\big]\Big\rangle_k\!
\tensor{\{\omega_{\text{A}_k}\}}{^\beta}
G_2\big(\tensor{\{J_k(\xi)\}}{^\mu_\nu}\big)
\tensor{\{\sigma_\text{S}\}}{^\nu}\\
&\nonumber
=2\hbar\sum_k p_k
\tensor{\varepsilon}{_\mu_\beta^\alpha}
\big\langle
\tensor{\{\sigma_{\text{A}_k}\}}{_\alpha}
\big\rangle_k
\tensor{\{\omega_{\text{A}_k}\}}{^\beta}
G_2\big(\tensor{\{J_k(\xi)\}}{^\mu_\nu}\big)
\tensor{\{\sigma_\text{S}\}}{^\nu}\\
&\nonumber
=2\hbar\sum_k p_k
\tensor{\varepsilon}{_\mu_\beta^\alpha}
\tensor{\{R_k\}}{_\alpha}
\tensor{\{\omega_{\text{A}_k}\}}{^\beta}
G_2\big(\tensor{\{J_k(\xi)\}}{^\mu_\nu}\big)
\tensor{\{\sigma_\text{S}\}}{^\nu}\\
&\nonumber
=\hbar
\tensor{\{\omega_2^{(1)}\}}{_\nu}
\tensor{\{\sigma_\text{S}\}}{^\nu},
\end{align}
where we have used 
\be
-\ii\Big\langle\big[
\tensor{\{\sigma_{\text{A}_k}\}}{_\mu},
\tensor{\{\sigma_{\text{A}_k}\}}{_\beta}
\big]\Big\rangle_k
=2\tensor{\varepsilon}{_\mu_\beta^\alpha}
\big\langle
\tensor{\{\sigma_{\text{A}_k}\}}{_\alpha}
\big\rangle_k
\ee and \eqref{sigmatoR}, and defined
\be
\tensor{\{\omega_2^{(1)}\}}{_\nu}
=2\sum_k p_k
\tensor{\varepsilon}{_\mu_\beta^\alpha}
\tensor{\{R_k\}}{_\alpha}
\tensor{\{\omega_{\text{A}_k}\}}{^\beta}
G_2\big(\tensor{\{J_k(\xi)\}}{^\mu_\nu}\big).
\ee
Finally, we compute $H_3^{(1)}$. For this calculation we temporarily drop the braces enclosing Pauli matrices, which separate the Hilbert space labels from the vector component labels. We have
\begin{align}
&H_3^{(1)}
=\sum_k p_k \ 
\Big\langle G_3\Big(
\frac{-\ii}{\hbar}
[H_{\text{SA}_k}(\xi_1),H_{\text{SA}_k}(\xi_2)]
\Big)\Big\rangle_k\\
&\nonumber
\!=\!
\sum_k p_k
\Big\langle G_3\Big(
\!\!-\!\ii\hbar
\big[
\tensor{\sigma_{\text{A}_k}{}}{_\mu}
\tensor{\{J_k(\xi_1)\}}{^\mu_\nu}
\tensor{\sigma_\text{S}{}}{^\nu}\!\!\!,
\tensor{\sigma_{\text{A}_k}{}}{_\alpha}
\tensor{\{J_k(\xi_2)\}}{^\alpha_\beta}
\tensor{\sigma_\text{S}{}}{^\beta}
\big]\Big)\Big\rangle_k\\
&\nonumber
\!=\!\hbar\!\sum_k p_k
G_3\Big(\!
\tensor{\{J_k(\xi_1)\}}{^\mu_\nu}
\tensor{\{J_k(\xi_2)\}}{^\alpha_\beta}\!
\Big)
\Big\langle \!\!-\!\ii
\big[\tensor{\sigma_{\text{A}_k}{}}{_\mu}
\tensor{\sigma_\text{S}{}}{^\nu}\!\!,
\tensor{\sigma_{\text{A}_k}{}}{_\alpha}
\tensor{\sigma_\text{S}{}}{^\beta}
\big]\Big\rangle_k.
\end{align}
To proceed with the calculation we make use of the identity
\begin{align}
\label{eq:ptrace_identity}
&\text{Tr}_\text{A}\big([A\otimes B,C\otimes D] \, (\boldsymbol{1}\otimes\rho_A)\big)\\
&\nonumber
=\frac{1}{2}\text{Tr}_\text{A}\big(
\{B,D\}\rho_A\big)[A,C]
+\frac{1}{2}\text{Tr}_\text{A}\big([B,D]\rho_A\big)\{A,C\}.
\end{align}
Using \eqref{eq:ptrace_identity}, we have
\begin{align}
\Big\langle& -\ii
\big[\tensor{\sigma_{\text{A}_k}{}}{_\mu}
\tensor{\sigma_\text{S}{}}{^\nu},
\tensor{\sigma_{\text{A}_k}{}}{_\alpha}
\tensor{\sigma_\text{S}{}}{^\beta}
\big]\Big\rangle_k\\
&\nonumber
=-\ii \, \TrAk\Big( \big[\tensor{\sigma_{\text{A}_k}{}}{_\mu}
\tensor{\sigma_\text{S}{}}{^\nu},
\tensor{\sigma_{\text{A}_k}{}}{_\alpha}
\tensor{\sigma_\text{S}{}}{^\beta}
\big]\rho_{\text{A}_k}\Big)\\
&\nonumber
=\frac{1}{2}\TrAk\Big( -\ii\big[\tensor{\sigma_{\text{A}_k}{}}{_\mu},
\tensor{\sigma_{\text{A}_k}{}}{_\alpha}
\big]\rho_{\text{A}_k}\Big)
\{\tensor{\sigma_\text{S}{}}{^\nu},\tensor{\sigma_\text{S}{}}{^\beta}\}\\
&\nonumber
+\frac{1}{2}\TrAk\Big( \{\tensor{\sigma_{\text{A}_k}{}}{_\mu},
\tensor{\sigma_{\text{A}_k}{}}{_\alpha}
\}\rho_{\text{A}_k}\Big)
\big(-\ii[\tensor{\sigma_\text{S}{}}{^\nu},\tensor{\sigma_\text{S}{}}{^\beta}]\big)\\
&\nonumber
=\TrAk\Big(2 
\tensor{\varepsilon}{^\gamma_\mu_\alpha}
\tensor{\sigma_{\text{A}_k}{}}{_\gamma}
\rho_{\text{A}_k}\Big) \ 
\tensor{\delta}{^\nu^\beta}\boldsymbol{1}
+\TrAk\Big(2 
\tensor{\delta}{_\alpha_\mu}\rho_{\text{A}_k}\Big) \ 
\tensor{\varepsilon}{_\eta^\nu^\beta}
\tensor{\sigma_\text{S}{}}{^\eta}
\\
&\nonumber
=2\tensor{\varepsilon}{^\gamma_\mu_\alpha} \, 
\tensor{\{R\}}{_\gamma} \, 
\tensor{\delta}{^\nu^\beta}\boldsymbol{1}
+2\tensor{\delta}{_\alpha_\mu}
\tensor{\varepsilon}{_\eta^\nu^\beta}
\tensor{\sigma_\text{S}{}}{^\eta},
\end{align}
where we have made use of 
\be
-\ii[\tensor{\sigma_\text{S}{}}{^\alpha},
\tensor{\sigma_\text{S}{}}{^\beta}]
=2\tensor{\varepsilon}{^\gamma^\alpha_\beta}
\tensor{\sigma_\text{S}{}}{^\gamma}
\ee 
and 
\be
\{\tensor{\sigma_\text{S}{}}{^\alpha},
\tensor{\sigma_\text{S}{}}{^\beta}\}
=2\tensor{\delta}{^\alpha^\beta}\boldsymbol{1},
\ee
as well as \eqref{sigmatoR} and the fact that $\text{Tr}_{\text{A}_k}(\rho_{\text{A}_k})=1$.

The term proportional to the identity does not contribute to the dynamics and can be dropped. Thus, reintroducing braces, we have
\begin{align}
H_3^{(1)}
=&\nonumber
2\hbar\sum_k p_k \ 
G_3\Big(
\tensor{\{J_k(\xi_1)\}}{^\mu_\nu}
\tensor{\{J_k(\xi_2)\}}{^\alpha_\beta}
\Big)
\tensor{\delta}{_\alpha_\mu}
\tensor{\varepsilon}{_\eta^\nu^\beta}
\tensor{\{\sigma_\text{S}\}}{^\eta}\\
=&\hbar
\tensor{\{\omega_3^{(1)}\}}{_\eta}
\tensor{\{\sigma_\text{S}\}}{^\eta},
\end{align}
where we define
\be
\tensor{\{\omega_3^{(1)}\}}{_\eta}
=2\sum_k p_k \ 
G_3\Big(
\tensor{\{J_k(\xi_1)\}}{^\mu_\nu}
\tensor{\{J_k(\xi_2)\}}{^\alpha_\beta}
\Big)
\tensor{\delta}{_\alpha_\mu}
\tensor{\varepsilon}{_\eta^\nu^\beta}
\ee
as in \eqref{omega3}. We have derived the expressions claimed for the subleading order Hamiltonian dynamics in \eqref{omega1}, \eqref{omega2}, and \eqref{omega3}.

Finally, we calculate the dissipative part of the dynamics. From \eqref{Ddef}, this is
\begin{align}
&\mathcal{D}[\,\cdot\,]
=\frac{1}{\hbar^2}
\big[H^{(0)},[H^{(0)},\,\cdot\,]\big]\\
\nonumber
&-\frac{1}{\hbar^2}\sum_k p_k \, \TrAk \Big(
\big[G_0(H_{\text{SA}_k}),[G_0(H_{\text{SA}_k}),
\,\cdot\,\otimes\rho_{\text{A}_k}]\big]\Big).
\end{align}
The $H^{(0)}$ term is easily calculated as
\begin{align}
&\frac{1}{\hbar^2}\big[H^{(0)},[H^{(0)},\rho_S(t)]\big]\\
&\nonumber
=\frac{1}{\hbar^2}
\big[\hbar \tensor{\{\omega^{(0)}\}}{_\alpha}
\tensor{\{\sigma_\text{S}\}}{^\alpha},
[\hbar \tensor{\{\omega^{(0)}\}}{_\beta}
\tensor{\{\sigma_\text{S}\}}{^\beta},
\rho_S(t)]\big]\\
&\nonumber
=\tensor{\{\omega^{(0)}\}}{_\alpha}
\tensor{\{\omega^{(0)}\}}{_\beta} \ 
\big[\tensor{\{\sigma_\text{S}\}}{^\alpha},
[\tensor{\{\sigma_\text{S}\}}{^\beta},
\rho_S(t)]\big]\\
&\nonumber
=\sum_k p_k
\tensor{\{R_k\}}{_\mu}
G_0\big(\tensor{\{J_k(\xi)\}}{^\mu_\alpha}\big)
\sum_l p_l
\tensor{\{R_l\}}{_\nu}
G_0\big(\tensor{\{J_l(\xi)\}}{^\nu_\beta}\big)\\
&\nonumber
\big[\tensor{\{\sigma_\text{S}\}}{^\alpha},
[\tensor{\{\sigma_\text{S}\}}{^\beta},
\rho_S(t)]\big]\\
&\nonumber
=\sum_{k,l}
G_0\big(\tensor{\{J_k(\xi)\}}{^\mu_\alpha}\big)
G_0\big(\tensor{\{J_l(\xi)\}}{^\nu_\beta}\big)\\
&\nonumber
p_k p_l
\tensor{\{R_k\}}{_\mu}
\tensor{\{R_l\}}{_\nu}
\big[\tensor{\{\sigma_\text{S}\}}{^\alpha},
[\tensor{\{\sigma_\text{S}\}}{^\beta},
\rho_S(t)]\big].
\end{align}
Next, we compute the trace term as 
\begin{align}
\sum_k p_k\frac{1}{\hbar^2}&\TrAk\!\Big(
\big[G_0(H_{SA_k}),
[G_0(H_{SA_k}),
\rho_S\otimes\rho_{A_k}]\big]\Big)\\
=&\nonumber
\sum_k p_k\TrAk\!\Big(
\big[G_0(\tensor{\sigma_{\text{A}_k}{}}{_\mu}
\tensor{\{J_k(\xi)\}}{^\mu_\nu}
\tensor{\sigma_\text{S}{}}{^\nu}),\\
&\nonumber
[G_0(\tensor{\sigma_{\text{A}_k}{}}{_\beta}
\tensor{\{J_k(\xi)\}}{^\beta_\alpha}
\tensor{\sigma_\text{S}{}}{^\alpha}),
\rho_S\otimes\rho_{A_k}]\big]\Big)\\
=&\nonumber
\sum_k p_k G_0\big(\tensor{\{J_k(\xi)\}}{^\mu_\nu}\big)
G_0\big(\tensor{\{J_k(\xi)\}}{^\beta_\alpha}\big)\\
&\nonumber
\TrAk\!\Big(
\big[\tensor{\sigma_{\text{A}_k}{}}{_\mu} \, 
\tensor{\sigma_\text{S}{}}{^\nu},
[\tensor{\sigma_{\text{A}_k}{}}{_\beta} \, 
\tensor{\sigma_\text{S}{}}{^\alpha},
\rho_\text{S}\otimes\rho_{\text{A}_k}]\big]\Big).
\end{align}
We then make use of the identity
\begin{align}
&\text{Tr}_\text{A}\Big(
\big[A\otimes B,
[C\otimes D,
(\rho_\text{S}\otimes\rho_\text{A})]
\big]\Big)\\
&\nonumber
=\frac{1}{2}\text{Tr}_\text{A}\Big(\{B,D\}\rho_A\Big)\big[A,[C,\rho_S]\big]\\
&\nonumber
+\frac{1}{2}\text{Tr}_\text{A}\Big([B,D]\rho_A\Big)\big[A,\{C,\rho_S\}\big].
\end{align}
Temporarily suppressing the Hilbert space labels again, the above identity yields
\begin{align}
&\text{Tr}_\text{A}\Big(
\big[\tensor{\sigma}{^a}\otimes 
\tensor{\sigma}{_b},
[\tensor{\sigma}{^c}\otimes 
\tensor{\sigma}{_d},
(\rho_\text{S}\otimes\rho_\text{A})]\big]\Big)\\
&\nonumber
=\frac{1}{2}\text{Tr}_\text{A}\big([
\tensor{\sigma}{_b},
\tensor{\sigma}{_d}
]\rho_\text{A}\big) \, 
\big[\tensor{\sigma}{^a},
\{\tensor{\sigma}{^c},\rho_\text{S}\}\big]\\
&\nonumber
+\frac{1}{2}\text{Tr}_\text{A}\big(\{
\tensor{\sigma}{_b},
\tensor{\sigma}{_d}
\}\rho_\text{A}\big) 
\big[\tensor{\sigma}{^a},
[\tensor{\sigma}{^c},\rho_\text{S}]\big]\\
&\nonumber
=\text{Tr}_\text{A}\big(\ii
\tensor{\varepsilon}{_b_d^n}
\tensor{\sigma}{_n}\rho_\text{A}\big) \, 
\big[\tensor{\sigma}{^a},
\{\tensor{\sigma}{^c},
\rho_\text{S}\}\big]\\
&\nonumber
+\text{Tr}_\text{A}(
\tensor{\delta}{_b_d}
\boldsymbol{1}\rho_\text{A}) \, 
\big[\tensor{\sigma}{^a},
[\tensor{\sigma}{^c},
\rho_\text{A}]\big]\\
&\nonumber
=\ii\tensor{\varepsilon}{_b_d^n} \, 
\tensor{\{R_k\}}{_n} 
[\tensor{\sigma}{^a},
\{\tensor{\sigma}{^c},
\rho_\text{S}\}]
+\tensor{\delta}{_b_d} \, \big[\tensor{\sigma}{^a},
[\tensor{\sigma}{^c},
\rho_\text{S}]\big]\\
&\nonumber
=2\ii\tensor{\varepsilon}{_b_d^n} \, 
\tensor{\{R_k\}}{_n} 
[\tensor{\sigma}{^a},
\tensor{\sigma}{^c}]
\text{Tr}_\text{S}(\rho_\text{S})
+\tensor{\delta}{_b_d} \, 
\big[\tensor{\sigma}{^a},
[\tensor{\sigma}{^c},
\rho_\text{S}]\big]\\
&\nonumber
=-2\tensor{\varepsilon}{_b_d^n} \, 
\tensor{\{R_k\}}{_n} 
\tensor{\varepsilon}{^a^c_m}
\tensor{\sigma}{^m}
+\tensor{\delta}{_b_d} \, \big[\tensor{\sigma}{^a},
[\tensor{\sigma}{^c},
\rho_\text{S}]\big],
\end{align}
where in the second-to-last step we have used the equality 
$[\tensor{\sigma}{^a},
\{\tensor{\sigma}{^c},X\}]
=[\tensor{\sigma}{^a},
\tensor{\sigma}{^c}]\ 
\text{Tr}(X)$ for any $2 \times 2$ Hermitian matrix $X$, which is easily verified: Taking $X=\tensor{\sigma}{^m}$ we find,
\be
[\tensor{\sigma}{^a}\!,
\{\tensor{\sigma}{^c},
\tensor{\sigma}{^m}\}]
=[\tensor{\sigma}{^a}\!,
\tensor{\delta}{^c^m}
\boldsymbol{1}]
=0
=[\tensor{\sigma}{^a}\!,
\tensor{\sigma}{^c}] 
\text{Tr}(\tensor{\sigma}{^m}).
\ee
Taking $X=\boldsymbol{1}$ we find
\be
[\sigma^a,\{\sigma^c,\boldsymbol{1}\}]
=2[\sigma^a,\sigma^c]
=[\tensor{\sigma}{^a},
\tensor{\sigma}{^c}]
\text{Tr}(\boldsymbol{1}).
\ee
The original claim follows from linearity.

Combining these partial results, we have for the trace term
\begin{align}
&\sum_k p_k G_0\big(\tensor{\{J_k(\xi)\}}{^\mu_\nu}\big)
G_0\big(\tensor{\{J_k(\xi)\}}{^\beta_\alpha}\big)\\
&\nonumber
\TrAk\!\Big(
\big[\tensor{\sigma_{\text{A}_k}{}}{_\mu} \, 
\tensor{\sigma_\text{S}{}}{^\nu},
[\tensor{\sigma_{\text{A}_k}{}}{_\beta} \, 
\tensor{\sigma_\text{S}{}}{^\alpha},
\rho_S\otimes\rho_{A_k}]\big]\Big)\\
&\nonumber
=\sum_k p_k G_0\big(\tensor{\{J_k(\xi)\}}{^\mu_\nu}\big)
G_0\big(\tensor{\{J_k(\xi)\}}{^\beta_\alpha}\big)\\
&\nonumber
\Big(-2\epsilon_{\mu \beta}{}^n \, 
\tensor{\{R_k\}}{_n} 
\epsilon^{\nu \alpha}{}_m 
\tensor{\sigma_\text{S}{}}{^m}
+\delta_{\mu \beta} \, \big[\tensor{\sigma_\text{S}{}}{^\nu},
[\tensor{\sigma_\text{S}{}}{^\alpha},
\rho_S]\big]\Big)\\
&\nonumber
=-2\bm{D}_0\cdot\bm{\sigma}_\text{S}\\
&\nonumber
+\sum_k p_k G_0\big(\tensor{\{J_k(\xi)\}}{^\mu_\nu}\big)
G_0\big(\tensor{\{J_k(\xi)\}}{^\beta_\alpha}\big)
\delta_{\mu \beta} \, \big[\tensor{\sigma_\text{S}{}}{^\nu},
[\tensor{\sigma_\text{S}{}}{^\alpha},
\rho_S]\big],
\end{align}
where 
\be
\tensor{\bm{D}_0{}}{_\eta}
\!=\!\!\sum_k p_k \, 
\tensor{\varepsilon}{^\mu^\nu_\eta}\!
\tensor{\{R_k\}}{_\gamma}
\tensor{\varepsilon}{^\gamma_\alpha_\beta}
G_0\Big(\!\tensor{\{J_k(\xi)\}}{^\alpha_\mu}\Big)
G_0\Big(\!\tensor{\{J_k(\xi)\}}{^\beta_\nu}\Big)
\ee
as in \eqref{qubitD0}.

Finally, we combine this with the  $H^{(0)}$ term to find
\begin{align}
\mathcal{D}[\,\cdot\,]
&=2\bm{D}_0\cdot\bm{\sigma}_\text{S}\\
&\nonumber
+\sum_{k,l}
G_0\big(\tensor{\{J_k(\xi)\}}{^\mu_\alpha}\big)
G_0\big(\tensor{\{J_l(\xi)\}}{^\nu_\beta}\big)\\
&\nonumber
\big(p_k\delta_{kl}\delta_{\mu\nu}
-p_k p_l
\tensor{\{R_k\}}{_\mu}
\tensor{\{R_l\}}{_\nu}
\big)
\big[\tensor{\{\sigma_\text{S}\}}{^\alpha},
[\tensor{\{\sigma_\text{S}\}}{^\beta},
\rho_S(t)]\big]\\
&\nonumber
=2\bm{D}_0\cdot\bm{\sigma}_\text{S}
+\tensor{\bm{\mathrm{D}}_{1}{}}{_\alpha_\beta}
\big[\tensor{\{\sigma_\text{S}\}}{^\alpha},
[\tensor{\{\sigma_\text{S}\}}{^\beta},
\rho_S(t)]\big],
\end{align}
where
\begin{align}
\tensor{\bm{\mathrm{D}}_{1}{}}{_\mu_\nu}
=\sum_{k,l} \Big(&p_k \, 
\tensor{\delta}{_k_l} \, 
\tensor{\delta}{_\alpha_\beta} \, 
-p_k \, p_l \, 
\tensor{\{R_k\}}{_\alpha} \, 
\tensor{\{R_l\}}{_\beta}\Big)\\
\nonumber
&\times G_0\Big(\tensor{\{J_k(\xi)\}}{^\alpha_\mu}\Big)
G_0\Big(\tensor{\{J_l(\xi)\}}{^\beta_\nu}\Big)
\end{align}
as in \eqref{qubitD1}. Thus we have confirmed all the terms in \eqref{QubitRhoEqs}, yielding the master equation
\begin{align}
\frac{\text{d}}{\text{dt}}\rho_\text{S}(t) 
&=-\ii \, \big[
\big(\bm{\omega}_\text{eff}^{(0)}
+\delta t \ \bm{\omega}_\text{eff}^{(1)}\big)
\bm{\sigma}_\text{S},
\rho_\text{S}(t)\big]\\
\nonumber
&+\delta t \, \bm{D}_0\bm{\sigma}_\text{S}
-\frac{\delta t}{2}
\tensor{D_{1}{}}{_\mu_\nu} \, 
\big[\sigma_\text{S}{}^\mu,
[\sigma_\text{S}{}^\nu,
\rho_\text{S}]\big].
\end{align}

Next, we translate the qubit master equation \eqref{QubitRhoEqs} in terms of the system's Bloch vector
\be
\bm{a}=\text{Tr}_\text{S}\big(\rho_\text{S}\bm{\sigma}_\text{S}\big).
\ee
The dynamics of the system's Bloch vector is described by
\be
\bm{a}'(t)
=\text{Tr}_\text{S}\big(\bm{\sigma}_\text{S}\rho_\text{S}'(t)\big),
\ee
which we compute term by term beginning with unitary part:
\begin{align}
&\text{Tr}_\text{S}\Big(
-\ii \, \big[
\bm{\omega}_\text{eff}
\cdot\bm{\sigma}_\text{S},
\rho_\text{S}(t)\big]
\bm{\sigma}_\text{S}\Big)\\
&\nonumber
=\frac{1}{2}\text{Tr}_\text{S}\Big(
-\ii \, \big[
\bm{\omega}_\text{eff}
\cdot\bm{\sigma}_\text{S},
\boldsymbol{1}+\bm{a}(t)\cdot\bm{\sigma}_\text{S}\big]
\bm{\sigma}_\text{S}\Big)\\
&\nonumber
=\frac{1}{2}\text{Tr}_\text{S}\Big(
-\ii \, \big[
\bm{\omega}_\text{eff}
\cdot\bm{\sigma}_\text{S},
\bm{a}(t)\cdot\bm{\sigma}_\text{S}\big]
\bm{\sigma}_\text{S}\Big)\\
&\nonumber
=\text{Tr}_\text{S}\Big(
\big((\bm{\omega}_\text{eff}\times
\bm{a}(t))\cdot\bm{\sigma}_\text{S}\big)
\bm{\sigma}_\text{S}\Big)\\
&\nonumber
=2\bm{\omega}_\text{eff}\times
\bm{a}(t).
\end{align}
Next, we calculate the effect of the affine part of the dynamics on the system Bloch vector:
\begin{align}
&\text{Tr}_\text{S}\Big(
\bm{D}_0\cdot
\bm{\sigma}_\text{S}\Big)
=\bm{D}_0{}^T
=\bm{b}.
\end{align}
where $\bm{b}=\bm{D}_0{}^T$. 

Finally, the last part of the dynamics gives
\begin{align}
&\text{Tr}_\text{S}\Big(
\tensor{D_{1}{}}{_\mu_\nu} \, 
\big[\sigma_\text{S}{}^\mu,
[\sigma_\text{S}{}^\nu,
\rho_\text{S}]\big]
\bm{\sigma}_\text{S}\Big)\\
&\nonumber
=\frac{1}{2}\text{Tr}_\text{S}\Big(
\tensor{D_{1}{}}{_\mu_\nu} \, 
\big[\sigma_\text{S}{}^\mu,
[\sigma_\text{S}{}^\nu,
\boldsymbol{1}+\bm{a}(t)\cdot\bm{\sigma}_\text{S}]\big]
\bm{\sigma}_\text{S}\Big)\\
&\nonumber
=\frac{1}{2}\tensor{D_{1}{}}{_\mu_\nu} \, 
\text{Tr}_\text{S}\Big(
\big[\sigma_\text{S}{}^\mu,
[\sigma_\text{S}{}^\nu,
\bm{\sigma}_\text{S}]\big]
\bm{\sigma}_\text{S}\Big)
\bm{a}(t)\\
&\nonumber
=2\tensor{D_1{}}{_\mu_\nu}
\tensor{\varepsilon}{^\beta^\mu_\alpha}
\tensor{\varepsilon}{^\nu^\alpha_\gamma}
\tensor{\bm{a}(t)}{^\gamma}\\
&\nonumber
=2\tensor{B}{^\beta_\gamma}
\tensor{\bm{a}(t)}{^\gamma}.
\end{align}
where $\tensor{B}{^\beta_\gamma}
=\tensor{D_1{}}{_\mu_\nu}
\tensor{\varepsilon}{^\beta^\mu_\alpha}
\tensor{\varepsilon}{^\nu^\alpha_\gamma}$.
Putting all of these together gives the Bloch dynamics \eqref{BlochDynamics}.
\bibliography{references}
\end{document}